\documentclass[format=acmsmall, review=false, screen=true]{acmart}

\usepackage{graphicx}
\usepackage{array}
\usepackage{url}
\usepackage{fancybox}
\usepackage{multirow}
\usepackage{amssymb}
\usepackage{color}

\usepackage{colortbl}
\usepackage{balance}
\usepackage{fancyhdr}

\usepackage{subfig}
\usepackage{diagbox}
\usepackage{bbding}
\usepackage{placeins}

\usepackage{booktabs}
\usepackage{enumitem}
\usepackage{xcolor,lipsum}
\usepackage[linesnumbered, ruled,vlined]{algorithm2e}

\newcommand{\todo}[1]{}
\renewcommand{\todo}[1]{{\color{red} TODO: {#1}}}

\definecolor{light-gray}{rgb}{.906,  .902,  .902}

\newcommand{\rev}[1]{\textcolor{black}{#1}}
\newcommand{\revv}[1]{\textcolor{black}{#1}}
\newcommand{\minor}[1]{\textcolor{black}{#1}}

\acmJournal{TOSEM}
\acmVolume{9}
\acmNumber{4}
\acmArticle{39}
\acmYear{2019}
\acmMonth{3}
\copyrightyear{2009}
\acmArticleSeq{9}

\setcopyright{acmlicensed}

\acmDOI{0000001.0000001}

\received{Mar 2019}
\received[revised]{?? 2019}
\received[accepted]{?? 2019}

\usepackage{bm}

\begin{document}
\title[]{I Know What You Are Searching For: Code Snippet Recommendation from Stack Overflow Posts}

\author{Zhipeng GAO}
\affiliation{%
  \institution{Shanghai Institute for Advanced Study of Zhejiang University}
  \city{Shanghai,}
  \country{China}
  }
\email{zhipeng.gao@zju.edu.cn}

\author{Xin Xia}
\authornote{This is the corresponding author}
\affiliation{%
  \institution{Huawei}
  \city{Hangzhou,}
  \country{China}
  }
\email{xin.xia@acm.org}

\author{David Lo}
\affiliation{%
  \institution{Singapore Management University}
  \city{Singapore,}
  \country{Singapore}
  }
\email{davidlo@smu.edu.sg}

\author{John Grundy}
\affiliation{%
  \institution{Monash University}
  \city{Melbourne,}
  \state{VIC}
  \postcode{3168}
  \country{Australia}
  }
\email{john.grundy@monash.edu}

\author{Xindong Zhang}
\affiliation{
    \institution{Alibaba Group}
    \city{Hangzhou,}
    \country{China}
}
\email{zxd139923@alibaba-inc.com}

\author{Zhenchang Xing}
\affiliation{%
  \institution{CSIRO's Data61 \& Australian National University}
  \city{College of Engineering and Computer Science, Canberra}
  \country{Australia}
  }
\email{Zhenchang.Xing@anu.edu.au}

\begin{abstract}
Stack Overflow has been heavily used by software developers to seek programming-related information.
More and more developers use Community Question and Answer forums, such as Stack Overflow, to search for code examples of how to accomplish a certain coding task. This is often considered to be more efficient than working from source documentation, tutorials or full worked examples.
However, due to the complexity of these online Question and Answer forums and the very large volume of information they contain, developers can be overwhelmed by the sheer volume of available information. This makes it hard to find and/or even be aware of the most relevant code examples to meet their needs.  
To alleviate this issue, in this work we present a query-driven code recommendation tool, named {\sc Que2Code}, that identifies the best code snippets for a user query from Stack Overflow posts. 
Our approach has two main stages: (i) semantically-equivalent question retrieval and (ii) best code snippet recommendation. 
During the first stage, for a given query question formulated by a developer, we first generate paraphrase questions for the input query as a way of query boosting, and then retrieve the relevant Stack Overflow posted questions based on these generated questions. 
In the second stage, we collect all of the code snippets within questions retrieved in the first stage and develop a novel scheme to rank code snippet candidates from Stack Overflow posts via pairwise comparisons. 
To evaluate the performance of our proposed model, we conduct a large scale experiment to evaluate the effectiveness of the semantically-equivalent question retrieval task and best code snippet recommendation task separately on Python and Java datasets in Stack Overflow. 
We also perform a human study to measure how real-world developers perceive the results generated by our model.
Both the automatic and human evaluation results demonstrate the promising performance of our model, and we have released our code and data to assist other researchers.

\end{abstract}

%
%

%
%


\maketitle

\renewcommand{\shortauthors}{Zhipeng GAO et al.}
\renewcommand{\shorttitle}{Code Snippet Recommendation from Stack Overflow Posts}

\section{Introduction}
\label{sec:intro}
To deliver high-quality software more effectively and efficiently, many developers frequently use community Question and Answer (Q\&A) sites, such as Stack Overflow, for solutions to their programming problems and tasks~\cite{xu2017answerbot}. 
Code search is one such task that plays an important role in software development. 
Software developers spend about 19\% of their development time searching for relevant code snippets on the web~\cite{brandt2009two}. 
To help in programming problem solving, searching for useful code snippets from Stack Overflow has become a common part of developer's daily work~\cite{xu2017answerbot}.
Typically, when developers encounter a technical problem, they formulate their problem as a query and use a search engine (e.g. Google or Stack Overflow's own search tool) to obtain a list of possible relevant posts that may contain useful solutions to their problem. After that, developers have to read answers with various levels of quality included in the returned posts to identify possible solutions. 

\revv{
While search engines such as Google are widely used for searching for required information in Stack Overflow, there are a significant number of user queries that can not be satisfied~\cite{gao2020deepans}. One primary reason for failed Stack Overflow searches are poorly constructed queries and/or low quality queries~\cite{rahman2018evaluating, xia2017developers, gao2020generating}.
Unsuccessful searches using low quality queries are difficult and ineffective due to the following challenges: }
\begin{itemize}
    \item \revv{ 
    {\em Query Mismatch}. 
    Low quality queries suffer from the \emph{query mismatch problem}.
    The query mismatch problem refers to different user expressions of the same problem.
    Because different developers often describe their problems in their own way, the queries formulated by different developers may be semantically related but expressed in a variety of different ways. 
    According to our empirical study, around 80\% of the lexical terms are expressed differently regarding a duplicate question pair. 
    Considering a general search engine heavily relies on whether similar posts exist and how similar the user search queries are, 
    \textbf{the query mismatch problem greatly hinders the performance of the standard information retrieval tools for searching relevant questions}.
    Therefore, there is a need by developers to find semantically relevant questions for a wide variety of user query expressions.
    }
    \item \revv{
    {\em Information Overload}. 
    Low quality queries can lead to an \emph{information overload problem}~\cite{xu2017answerbot, xia2017developers}. 
    The information overload problem refers to the huge amount of search results returned by a general search engine, where the expected result can be easily buried in. 
    Searching Stack Overflow using low quality queries often brings in a large number of irrelevant results and developers can easily get lost in the massive amount of contents and thus fail to locate their desired result. 
    Xu et al.~\cite{xu2017answerbot} conducted a survey with 72 developers, according to their interview and survey-based study, there is too much noisy and redundant information online, and developers often wasted their time on reading irrelevant posts, which was really time-consuming. 
    Similarly, Xia et al.~\cite{xia2017developers} surveyed 235 developers from 21 countries, they found it is hard for developers to get the desired solutions for the search tasks due to too many noisy results, and the best solution for a problem is often ranked low in the results of search engines. 
    Therefore, there is a need by developers to \textbf{receive the most suitable code solutions for their current programming tasks high up in their search results list}. 
    }
\end{itemize}

\revv{
We focus on Stack Overflow in this work due to the large number of posts in Stack Overflow that provide variable user descriptions about different types of problems. 
Moreover, a tremendous number of reusable code snippets archived in Stack Overflow enable information seekers to directly get solutions from the repositories. 
In this study, we aim not to replace Google searches for Stack Overflow solutions, but to help developers to search for the best code fragments in Stack Overflow that more closely match diverse developer user queries. 
}

We formulate this task as a \emph{query-driven code recommendation} task for a given input question to Stack Overflow. 
Given an input question, instead of naively choosing the answers from relevant questions, we present a novel model and tool, named {\sc Que2Code}, to achieve this goal of searching for the best code snippet to answer a user query.  
We use a two-stage model: in the first stage, we use a query rewriter to tackle the {\em query mismatch} challenge. 
The idea is to use rewritten version of a query question to cover different forms of semantically equivalent expressions. 
In the second stage, we use a code selector to tackle the {\em information overload} challenge. 
We extract all the code snippets from the collected answers to construct a candidate pool, and then train a Pairwise Learning to Rank neural network by automatically establishing positive and negative training samples. 
We then select the best code snippet from the code snippet candidates via pairwise comparisons.
We conduct extensive experiments to evaluate our {\sc Que2Code} model.
To evaluate the first stage of semantically-equivalent question retrieval, we collect duplicate question pairs of Python and Java from Stack Overflow and verify the effectiveness of our approach for identifying the semantically-equivalent question for a given user query question. 
To evaluate the second stage of best code snippet recommendation, we collect more than 218K QC (question-code snippet) pairs for Python and more than 270K QC pairs for Java. 
We then evaluate the effectiveness of our approach for choosing the right code snippet in the code snippet candidate pool.
Our experimental results show that our proposed {\sc Que2Code} model outperforms several baselines in both stages. 

This paper makes the following three main contributions:
\begin{itemize}
    \item All previous studies of question routing in CQA systems~\cite{wang2009syntactic, cao2010generalized, ganguly2015partially, ye2014interrogative, zou2015learning, xu2018domain} work on finding similar questions. However, it is hard to measure the relevance between different questions automatically and experts are often asked to manually rate the relevance score~\cite{xu2017answerbot, xu2018domain}.
    In our study, we propose a new task of semantically-equivalent question retrieval. 
    By utilizing duplicate question pairs archived in Stack Overflow, we present a novel model and an evaluation method to automatically evaluate the semantically-equivalent questions without a labor-intensive labeling process.
    \item All current studies that have investigated code snippet searching~\cite{gu2018deep, cambronero2019deep, sachdev2018retrieval, ye2016word} rely on calculating a matching score between a query and a code snippet. 
    We argue that code snippet recommendation is more about predicting  relative orders rather than precise relevance scores.
    Hence, we propose a novel pairwise learning to rank model to recommend code snippet from Stack Overflow posts, and we first use the BERT model for searching for code snippets in Stack Overflow. 
    \item \revv{Our experimental results show that our {\sc Que2Code} is more effective for code snippet recommendation than several state-of-the-art baselines, and performs better than Google search engine in identifying the duplicate questions for the low quality user queries.} 
    We have released the source code of our {\sc Que2Code} and our dataset\footnote{\url{https://github.com/beyondacm/Que2Code}} to help other researchers replicate and extend our study.
\end{itemize}

The rest of the paper is organized as follows.
Section~\ref{sec:motiv} presents a motivating example and user scenario of our approach.
Section~\ref{sec:approach} presents details of our approach for semantically-equivalent question retrieval and best code snippet recommendation. 
Section~\ref{sec:eval} presents the automatic experimental results of our approach with respect to the two stages separately.
Section~\ref{sec:human_eval} presents the results of our approach on human evaluation. 
Section~\ref{sec:discussion} discusses the strengths and of our approach and threads to validity. 
Section~\ref{sec:related} presents key related work associated with our study.
Section~\ref{sec:con} concludes the paper. 

\section{Motivation}
\label{sec:motiv}
\begin{figure}
\vspace{0.0cm}
\centerline{\includegraphics[width=0.99\textwidth]{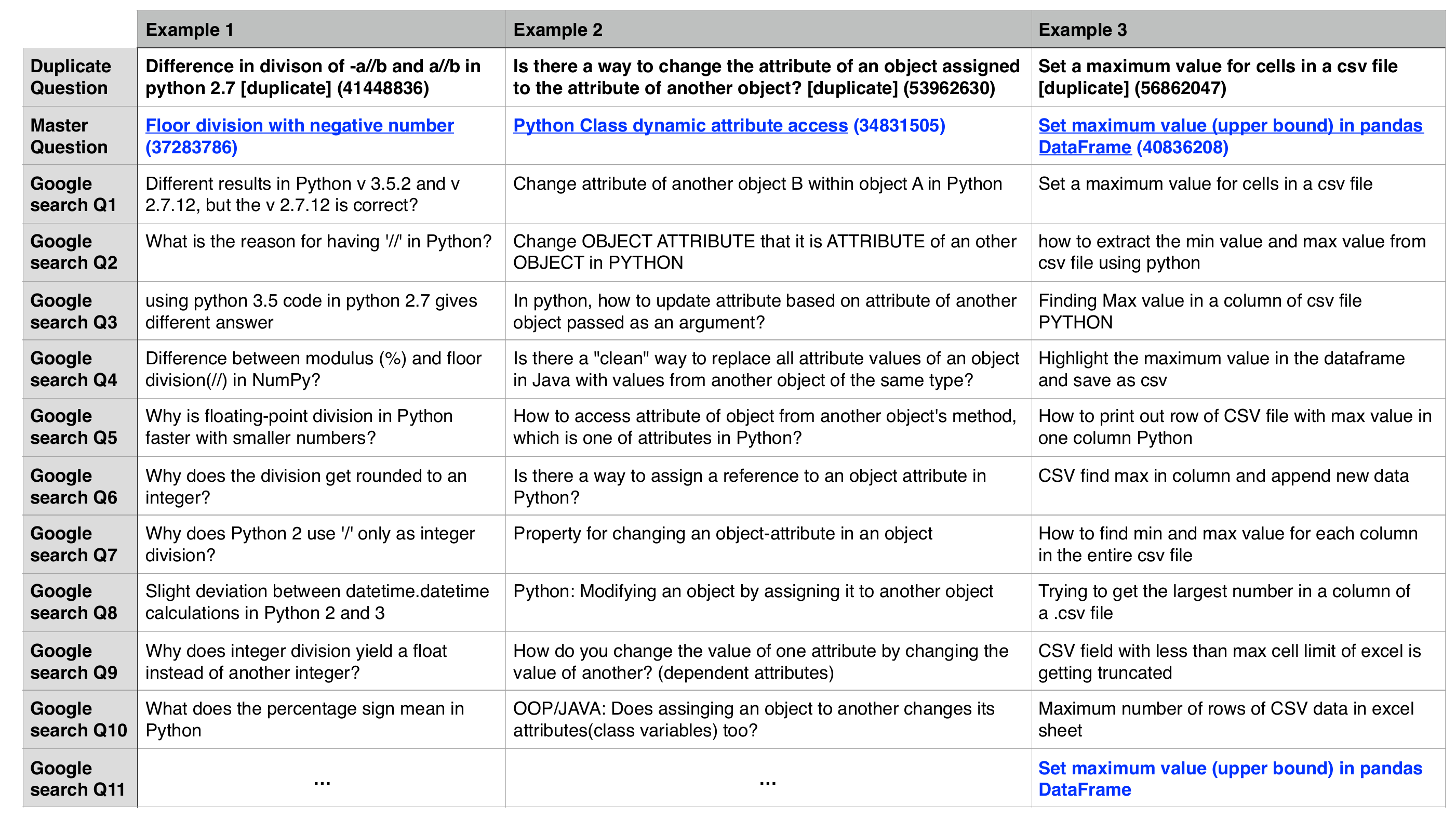}}
\vspace*{-0pt}
\caption{Motivating Examples}
\label{fig:motivation}
\vspace{-0.0cm}
\end{figure}

\revv{
In this section, we first show motivating examples from Stack Overflow of the sorts of problems mentioned above (i.e., \textit{query mismatch} and \textit{information overload}), we then present the user scenarios of employing our approach which can help developers to address these problems. 
Fig.~\ref{fig:motivation} shows three motivating examples of user query questions in Stack Overflow and their corresponding top results in appearance order returned by the Google search engine (the screenshots of the Google search results are saved in our replication package).
From the figure, we can observe the aforementioned two challenges with respect to the low-quality questions: 
\begin{enumerate}
    \item The \textit{query mismatch} challenge for low quality questions. 
    Consider Example 1 -- the objective of the user question (i.e., duplicate question) is related to floor mechanism in python. 
    However, due to lack of knowledge or terminology of the problem, the developer was unable to summarize the key point, therefore he/she expressed this problem in his/her own way, i.e., ``\textit{difference in division of -a//b and a//b in python2.7}''. 
    This duplicate question shares only one lexical unit (i.e., division) with the target master question (i.e., ``\textit{Floor division with negative number}''). 
    Another example is shown in Example 2, the user query question (i.e., ``\textit{Is there a way to change the attribute of an object assigned to the attribute of another object}'') and its duplicate question (i.e., ``\textit{Python Class dynamic attribute access}'') also varies considerably. 
    Such a query mismatch problem often results in low-quality questions due to missing key information or important technical details. 
    We manually checked the top-50 posts returned by Google search engine and found that the target post (i.e., master question) can not be successfully retrieved based on taking user query questions (i.e., duplicate question) as input, which verifies the Google search engine lacks the capacity to retrieve duplicate questions due to query mismatch challenge. 
    Therefore, it is necessary to have an approach to fill the gap between diverse user expressions. 
    \item The \textit{information overload} challenge for low quality questions.
    The information overload problem can be detrimental to the developers. 
    Searching with low quality user queries often bring in irrelevant and useless results~\cite{xia2017developers}, if the search engine fails to return the desired result (as shown in Example 1 and Example 2), the developer has to spend significant time and effort browsing useless posts. 
    Even if the search engine successfully retrieves the target post, if the target post is not ranked higher up among other searching results then the potential solution to the programming task can be easily buried in the overwhelming amount of information. 
    For example, as shown in Example 3, the Google search engine returns the target post at the 11th position.
    There are 10 irrelevant posts in front of the target post, each post often includes multiple answers with many not useful code examples. 
    Even just reading all these answers can cost enormous amount of time for developers, not to mention digesting the associated undesirable code solutions. 
    For such cases, the developers may lose interest in browsing the large number of the irrelevant posts before finding the correct code solution, it is thus beneficial to have an approach to rank the desirable code snippet to a higher position among the searching results. 
\end{enumerate}
}

We illustrate the usage scenario of our proposed tool, {\sc Que2Code}, as follows:

\textbf{Without Our Tool:}
Consider Bob is a developer. Daily, Bob encounters a technical problem and wants a code snippet to help solve it. He tries his best to write a query to summarize his problem  and searches related questions on Stack Overflow.
However, due to lack of the knowledge and terminology about the problem, the query formulated by Bob does not match any post with potential answers. 
Furthermore, Bob has to painstakingly browse a lot of low-quality and/or irrelevant posts to identify any possible solutions. 
Therefore, Bob loses interest and is unsatisfied with the overwhelming number of seemingly irrelevant posts.
As a result, he posts a duplicate question on Stack Overflow.  

\textbf{With Our Tool:}
Now consider Bob adopts our {\sc Que2Code}. When Bob types in his query question, our \textit{QueryRewriter} first generates a list of paraphrase questions for his problem, which increases the likelihood of retrieving semantically-equivalent questions in Stack Overflow. Following that, instead of naively returning a massive amount of similar questions, our \textit{CodeSelector} sorts the returned code snippet candidates and recommends the most relevant ones that may contain possible solutions. With the help of our tool, Bob can quickly get answers for his problem without spending much time on reviewing and digesting the low-quality and/or irrelevant information. This time, Bob quickly identifies a useful code snippet from an existing Stack Overflow post for his problem by using our tool.

\section{Approach}
\label{sec:approach}

We present a novel query-driven code recommendation system. 
{\sc Que2Code} consists of two stages: \textit{Semantically-Equivalent Question Retrieval} and \textit{Best Code Snippet Selection}. 
Our approach takes in a technical question as a query from a developer, and recommends a sorted list of code snippets. 

\subsection{Overview of Approach}

\begin{figure}
\vspace{0.0cm}
\centerline{\includegraphics[width=0.85\textwidth]{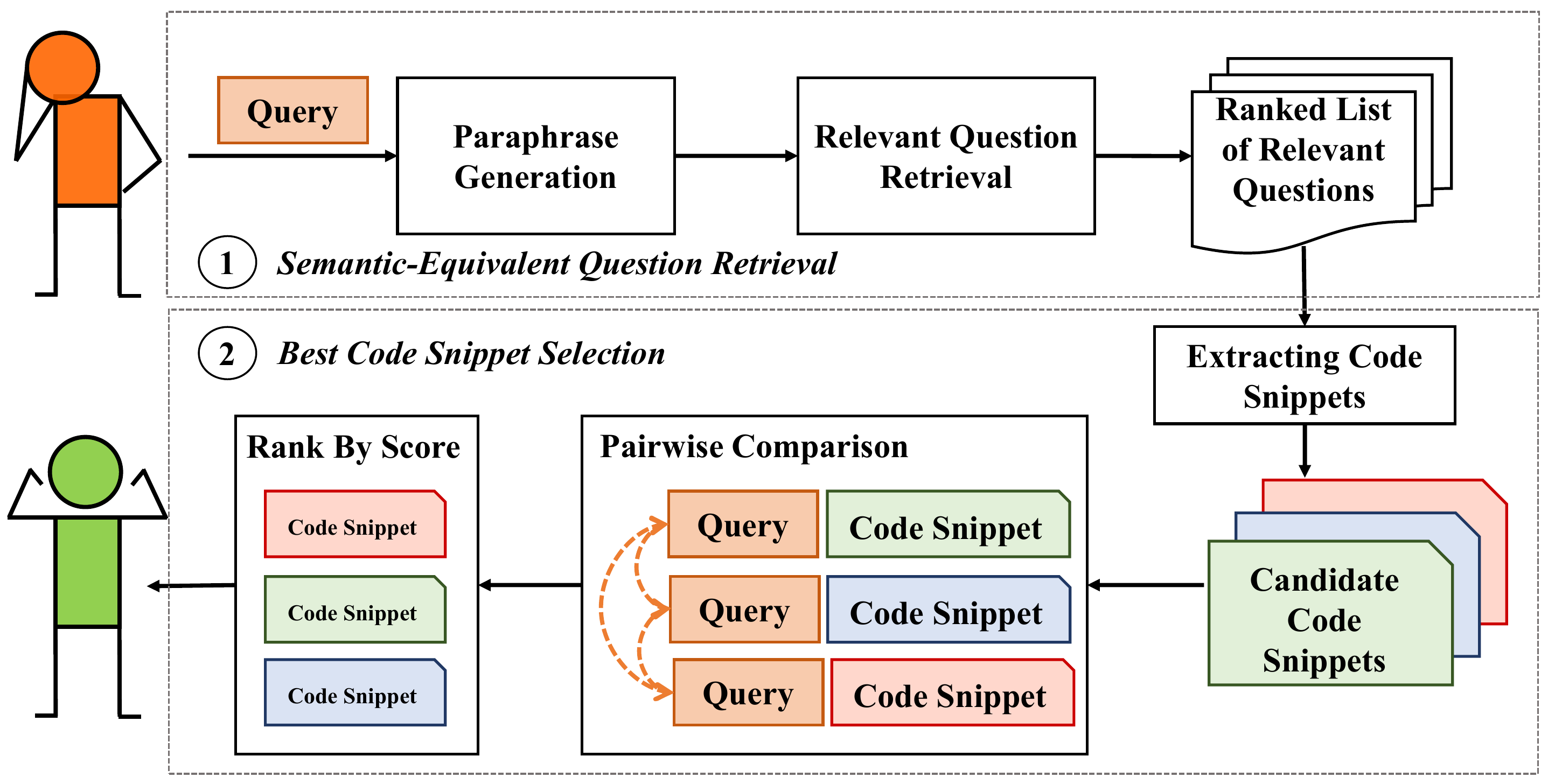}}
\caption{Workflow of Que2Code}
\label{fig:workflow}
\vspace{-0.0cm}
\end{figure}

Fig.~\ref{fig:workflow} demonstrates the workflow of {\sc Que2Code}. 
Our model contains two stages: (i) semantically-equivalent question retrieval and (ii) best code snippet recommendation.
It has two sub-components, i.e., \emph{QueryRewriter} and \emph{CodeSelector}. 
The first can qualitatively retrieve semantically-equivalent questions, and the second can quantitatively rank the most relevant code snippets to the top of the recommendation candidates. 

In the first stage, our \emph{QueryRewriter} component tackles the {\em query mismatch} challenge.
\rev{
To bridge the gap between different expressions of semantically-equivalent questions, we introduce the idea of \emph{query rewriting}.
}
The idea is to use a rewritten version of a query question to cover a variety of different forms of semantically equivalent expressions. 
In particular, we first collect the duplicate question pairs from Stack Overflow, because duplicate questions can be considered as semantically-equivalent questions of various user descriptions.
We then frame this problem as a sequence-to-sequence learning problem, which directly maps a technical question to its corresponding duplicate question. 
We train a text-to-text transformer, named \emph{QueryRewriter}, by using the collected duplicate question pairs. 
After the training process, for any given query question, \emph{QueryRewriter} outputs semantically equivalent paraphrased questions of the input query. 
Following that, the query question and its generated paraphrased questions are encoded by \emph{QueryRewriter} to measure their relevance with other question titles.   

In the second stage, our \textit{CodeSelector} component tackles the {\em information overload} challenge. 
\rev{To do this,} we first collect all the answers of the semantic relevant questions retrieved in the first stage. 
\rev{We then extract all the code snippets from the collected answer posts to construct a candidate code snippets pool.} 
For the given query question, we pair it with each of the code snippet candidates. We then fit them into the trained \emph{CodeSelector} to estimate their matching scores and judge the preference orders.
\emph{CodeSelector} can then select the best code snippet from the code snippet candidates via pairwise comparison. 
Our approach is fully data-driven and does not rely on hand-crafted rules.

\subsection{Semantically-Equivalent Question Retrieval}
\label{sec:que_ret}
In this stage, given a technical problem formulated as a query, we propose a \textit{QueryRewriter} to generate paraphrase questions and retrieve the semantically-equivalent questions in Stack Overflow. 
Fig.~\ref{fig:querywriter} demonstrates the workflow of \textit{QueryRewriter}.
\textit{QueryRewriter} has three steps: paraphrase generation, question embedding and questions retrieval.

\subsubsection{Paraphrase Generation}
To obtain the features of semantically-equivalent questions for a given user query, we utilize the historical archives of duplicate questions in Stack Overflow, which are manually marked by users and moderators.
These duplicate question pairs can be viewed as questions of same intent but written in different ways. 
In this step, we first automatically generate paraphrase questions for the query question to represent different forms of user expressions.
The underlying idea is that by adding these paraphrase questions, we are more likely to find the relevant question that match the intent expressed in the user query. 

In this step, we model the task of paraphrase question generation as a sequence-to-sequence learning problem, where the question title is viewed as a sequence of tokens and its corresponding duplicate question as another sequence of tokens. 
We adopt the Seq2Seq Transformer architecture~\cite{vaswani2017attention}, which includes an Encoder Transformer and a Decoder Transformer. 
Both the Encoder and the Decoder Transformers
have multiple layers and each layer contains a multi-head attentive sub-layers followed by a fully connected sub-layer with residual connections~\cite{he2016deep} and and layer normalization~\cite{ba2016layer}. 

\begin{figure}
\centerline{\includegraphics[width=0.97\textwidth]{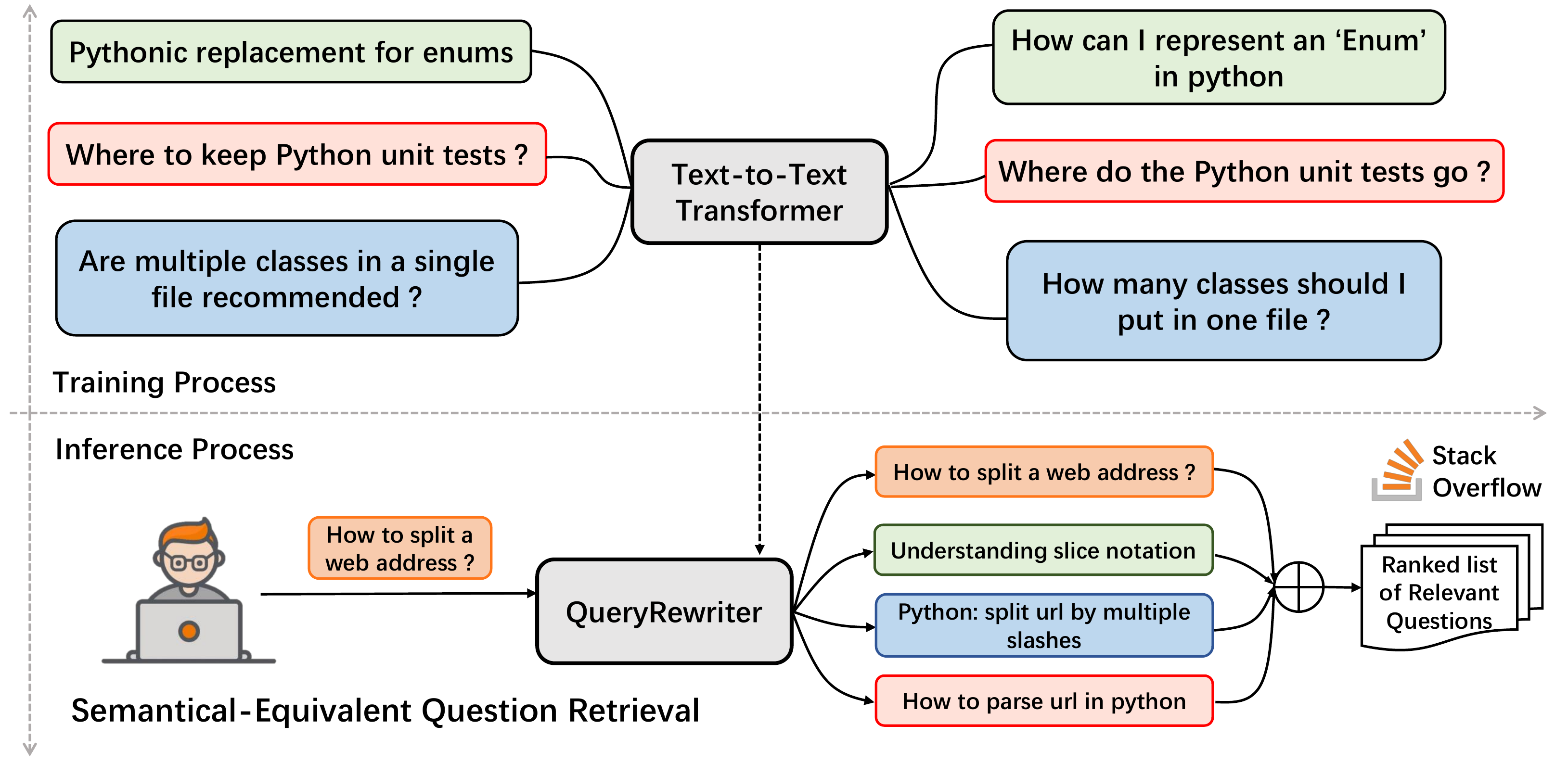}}
\caption{QueryRewriter Workflow}
\label{fig:querywriter}
\vspace{-0.0cm}
\end{figure}

More formally, 
given a question $\mathbf{X}$ as a sequence of tokens $(x_1, x_2, ..., x_{M})$ of length $M$, 
and its duplicate question $\textbf{Y}$ as a sequence of tokens $(y_1, y_2, ..., y_{N})$ of length $N$. 
The encoder takes the question $\mathbf{X}$ as input and transforms it to its contextual representations.
The decoder learns to generate the corresponding duplicate question $\mathbf{Y}$ one token at a time based on the contextual representations and all preceding tokens that have been generated so far. 
Mathematically, the paraphrase generation task is defined as finding $\overline{y}$, such that:
\begin{equation}
    \overline{y} = argmax_{\mathbf{Y}}P_{\theta}(\mathbf{Y}|\mathbf{X})
\end{equation}
where $P_{\theta}(\mathbf{Y}|\mathbf{X})$ is defined as:
\begin{equation}
    P_{\theta}(\mathbf{Y}|\mathbf{X}) = \prod^{L}_{i=1}P_{\theta}(y_i|y_1, ..., y_{i-1};x_1, ..., x_M)
\end{equation}
$P_{\theta}(\mathbf{Y}|\mathbf{X})$ can be seen as the conditional log-likelihood of the predicted duplicate question $\mathbf{Y}$ given the input question $\mathbf{X}$. 
This model can be trained by minimizing the negative log-likelihood of the training question duplicate question pairs.
\rev{
Once the model is trained, we do inference using beam search~\cite{koehn2004pharaoh}. 
Beam Search returns a list of most likely output sequences (i.e., paraphrase questions). 
It searches question tokens produced at each step one by one.  
At each time step, it selects $b$ tokens with the least cost, where $b$ is the beam wise. 
It then prunes off the remaining branches and continues selecting the possible tokens that follow on until it meets the end-of-sequence symbol. 
We repeat the process and generate the top-N most likely paraphrase questions for our study. 
}

A working example is demonstrated in Fig.~\ref{fig:querywriter}, when we input the user query \textit{``how to split a web address?''} described in the motivation example into our trained model, the \textit{QueryRewriter} automatically generates paraphrase questions of different expressions, such as \textit{``understanding slice notation''}, \textit{``python: split url by multiple slashes''}, \textit{``how to parse url in python''}. 
Since the user query question (i.e., ``\textit{how to split a web address}'') and the target question (i.e., ``\textit{slicing url with python}'') do not share any lexical units, incorporating the aforementioned paraphrase questions can successfully bridge this gap and solve the {\em query mismatch} problem.


\subsubsection{Question Embedding}
After the text-to-text transformer is trained, we can generate multiple paraphrase questions for a newly posted user query.
We use these paraphrase questions to boost the user query for the downstream task of question retrieval.
By incorporating the paraphrase questions, we can alleviate the {\em query mismatch} problem by covering the different forms of duplicate question expressions.  
Thus, we have a better chance to retrieve semantically-equivalent questions in Stack Overflow that are intended to solve the same problem. 

For a given user query question $q_{u}$, we first construct $\mathbf{Q_{u}}=\{q_{u}, pq_{k}\} (1 \leqslant k \leqslant N)$, where $q_{u}$ is the user query question and $pq_{k} (1 \leqslant k \leqslant N)$ are the top-N generated paraphrase questions.
To capture the overall features and semantics of the user query $q_{u}$ we embed each question $q_{i}$ in $\mathbf{Q_{u}}$ (including the user query question and the paraphrase questions) to a fixed dimensional vector $e_{q_{i}}$ via the Encoder Transformer. We then use the average embeddings of all possible questions as the final representation for the user query question. 
More formally, the question embedding step is defined as follows:

\begin{equation}
e_{q_{i}} = Encoder(q_{i}), q_{i} \in \mathbf{Q_{u}}
\label{eq3}
\end{equation}
\begin{equation}
e_{q_{u}} = \frac{1}{N}\sum e_{q_{i}}, q_{i} \in \mathbf{Q_{u} }
\label{eq:eq4}
\end{equation}

\subsubsection{Question Retrieval}
Given a newly posted user query question $q_{u}$ and a question title $q$ in the Stack Overflow repository, \minor{we use Euclidean distance to estimate the semantic distance because Euclidean distance has shown to be effective for different software engineering tasks~\cite{huang2020poster, gao2020checking}, especially when the candidates are similar to each other. 
The definition of distance as well as the relevance between two questions are as follows:}


\begin{equation}
Distance(q_{u}, q) = \frac{Euclidean(e_{q_{u}}, e_{q})}{ |e_{q_{u}}| + |e_{q}|}
\end{equation}
\begin{equation}
Relevance(q_{u}, q)= 1 - Distance(q_{u}, q) 
\label{eq6}
\end{equation}

\rev{
For a given user query, we can easily compute a relevance score between the user query and any candidate question in our database. 
Following that, all the relevance scores are sorted and the top-K ranked questions are returned as the most semantically-relevant questions for the query.
\revv{
Considering the number of paraphrase questions $N$ is the key parameter added to our approach, we further investigated the optimal parameter settings of $N$ for the \textit{QueryRewriter} by performing a parameter tuning analysis. 
We vary $N$ from 0 to 20 with step size 1 to select the optimal parameter, we find that: (i) the performance of our model rapidly increases as $N$ is increased from 0 to 3, (ii) achieves its best performance when $N$ reaches around 5, (iii) the overall performance of our model reaches a plateau after achieving the best performance. 
We thus recommend setting the number of generated paraphrase questions to 5, which is close to the optimal settings of $N$ in our approach.
}
}

\begin{figure}[t]
\vspace{0.0cm}
\centerline{\includegraphics[width=0.95\textwidth]{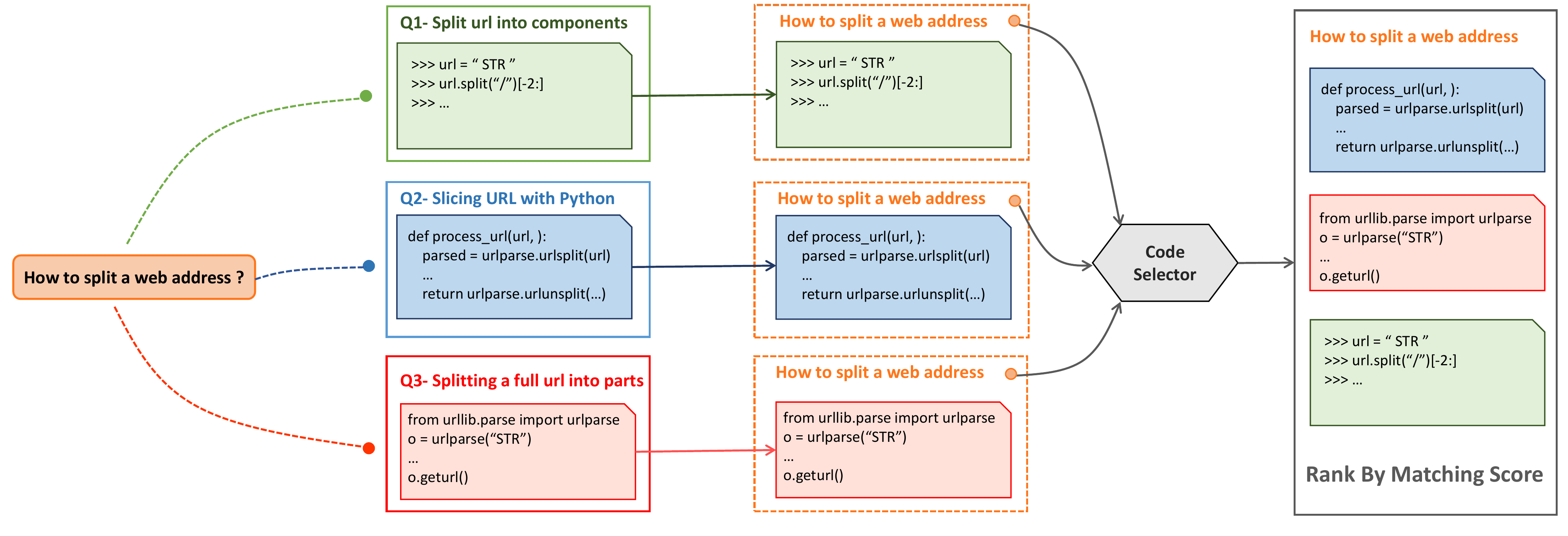}}
\caption{\rev{CodeSelector Workflow}}
\label{fig:codeselector-workflow}
\vspace{-0.0cm}
\end{figure}

\subsection{Best Code Snippet Recommendation}
\label{sec:code_sel}
\rev{
Theoretically, after retrieving the semantically-equivalent questions for a given user query, we can naively choose the code snippet from the top ranked questions and recommend the solution to the developers.
However, we argue this is not sufficient regarding the following reasons:
(i) The technical queries submitted by developers are complicated or sometimes inaccurate~\cite{gao2020generating}, there is no guarantee that the top ranked solution is correct and can satisfy the needs of developers. 
Therefore, we need to collect as many as possible relevant potential code snippet candidates. 
(ii) Although the search engine can return a list of relevant questions to their problems, the large number of relevant posts and the sheer amount of information in them makes it difficult for developers to find the most needed answer~\cite{xu2017answerbot}. 
Therefore, how to pick the best solution from the massive amounts of information is a non-trivial task.
}

\rev{
To address this time-consuming task of online code searching, we propose \textit{CodeSelector} to help developers effectively select the most relevant and suitable code snippet for a specific query question.
Particularly, the \textit{CodeSelector} reranks all the code snippets by comparing the matching score among different QC (query-code) pairs.
Snippets ranked in the top of the final result indicate that these code snippets are more likely and suitable for the programming task.
Fig.~\ref{fig:codeselector-workflow} demonstrates the workflow of our \textit{CodeSelector}. 
For a given query question, a list of relevant questions are retrieved from stage one. 
After that, all the code snippets associated with these questions are collected, and each code snippet is paired with the given query to make a QC pair. 
Then the \textit{CodeSelector} reranks all the code snippet candidates by conducting pairwise comparison among QC pairs. 
To build the \textit{CodeSelector}, two steps are performed: Preference Pairs Construction and Pairwise Comparison.
}

\subsubsection{Preference Pairs Construction}
In this step, we use the available crowdsourced data on Stack Overflow for preference pairs construction.
We propose three heuristic rules that can automatically establish the preference pairs and construct the training sets. 
Our approach is fully data driven and it does not need manual effort.
Our three heuristic rules are as follows:
\begin{itemize}
    
    \item \textit{For a given question, its best code snippet is preferable to a non-relevant code snippet.} We define the \emph{best code snippet} for a question as the code snippet associated with an accepted answer to the question, or the one associated with the highest-vote answer if there are multiple answers to the question.
    A non-relevant code snippet is randomly selected from the repository. This rule suggests that the quality of the best code snippet is better than the non-relevant ones.
    
    \item \textit{For a given question, its non-best code snippet is preferable to a non-relevant code snippet.} 
    We define the \emph{non-best code snippet} as other code snippets apart from the best one within the same question thread. This rule suggests that a question prefers the code snippets associated with answers to itself to those of others.

    \item \textit{For a given question, its best code snippet is preferable to its non-best code snippet.} 
    Even though an individual user vote may not be very reliable, the aggregation of a great number of user votes can provide a very powerful indicator of relevance preference. 
    We argue that the \emph{best code snippet} from an answer post for a question is better than the \emph{non-best ones} in different answer posts for the same question, in most cases.
\end{itemize}

\begin{figure}[t]
\vspace{0.0cm}
\centerline{\includegraphics[width=0.95\textwidth]{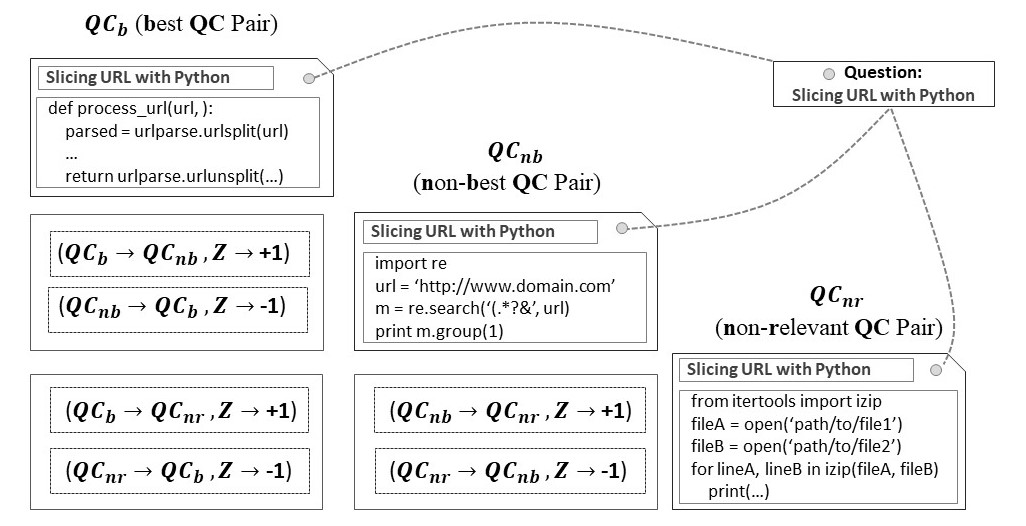}}
\caption{\revv{Preference-Pairs Construction}}
\label{fig:labelestablish}
\vspace{-0.0cm}
\end{figure}

\rev{
According to the above heuristic rules, for each given question, three query-code (QC) pairs can be generated: we pair it with its best code snippet as the $QC_{b}$ (\textbf{\underline{b}}est \textbf{QC} pair), we pair it with its non-best code snippet as the $QC_{nb}$ (\textbf{\underline{n}}on-\textbf{\underline{b}}est \textbf{QC} pair), we pair it with a non-relevant code snippet as the $QC_{nr}$ (\textbf{\underline{n}}on-\textbf{\underline{r}}elevant \textbf{QC} pair).
\revv{
We then automatically construct the training samples $\langle QC_{1}, QC_{2}, Z \rangle$ for this study. 
In particular, a training sample contains three parts: two QC pairs (i.e., $QC_{1}$ and $QC_{2}$) and a label (i.e., $Z$), the label is automatically determined by the preference relationship between the two QC pairs. 
Fig.~\ref{fig:labelestablish} demonstrates our data labeling process. 
Given the query question ``\textit{Slicing URL with Python}'', we first make three QC pairs, i.e., $QC_{b}$, $QC_{nb}$ and $QC_{nr}$ as mentioned above. 
Then a Pairwise comparison between any two QC pairs can establish a label $Z$ for this training sample. 
It is worth emphasizing that the comparison order of the two QC pairs matters. For example, a comparison between $QC_{b}$ and $QC_{nb}$ will be labelled as positive, while a comparison between $QC_{nb}$ and $QC_{b}$ will be labelled as negative.
}}

\rev{
There are several advantages of employing this data labeling process: 
(i) due to the professionality of technical queries, only experts with domain knowledge are qualified to judge the usefulness of a code snippet to a query question. 
Therefore, manually labeling the relevance scores for all code snippets is very time-consuming and requires a substantial effort~\cite{jiang2016rosf, gao2020deepans}. 
Our heuristic rules can automate the labeling process without any human efforts.  
(ii) By using these heuristic rules, we gather more training samples, which can provide enough data points for training a deep learning based model. 
}

\subsubsection{Pairwise Comparison}

\begin{figure}[t]
\vspace{0.0cm}
\centerline{\includegraphics[width=0.85\textwidth]{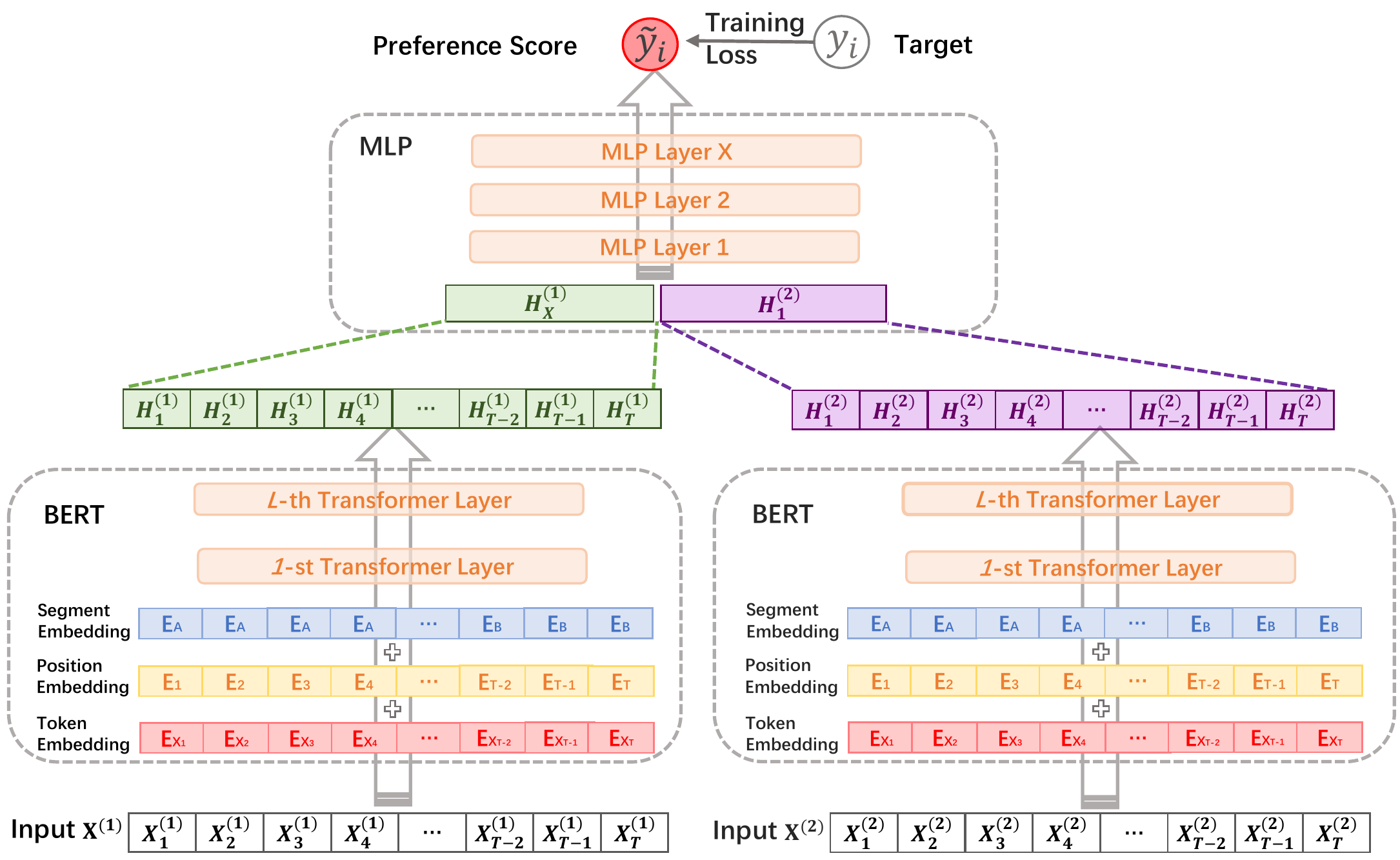}}
\caption{CodeSelector Workflow}
\label{fig:codeselector}
\end{figure}

After collecting large amounts of labeled training data via preference pairs construction, we develop a learning-to-rank model to sort all the code snippets for a given query question.
We design a novel semi-supervised network for ranking query-code (QC) pairs.
Fig~\ref{fig:codeselector} demonstrates the architecture of our proposed model.
The input to \textit{CodeSelector} are two QC pairs. \textit{CodeSelector} then learns to judge the preference relationship between two QC pairs based on the positive and negative training samples.
In other words, we not only consider the program semantics between a query and a code snippet, but also investigate the relevance preference among different QC pairs.
\begin{itemize}
    \item \textbf{BERT Embedding Layer.} 
    We use the modeling power of BERT~\cite{devlin2018bert}, which is one of the most popular pre-trained models trained using Transformers~\cite{vaswani2017attention}. 
    BERT consists of 12-layer transformers, each of the transformers being composed of a self-attention sub-layer with multiple attention heads. 
    Since BERT has been proven to be effective for capturing semantics and context information of sentences in other work, we use BERT as the feature extractor for our task. 
    The input to the BERT embedding layer are two parallel QC pairs.
    Given each QC pair as a sequence of tokens $\mathbf{x}=\{x_{1},...,x_{T}\}$ of length $T$, BERT takes the tokens as input and calculate the contextualized representations $\mathbf{H}^{l}=\{h_{1}^{l},...,h_{T}^{l}\} \in \mathbb{R}^{T \times D}$ as output, where $l$ denotes the $l$-th transformer layer and $D$ denotes the dimension of the representation vector.  
    The underlying sub-components work in parallel, mapping each QC pair to its distributional vectors $\mathbf{h}^{(1)}$ and $\mathbf{h}^{(2)}$ respectively, which are then used to perform the predictions for the downstream task.
    \item \textbf{Multi-Layer Perceptron.}
    After obtaining the BERT representations, we add a Multi-Layer Perceptron (MLP) on top of BERT embedding layer to calculate the preference score between the input two QC pairs. 
    Since \textit{CodeSelector} adopts BERT to model two QC pairs respectively, it is intuitive to combine the features of two pathways by concatenating them. This design has been widely adopted in other deep learning work~\cite{he2017neural, gao2020deepans}. 
    To further capture the preference between the latent features of $\mathbf{h}^{(1)}$ and $\mathbf{h}^{(2)}$, we add a standard MLP on the concatenated vector. 
    In this sense, we can endow the model a large level of flexibility and non-linearity to learn the interactions between the two QC pairs. 
    The contextualized representations (i.e., $\mathbf{h}^{(1)}$ and $\mathbf{h}^{(2)}$) are fed to the MLP layer to predict the final preference $\mathbf{y}=\{0, 1\}$.
    More precisely, the MLP is defined as follows:
    \begin{equation}
    \label{eq:log_loss}
    \begin{split}
    &\mathbf{z}_{1} = \phi_{1} ( \mathbf{h}^{(1)}, \mathbf{h}^{(2)} ) =  \left[ \begin{array}{l} 
    \mathbf{h}^{(1)} \\ 
    \mathbf{h}^{(2)} \end{array}
    \right] \\
    &\mathbf{z}_{2} = \phi_{2} ( \mathbf{z}_{1} ) = \mathbf{a}_{2} ( \mathbf{W}_{2}^{T} \mathbf{z}_{1} + \mathbf{b}_{2} ) \\
    &... \\
    &\mathbf{z}_{L} = \phi_{L} ( \mathbf{z}_{L-1} ) = \mathbf{a}_{L} ( \mathbf{W}_{L}^{T} \mathbf{z}_{L-1} + \mathbf{b}_{L} ) \\
    & P ( \mathbf{y} = j | \mathbf{x}^{(1)}, \mathbf{x}^{(2)} ) = \sigma ( \mathbf{z}_{L} ) \\ 
    \end{split}
    \end{equation}
\end{itemize}

$\mathbf{W}_{x}$, $\mathbf{b}_{x}$, and $\mathbf{a}_{x}$ denote the weight matrix, bias vector, and activation function for the $x$-layer's perceptron respectively. $\sigma$ is the sigmoid function $\sigma(x) = 1/(1+e^{-x})$ which will output the final preference score between 0 and 1. 
For the preference score, we want this score to be high if the first QC pair is preferable to the second one (i.e., $\mathbf{x}^{(1)} \succ  \mathbf{x}^{(2)}$), and to be low if the second QC pair is preferable to the first one (i.e., $\mathbf{x}^{(2)} \succ  \mathbf{x}^{(1)}$).

\subsection{\rev{Experimental Settings}}
We implemented our system in Python using the Pytorch framework. 
For the \textit{QueryWriter}, we trained the text-to-text transformer on the duplicate question pairs, we follow the parameter settings from ~\cite{raffel2019exploring}t, which has achieved state-of-the-art results on many benchmarks covering summarization, question answering, text classification, and more. 
For the \textit{CodeSelector}, we use the pre-trained BERT model released by ~\cite{devlin2018bert} as our feature extractor. 
We use the ReLu as the activation function and employ three hidden layers for MLP. 
The size of the first hidden layer in MLP is equal to the size of the joint vector obtained after concatenating two QC vectors from the BERT model. 
We fix the parameters of the BERT model and fine tune the MLP parameters for our task, and the \textit{CodeSelector} is learnt by optimizing the log loss of Equation.~\ref{eq:log_loss}. 


\section{Automatic Evaluation}
\label{sec:eval}
To recommend the code snippets for developers in Stack Overflow, our {\sc Que2Code} is divided into two stage: semantically-equivalent question retrieval and best code snippet recommendation. 
We wanted to evaluate the performance of the proposed \textit{QueryRewriter} to address the {\em query mismatch} problem in the first stage,  and \textit{CodeSelector} to address the  \emph{information overload} problem in the second stage. We want to answer the following key research questions:
\begin{itemize}
    \item RQ-1: How effective is our \textit{QueryRewriter} for semantically-equivalent question retrieval?
    \revv{
    \item RQ-2: How effective is our \textit{QueryRewriter} compared with Google search engine?  
    }
    \rev{
    \item \revv{RQ-3}: How effective is our \textit{QueryRewriter} for capturing the domain-pecific contextual information? 
    }
    \item \revv{RQ-4}: How effective is the paraphrase generation added to our \textit{QueryRewriter}?
    \item RQ-5: How effective is our \textit{CodeSelector} for best code snippet selection? 
    \item RQ-6: How effective is the BERT and preference pairs added to our \textit{CodeSelector}? 
    \item RQ-7: How robust is our \textit{CodeSelector} with different parameter settings? 
\end{itemize}

\subsection{RQ-1: Effectiveness of QueryRewriter}
We want to identify the best code snippet from a list of semantically-equivalent questions for a given query question.
If the retrieved questions are not relevant to the query question, it is unlikely that our tool is able to find the suitable code snippets to solve the target problem.
In this study, we consider the duplicate questions in Stack Overflow as semantically-equivalent question pairs.
After training with the duplicate question pairs archived in Stack Overflow, \textit{QueryRewriter} is able to generate paraphrase questions for a given user query question. 
By jointly embedding the user query question and the generated paraphrase questions, \textit{QueryRewriter} retrieves the most relevant questions in the repository.
We want to investigate the effectiveness of our \textit{QueryRewriter} for retrieving semantically-equivalent questions in Stack Overflow.

\subsubsection{Data Preparation}
\begin{figure}
\vspace{0.0cm}
\centerline{\includegraphics[width=0.75\textwidth]{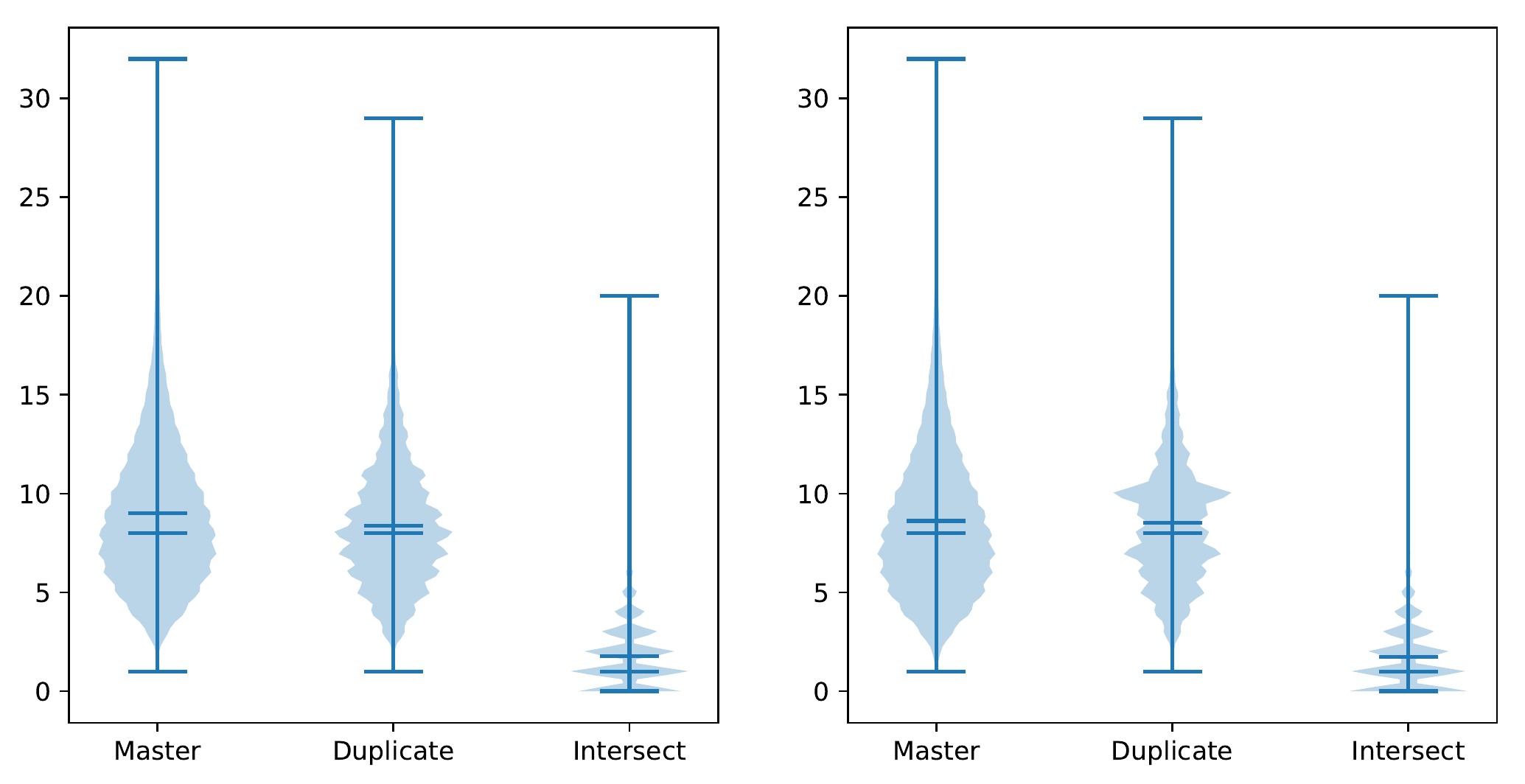}}
\caption{
\rev{Volinplots of Question Distribution for Python(left) and Java(right) Dataset}
}
\label{fig:query-mismatch}
\end{figure}

\rev{
We first downloaded the official data dump of Stack Overflow from the StackExchange\footnote{\url{https://archive.org/download/stackexchange}} website.
The raw data dump contains timestamped information about the \textit{Posts}, \textit{Comments}, \textit{Users}, \textit{Tags}, \textit{Postlinks} etc,.
We extracted the duplicate question pairs as follows: 
We first parsed the \textit{PostLinks} and \textit{Posts} database files.
Because duplicate questions are marked with a special marker in \textit{PostLinks} database, we can easily identify the \texttt{PostId} of the source post and the target post if they are duplicate question pairs. 
After that, we extracted the question title of the source post and the target post by checking the \texttt{PostId} in \textit{Posts} database. 
We regard the question title of the source post as a duplicate question and the question title of the target post as a master question.
We then paired each master question $q_{m}$ and duplicate question $q_{d}$ as $\langle q_{m}, q_{d} \rangle$ pair.}
After that, these collected $\langle q_{m}, q_{d} \rangle$ pairs are fed into the text-to-text Transformer to train our \textit{QueryRewriter}. 
In this study, we only focused on Python and Java programming languages for our experiment.
\rev{
As a result, we obtained more than 47K $\langle q_{m}, q_{d} \rangle$ pairs for Python and more than 56K pairs for the Java dataset. 
We further investigated the \textit{query mismatch} problem between the master questions and its corresponding duplicate questions.
Specifically, we counted the number of tokens of the master question $q_{m}$ and its duplicate question $q_{d}$, and then we counted the common lexical tokens between the master question and the duplicate question.
The violin plot for Python and Java is demonstrated in Fig.~\ref{fig:query-mismatch}, we can see that the overlap tokens between the master question and duplicate question are small.
For example, for the Python dataset, the average number of tokens of the master question and duplicate question are 9.0 and 8.4 respectively, while the average number of overlap tokens is 1.8.
Even though the master question and its duplicate question are semantic-equivalent, they only share a few tokens in common. 
This also justifies our assumption that the \textit{query mismatch} problem frequently occurs when developers submit their search queries. 
Therefore, it is necessary to propose a model to address the \textit{query mismatch} problem.
Table~\ref{tab:dup-dataoverview} shows the statistics of our collected datasets of Python and Java duplicate questions.
We randomly sampled 2,000 $\langle q_{m}, q_{d} \rangle$ pairs for validation and 2,000 $\langle q_{m}, q_{d} \rangle$ pairs for testing, and kept the rest for training.
We clarify that different duplicate questions can point to the same master question and this is the reason why the number of duplicate questions is more than the master questions. 
}

                                  

\begin{table*} \vspace{-0.0cm}
\caption{\rev{Duplicate Questions Statistics}}
\rev{
\begin{center}
\vspace{-0.2cm}\begin{tabular}{||c|l|c|l|c||}
    \hline
    \multirow{4}{*}{Python}  
                              & \# Duplicate Questions & 47,170 & 
                                \# Avg. Tokens (Duplicate) & 8.4 \\ \cline{2-5}
                              & \# Master Questions & 20,430 & 
                                \# Avg. Tokens (Master) & 9.0 \\ \cline{2-5}
                              & \# $\langle q_{m}, q_{d} \rangle$ Pairs (Train) & 43,170 &
                              \# Avg. Tokens (Intersect) & 1.8 \\\cline{2-5}
                              & \# $\langle q_{m}, q_{d} \rangle$ Pairs (Val) & 2,000 & 
                                \# $\langle q_{m}, q_{d} \rangle$ Pairs (Test) & 2,000 \\\cline{2-5}
    \hline\hline
    \multirow{4}{*}{Java}   
                             & \# Duplicate Questions & 56,938 & 
                                \# Avg. Tokens (Duplicate) & 8.5 \\ \cline{2-5}
                              & \# Master Questions & 23,889 & 
                                \# Avg. Tokens (Master) & 8.6 \\ \cline{2-5}
                              & \# $\langle q_{m}, q_{d} \rangle$ Pairs (Train) & 52,938 &
                              \# Avg. Tokens (Intersect) & 1.7 \\\cline{2-5}
                              & \# $\langle q_{m}, q_{d} \rangle$ Pairs (Val) & 2,000 & 
                                \# $\langle q_{m}, q_{d} \rangle$ Pairs (Test) & 2,000 \\\cline{2-5}
    \hline
\end{tabular}
\label{tab:dup-dataoverview}
\end{center}
}
\vspace{-0.0cm}
\end{table*}

\subsubsection{Experimental Setup}
To determine the effectiveness of our \textit{QueryRewriter} for retrieving semantic-equivalent questions, we designed the following experiment: 
\rev{we first build an index with Lucene using all the question titles in testing set, the generated index is later used to retrieve the most relevant questions against a given query. In such a way,
for each duplicate question $q_{d}$ in the testing set, we first obtained a set of top-K similar questions $\mathcal{Q}=\{q_{i}\} (1 \leqslant i \leqslant k)$ via the Apache Lucene.
}
We then added its corresponding master question $q_{m}$ to $\mathcal{Q}$, where now $\mathcal{Q}=\{q_{m}, q_{i}\} (1 \leqslant i \leqslant k, q_{m} \neq q_{i})$.
For a given testing question $q_{d}$, we can ensure its corresponding master question $q_{m}$ is in evaluation candidate pool $\mathcal{Q}$. 
Since $\mathcal{Q}$ includes the semantic-equivalent question $q_{m}$ for $q_{d}$. One way to evaluate our approach is to look at how often the master question $q_{m}$ can be retrieved successfully among other members of $\mathcal{Q}$. 
\rev{
Thus we adopted the metric, $P@K$ and $DCG@K$~\cite{Manning2005IntroductionTI}
which are widely-used in previous studies~\cite{xu2017answerbot, jiang2016rosf}, to measure the ranking performance of the approach in our study.
} 
The evaluation metrics are defined as follows:
\begin{itemize}
    \item $P@K$ is the precision of the master question in top-K candidate questions. 
    Given a question, if one of the top-K ranked questions includes the master question, we consider the recommendation to be successful and set $success (q_{m} \in topK)$ to 1, otherwise, we consider the recommendation to be unsuccessful and set $success(q_{m} \in topK)$ to 0. 
    The $P@K$ metric is defined as follows:
    \begin{equation}
    P@K = \frac{1}{N}\sum_{i=1}^N[success(q_{m} \in topK)]
    \end{equation}
    
    \item $DCG@K$ is another popular top-K accuracy metric that measures a recommender system performance based on the graded relevance of the recommended items and their positions in the candidate set. Different from $P@K$, the intuition of $DCG@K$ is that highly-ranked items are more important than low-ranked items. According to this metric, a recommender system gets a higher reward for ranking the correct answer at a higher position. 
    The $success(q_{m} \in topK)$ is same with the previous definition, while the $rank_{q_{m}}$ is the ranking position of the master question $q_{m}$. The $DCG@K$ is defined as follows:
    \begin{equation}
    DCG@K = \frac{1}{N}\sum_{i=1}^N\frac{[ success(q_{m} \in topK) ]}{\log_2(1+rank_{q_{m}})}
    \end{equation}
\end{itemize}

\subsubsection{Experimental Baselines}
To demonstrate the effectiveness of our proposed method for relevant question retrieval task, we compared it to the following baselines:
\begin{itemize}
    \item \textbf{IR} stands for the information retrieval baseline. 
    For a given testing question $q_{i}$, it retrieves the question that is closest to $q_{i}$ from the testing set. 
    We use the traditional TF-IDF metric in our experiment, which is often used to calculate the relevance between a document and a user query in software engineering tasks, such as question retrieval~\cite{xia2017developers, cao2010generalized} and code search~\cite{sachdev2018retrieval}. 
    
    \item \textbf{Word2Vec} is a model that embeds words in a high dimensional vector space using a shallow neural network~\cite{mikolov2013distributed}.
    This word embedding technique has provided a strong baseline for information retrieval tasks. 
    Yang et al.~\cite{yang2016combining} used average word embeddings of words in a document as the vector representation for a document. 
    The average word embeddings can be used to calculate the relevance between two question titles in our task.
    
    \item \rev{
    \textbf{FastText} is another word embedding model proposed by ~\cite{bojanowski2017enriching}. Similar to the Word2Vec model, each word is represented as a high dimensional vector in such a way that similar words have similar vector representations. Same with the Word2Vec baseline, the average FastText word embeddings are used to estimate the similarity between different question titles. 
    }
    \item \textbf{Sent2Vec} Sent2Vec method, which is also known as para2vec or sentence embedding~\cite{le2014distributed}. 
    This method modifies the Word2Vec algorithm to generate semantic embeddings of longer pieces of text (e.g., sentences or paragraphs) via unsupervised learning.
    The generated sentence embeddings have been applied in textual similarity tasks~\cite{le2014distributed}. 
    With the help of publicly accessible tool~\cite{rehurek_lrec}, we train the Sent2Vec model using the duplicate question corpus and obtain the sentence embeddings for each question title.
    
    \item \textbf{AnswerBot} Xu et al.~\cite{xu2017answerbot} proposed a three-stage framework called AnswerBot to generate an answer summary for a non-factoid technical question (\rev{i.e., non-factoid questions are defined as open-ended questions that require complex answers, like descriptions, opinions, or explanations, and technical questions are often non-factoid questions. ~\cite{song2017summarizing, hashemi2020antique, xu2017answerbot}}).
    In their first stage, they combined the word embeddings with the traditional traditional IDF metrics for retrieving relevant questions. 
    Their method has been proven to be effective in the task of relevant question retrieval compared with a set of baselines. 
    
    \rev{
    \item \textbf{CROKAGE} 
    More recently, 
    Da et al.~\cite{da2020crokage} proposed a model named CROKAGE to recommend relevant solutions from Stack Overflow for a searching query. 
    They aimed to address the lexical gap problem between the query and the solutions via using a multi-factor relevance mechanism. 
    To be more specific, they calculated the final relevance score by combining four types of scores (\textit{lexical score}, \textit{semantic score}, \textit{API method score}, \textit{API class score}).
    We adapt their approach for our task of retrieving semantically-equivalent questions. 
    Particularly, we only retain the \textit{lexical scores} and \textit{semantic scores} for this research question since question titles usually don't contain API classes.
    }
\end{itemize}


\begin{table*}[htb]
\centering
\caption{ \rev{Effectiveness evaluation (Python)} }
\label{tab:effect_eval_py}
\rev{
\resizebox{0.99\textwidth}{!}{
    \begin{tabular}{||l|cccc|cccc||} 
      \hline
      Model & P@1  & P@2 & P@3 & P@4 & DCG@2 & DCG@3 & DCG@4 & DCG@5 \\
      \hline
      IR & $32.1\pm2.4\%$ 
                  & $44.6\pm2.6\%$ 
                  & $54.0\pm3.6\%$ 
                  & $64.8\pm4.8\%$ 
                  & $40.0\pm2.1\%$ 
                  & $44.7\pm2.5\%$ 
                  & $49.3\pm2.9\%$ 
                  & $63.0\pm1.3\%$ \\
      Word2Vec & $23.9\pm2.8\%$ 
            & $40.6\pm1.9\%$ 
            & $56.1\pm1.9\%$ 
            & $72.0\pm2.0\%$ 
            & $34.4\pm1.9\%$ 
            & $42.2\pm1.8\%$ 
            & $49.0\pm1.5\%$
            & $59.9\pm1.2\%$ \\ 
      FastText & $28.4\pm3.6\%$ 
            & $42.1\pm2.9\%$ 
            & $53.9\pm3.1\%$ 
            & $66.7\pm2.3\%$ 
            & $37.0\pm2.9\%$ 
            & $42.9\pm3.0\%$ 
            & $48.4\pm2.6\%$
            & $61.3\pm1.7\%$ \\         
      Sent2Vec & $26.0\pm2.6\%$ 
                  & $45.9\pm2.5\%$ 
                  & $61.9\pm3.2\%$ 
                  & $78.6\pm3.1\%$ 
                  & $38.6\pm2.1\%$ 
                  & $46.5\pm2.0\%$ 
                  & $53.7\pm2.2\%$ 
                  & $62.0\pm1.2\%$ \\
      AnswerBot & $33.9\pm2.7\%$ 
            & $54.5\pm4.2\%$ 
            & $68.9\pm3.2\%$ 
            & $81.4\pm2.6\%$ 
            & $46.9\pm3.4\%$ 
            & $54.1\pm2.9\%$ 
            & $59.5\pm2.4\%$
            & $66.7\pm1.7\%$ \\
      CROKAGE & $36.7\pm2.6\%$ 
            & $51.5\pm3.1\%$ 
            & $64.4\pm1.9\%$ 
            & $78.4\pm2.5\%$ 
            & $46.0\pm2.7\%$ 
            & $52.4\pm2.0\%$ 
            & $58.5\pm1.6\%$
            & $66.8\pm1.3\%$ \\
      \hline
      \textbf{Ours}  & $\mathbf{45.8\pm3.2\%}$ 
                     & $\mathbf{60.5\pm2.5\%}$ 
                     & $\mathbf{72.3\pm2.7\%}$ 
                     & $\mathbf{85.1\pm2.6\%}$ 
                     & $\mathbf{55.0\pm2.6\%}$ 
                     & $\mathbf{61.0\pm2.5\%}$
                     & $\mathbf{66.5\pm2.1\%}$
                     & $\mathbf{72.2\pm1.6\%}$ \\ 
      \hline
\end{tabular}
}
}
\end{table*}

\begin{table*}[htb]
\centering
\caption{ \rev{Effectiveness evaluation (Java)} }
\label{tab:effect_eval_java}
\rev{
\resizebox{0.99\textwidth}{!}{
    \begin{tabular}{||l|cccc|cccc||} 
      \hline
      Model & P@1  & P@2 & P@3 & P@4 & DCG@2 & DCG@3 & DCG@4 & DCG@5 \\
      \hline
      IR & $31.1\pm2.0\%$ 
                  & $43.1\pm2.5\%$ 
                  & $53.7\pm2.4\%$ 
                  & $62.8\pm2.5\%$ 
                  & $38.6\pm2.1\%$ 
                  & $43.9\pm2.1\%$ 
                  & $47.9\pm1.8\%$ 
                  & $62.3\pm1.2\%$ \\
      Word2Vec & $26.4\pm3.2\%$ 
            & $37.7\pm4.3\%$ 
            & $51.7\pm4.1\%$ 
            & $66.3\pm3.3\%$ 
            & $33.6\pm3.7\%$ 
            & $40.5\pm3.6\%$ 
            & $46.8\pm3.0\%$
            & $59.9\pm2.0\%$ \\ 
      FastText & $29.8\pm1.9\%$ 
            & $42.7\pm2.6\%$ 
            & $53.0\pm5.2\%$ 
            & $65.8\pm3.8\%$ 
            & $38.0\pm2.2\%$ 
            & $43.1\pm3.5\%$ 
            & $48.6\pm2.7\%$
            & $61.9\pm1.4\%$ \\          
      Sent2Vec & $24.8\pm2.2\%$ 
                  & $45.8\pm3.4\%$ 
                  & $62.0\pm2.8\%$ 
                  & $80.8\pm2.6\%$ 
                  & $38.0\pm2.9\%$ 
                  & $46.1\pm2.5\%$ 
                  & $54.2\pm1.8\%$ 
                  & $61.7\pm1.4\%$ \\
      AnswerBot & $33.3\pm3.0\%$ 
            & $48.1\pm2.4\%$ 
            & $61.1\pm2.5\%$ 
            & $75.0\pm2.3\%$ 
            & $42.6\pm2.4\%$ 
            & $49.1\pm2.5\%$ 
            & $55.1\pm2.3\%$
            & $64.8\pm1.6\%$ \\
      CROKAGE & $38.9\pm4.3\%$ 
            & $53.4\pm5.5\%$ 
            & $65.3\pm4.8\%$ 
            & $78.0\pm2.8\%$ 
            & $48.0\pm5.0\%$ 
            & $54.0\pm4.4\%$ 
            & $59.4\pm3.3\%$
            & $68.0\pm2.6\%$ \\
      \hline
      \textbf{Ours}  & $\mathbf{58.6\pm4.3\%}$ 
                     & $\mathbf{73.9\pm1.4\%}$ 
                     & $\mathbf{84.3\pm2.0\%}$ 
                     & $\mathbf{92.9\pm1.8\%}$ 
                     & $\mathbf{68.2\pm2.3\%}$ 
                     & $\mathbf{73.4\pm2.0\%}$
                     & $\mathbf{77.1\pm1.4\%}$
                     & $\mathbf{79.9\pm1.7\%}$ \\ 
      \hline
\end{tabular}
}
}
\end{table*}

\subsubsection{Experimental Results}
The experimental results of our \textit{QueryRewriter} compared to the above baselines for Python and Java are summarized in Table~\ref{tab:effect_eval_py} and Table~\ref{tab:effect_eval_java} respectively. 
We do not report $P@5$ and $DCG@1$ in our tables, since 
$P@5$ is always equal to 1 and $DCG@1$ is always equal to $P@1$, both can be easily inferred from the tables.
The best performing system for each column is highlighted in boldface. 
From the table, several points stand out: 
\begin{enumerate}
    \item 
    \rev{
    It is a little surprising that the \textbf{word embedding-based approaches (e.g., Word2Vec, FastText and Sent2Vec) achieve the worst performance} regarding $P@1$. 
    }
    This indicates that retrieving semantically-equivalent questions from a set of similar questions is a non-trivial task. 
    Word2Vec and Sent2Vec map each question to a fixed-length vector, so the vectors of similar questions are also close together in vector space. 
    However, due to the reason that all candidate questions in $\mathcal{Q}$ are similar to each other, it is thus very hard for Sent2Vec and Word2Vec approach to distinguish the duplicate questions from a list of similar questions. 
    This is the reason why their performance is weak and ineffective for retrieving semantically-equivalent questions.
    
    \item 
    \rev{Regarding the $P@1$ score, the traditional \textbf{IR method performs better} than the word embedding-based methods (i.e., Word2Vec and Sent2Vec). 
    }
    For the IR based approach, it relies heavily on how similar the testing question and its duplicate question are. 
    Considering that many duplicate questions may share same lexical units with the testing questions, these duplicate questions can be easily retrieved by the IR-based approach. 
    \minor{However, there is still a large number of duplicate questions that are semantically-equivalent with only a few common words or without at all (e.g., as shown in Fig.~\ref{fig:motivation})}, the IR-based approach are unable to to retrieve these questions correctly solely based on the similarity between words or tokens.
    This may also explain its surprisingly low score as $K$ increases.

    \item 
    \rev{
    AnswerBot and CROKAGE perform better than other baselines excluding our proposed model. 
    This is because both AnswerBot and CROKAGE combines the lexical-based model (i.e., IR) and semantic-based model (e.g, Word2Vec, FastText and Sent2Vec) for modeling the question titles, which also signals that solely based on the lexical features or semantic features is not sufficient for our task.
    It is also notable that the performance of CROKAGE is better than the AnswerBot approach.
    We attribute this to the different word embedding techniques they employed. 
    AnswerBot combines IDF metrics with Word2Vec model, while CROKAGE combines IDF metrics with FastText model. 
    This signals that the FastText model has its advantage as compared to Word2Vec model for modeling the duplicate question titles. 
    }
    
    \item It is clear that \textbf{our model outperforms all the other methods by a large margin} in terms of $P@K$ and $DCG@K$ scores of different depth. We attribute this to the following reasons:
    Firstly, \textit{QueryRewriter} generates multiple paraphrase questions for a given testing question.
    Adding these paraphrase questions can reduce the lexical gap between the testing questions and its corresponding duplicate questions and alleviate the {\em query mismatch} problem for different developers. 
    Secondly, all of the baseline methods including our approach can be viewed as variants of embedding algorithm(s), which can map the questions into vectors of a high-dimensional space and then calculate the relevance score between vectors.
    Hence the key of retrieving semantically-equivalent questions relies on how good the embeddings are for capturing the semantics of different duplicate questions. 
    Our approach has its advantage as compared to other baselines because our \textit{QueryRewriter} trains the historical duplicate question pairs in Stack Overflow by using a text-to-text transformer.
    As a result, the embeddings generated by our approach are more suitable for identifying the semantically-equivalent questions.
    The superior performance of our approach also verifies the embeddings generated by our approach convey a lot of valuable information.
\end{enumerate}

\noindent
\framebox{\parbox{\dimexpr\linewidth-2\fboxsep-2\fboxrule}{
\textbf{Answer to RQ-1: How effective is our approach for retrieving semantically-equivalent questions? -- we conclude that our approach is highly effective for semantically-equivalent question retrieval in Stack Overflow.}}} 

\subsection{\revv{RQ-2: Google Search Results Analysis}}
\revv{
Google search engine has been widely used by developers to search online resources and improve their daily productivity. 
Typically, when developer encounters a technical problem, he or she usually use a web search engine (e.g., Google) to obtain a list of relevant posts in Stack Overflow.
Therefore, the Google search engine can be considered as a baseline for searching duplicate questions intuitively. 
In this research question, for a given question, we would like to investigate whether our \textit{QueryRewriter} can rank its duplicate question higher up among other question candidates compared with Google search. 
}

\begin{table}
\caption{\revv{QueryRewriter vs. Google (Python)}}
\label{tab:python_google_eval}
\vspace*{-10pt}
\begin{center}
\revv{
\begin{tabular}{||c|c|c|c|c|c|c||}
    \hline
    \multirow{3}{*}{Dataset} & \multicolumn{2}{c|}{\bf Overall Dataset} & \multicolumn{2}{c|}{\bf Successful Dataset} & \multicolumn{2}{c|}{\bf Failed DataSet} \\\cline{2-7} 
    & {Count} & {2,000} & {Count} & {393} & {Count} & {1607} \\
    \cline{2-7} 
    & {Avg. Score} & {3.01} & {Avg. Score} & {6.99} & {Avg. Score} & {2.05} \\
    \hline\hline
    Approch & {Google} & {Ours} & {Google} & {Ours} & {Google} & {Ours} \\
    \hline
    P@1 & $20.0\%$ & $\mathbf{35.6\%}$ & $59.3\%$ & $\mathbf{47.0\%}$ & $10.4\%$ & $\mathbf{32.9\%}$ \\
    P@2 & $36.4\%$ & $\mathbf{54.2\%}$ & $78.4\%$ & $\mathbf{67.9\%}$ & $26.1\%$ & $\mathbf{50.9\%}$ \\
    P@3 & $44.6\%$ & $\mathbf{65.4\%}$ & $89.3\%$ & $\mathbf{79.9\%}$ & $33.6\%$ & $\mathbf{61.9\%}$ \\
    P@4 & $47.1\%$ & $\mathbf{76.2\%}$ & $97.7\%$ & $\mathbf{92.6\%}$ & $34.7\%$ & $\mathbf{72.2\%}$ \\
    P@5 & $75.4\%$ & $\mathbf{90.9\%}$ & $100.0\%$ & $\mathbf{99.0\%}$ & $69.4\%$ & $\mathbf{90.0\%}$ \\
    \hline
    DCG@1 & $20.0\%$ & $\mathbf{35.6\%}$ & $59.3\%$ & $\mathbf{47.0\%}$ & $10.4\%$ & $\mathbf{32.9\%}$ \\
    DCG@2 & $30.3\%$ & $\mathbf{47.4\%}$ & $71.3\%$ & $\mathbf{60.2\%}$ & $20.3\%$ & $\mathbf{44.2\%}$ \\
    DCG@3 & $34.4\%$ & $\mathbf{53.0\%}$ & $76.8\%$ & $\mathbf{66.2\%}$ & $24.0\%$ & $\mathbf{49.7\%}$ \\
    DCG@4 & $35.5\%$ & $\mathbf{57.6\%}$ & $80.4\%$ & $\mathbf{71.7\%}$ & $24.5\%$ & $\mathbf{54.2\%}$ \\
    DCG@5 & $46.5\%$ & $\mathbf{63.3\%}$ & $81.3\%$ & $\mathbf{74.2\%}$ & $37.9\%$ & $\mathbf{60.7\%}$ \\
    \hline
\end{tabular}
}
\end{center}
\end{table}

\begin{table}
\caption{\revv{QueryRewriter vs. Google (Java)}}
\label{tab:java_google_eval}
\vspace*{-10pt}
\begin{center}
\revv{
\begin{tabular}{||c|c|c|c|c|c|c||}
    \hline
    \multirow{3}{*}{Dataset} & \multicolumn{2}{c|}{\bf Overall Dataset} & \multicolumn{2}{c|}{\bf Successful Dataset} & \multicolumn{2}{c|}{\bf Failed Dataset} \\\cline{2-7} 
    & {Count} & {2,000} & {Count} & {410} & {Count} & {1590} \\
    \cline{2-7} 
    & {Avg. Score} & {2.52} & {Avg. Score} & {6.93} & {Avg. Score} & {1.38} \\
    \hline\hline
    Approch & {Google} & {Ours} & {Google} & {Ours} & {Google} & {Ours} \\
    \hline
    P@1 & $21.5\%$ & $\mathbf{39.0\%}$ & $62.7\%$ & $\mathbf{48.5\%}$ & $10.9\%$ & $\mathbf{35.5\%}$ \\
    P@2 & $39.6\%$ & $\mathbf{59.8\%}$ & $80.7\%$ & $\mathbf{71.9\%}$ & $29.0\%$ & $\mathbf{56.7\%}$ \\
    P@3 & $48.8\%$ & $\mathbf{70.5\%}$ & $92.1\%$ & $\mathbf{82.4\%}$ & $37.7\%$ & $\mathbf{67.5\%}$ \\
    P@4 & $51.2\%$ & $\mathbf{79.6\%}$ & $98.0\%$ & $\mathbf{94.1\%}$ & $39.1\%$ & $\mathbf{75.8\%}$ \\
    P@5 & $76.0\%$ & $\mathbf{91.6\%}$ & $99.5\%$ & $\mathbf{98.6\%}$ & $70.0\%$ & $\mathbf{89.8\%}$ \\
    \hline
    DCG@1 & $21.5\%$ & $\mathbf{39.0\%}$ & $62.7\%$ & $\mathbf{48.5\%}$ & $10.9\%$ & $\mathbf{36.5\%}$ \\
    DCG@2 & $33.0\%$ & $\mathbf{52.1\%}$ & $74.0\%$ & $\mathbf{63.3\%}$ & $22.4\%$ & $\mathbf{49.2\%}$ \\
    DCG@3 & $37.6\%$ & $\mathbf{57.5\%}$ & $79.8\%$ & $\mathbf{68.6\%}$ & $26.7\%$ & $\mathbf{54.7\%}$ \\
    DCG@4 & $38.6\%$ & $\mathbf{61.4\%}$ & $82.3\%$ & $\mathbf{73.6\%}$ & $27.3\%$ & $\mathbf{58.2\%}$ \\
    DCG@5 & $48.2\%$ & $\mathbf{66.0\%}$ & $82.9\%$ & $\mathbf{75.3\%}$ & $39.2\%$ & $\mathbf{63.7\%}$ \\
    \hline
\end{tabular}
}
\end{center}
\end{table}

\subsubsection{\revv{Experimental Setup}}
\revv{
Regarding the Google search results analysis, for each duplicate question pair $\langle q_{d}, q_{m} \rangle$ in the testing set, we first use the question title of $q_{d}$ as a searching query and feed to the Google search engine, we then crawl the first page returned by the Google search engine and extract all the Stack Overflow related posts as the Google search results, the Google search results on our testing set are also saved in our replication package. 
Similar to RQ-1, for each testing question $q_{d}$, we obtain a set of relevant questions $\mathcal{G}=\{q_{i}\}(1 \leqslant i \leqslant k)$, where $k$ is the number of relevant questions returned by the Google search engine. 
We remove $q_{d}$ from $\mathcal{G}$ if $\mathcal{G}$ contains $q_{d}$, and add $q_{m}$ to $\mathcal{G}$ if $q_{m}$ is not within $\mathcal{G}$. 
In the light of this, for the given testing question $q_{d}$, we can ensure its master question $q_{m}$ is in the evaluation candidate pool $\mathcal{G}$. 
Hereafter, we pair the given testing question with each of the candidate question in $\mathcal{G}$ and utilize our model to generate a ranking list of the candidate questions by estimating their relevance scores.
We evaluate our approach and Google search engine with $\mathcal{G}$ in terms $P@K$ and $DCG@K$. 
}
\subsubsection{\revv{Experimental Results.}}
\revv{
The experimental results of \textit{QueryRewriter} and Google search engine for Python and Java are presented in Table~\ref{tab:python_google_eval} and Table~\ref{tab:java_google_eval} respectively. 
To have a deeper understanding about the performance of the Google search engine and our approach, we further split our testing dataset into \textit{successful dataset} and \textit{failed dataset} based on whether the duplicate question can be successfully retrieved by the Google search engine. 
To be more specific, for a given testing question, if Google search engine successfully finds its duplicate question, then we add this testing question to the \textit{successful dataset}, otherwise it will be added to the \textit{failed dataset}. 
From the tables, we can observe the following points: 
\begin{enumerate}
    \item The Google search achieves a poor performance for retrieving the duplicate questions on the testing set. 
    Regarding the 2,000 testing questions, the Google search engine only finds 393 duplicate questions successfully for the Python and 410 duplicate questions for Java dataset, which means that around 80\% of the duplicate questions can not be retrieved effectively by Google search engine.
    Since we manually added the master question to the \textit{failed dataset} for evaluation, the $P@K$ and $DCG@K$ are not equal to zero for Google search engine. 
    \item
    The quality of the questions within the \textit{successful dataset} is much higher than the \textit{failed dataset}.
    The successfully retrieved questions are grouped together into the \textit{successful dataset}, while the other questions are grouped together into the \textit{failed dataset}. 
    We then compute the average question scores with respect to these two datasets, which has been widely used for measuring a post quality in Stack Overflow~\cite{yang2016security, ahmed2018concurrency, bajaj2014mining, rosen2016mobile}.
    For the questions within the \textit{successful dataset}, the average score is 6.99 for Python and 6.93 for Java, much higher than the questions of the \textit{failed dataset} (e.g., 2.05 for Python and 1.38 for Java). 
    The large number of low quality posts lead to the comparatively suboptimal performance of Google search engine for retrieving duplicate questions. 
    \item Our approach performs better than the Google search engine on \textit{failed dataset}, while the Google search engine outperforms our approach on \textit{successful dataset}. 
    The superior performance of the Google search engine on the \textit{successful dataset} indicates its capability of handling the high quality posts. 
    Considering these high quality posts often obtain more attentions (e.g., user views and clicks), it is not surprising that the Google search engine can easily record and index them for recommendation. 
    However, the Google search engine fails to return the desired results on the \textit{failed dataset}, which suggests the Google search engine lacks the ability to deal with low quality posts. 
    \item Our approach outperforms the Google search engine by a large margin on the overall testing dataset. 
    This is because our approach stably and substantially performs better than the Google search baseline on the \textit{failed dataset}, which accounts for the majority part of the testing questions. 
    When dealing with the low quality posts in the \textit{failed dataset}, our approach can successfully rank the desired results (i.e., duplicate question) higher up among other question candidates. 
    The advantages of our approach for this scenario are due to the following reasons: First, without relying on the feedback from online users, our approach is trained with the historical duplicate question pairs, the various quality of duplicate questions can increase the ability and generality for handling the low quality posts. 
    Second, our approach can map the semantic-equivalent questions into the vector space that are close to each other, the lexical gap between duplicate question pairs can be filled by using the question embeddings. 
\end{enumerate}
}

\noindent
\framebox{\parbox{\dimexpr\linewidth-2\fboxsep-2\fboxrule}{
\revv{
\textbf{Answer to RQ-2: How effective is our approach compared with Google search engine? -- we conclude that our approach performs better than Google search engine when dealing with low quality posts in Stack Overflow.}}}
}

\subsection{\rev{\revv{RQ-3:} Context Analysis}}
\rev{
As the technical Q\&A sites (i.e., Stack Overflow) are used by developers and professional experts, the questions in these Q\&A communities are, more often than not, very professional with specific domain context.
For example, these questions often include software-specific entities (e.g., software libraries/frameworks, software-specific concepts).
To investigate whether the domain-specific context could influence the performance of our approach, or in other words, whether our model can learn the domain-specific context features from the training corpus, we perform a context analysis for this study.
}
\subsubsection{\rev{Experimental Setup}}
\rev{
For the context analysis, we create a training set without domain-specific context and use the same testing set for evaluation. 
To do this, we check each token in the training corpus if the token is a normal English word (using the NLTK package), and we only keep the English words and remove the other non proper English words.
As a result, the software-specific terms within the master questions and duplicate questions are deleted. 
After that, We retrain our \textit{QueryRewriter} and all the baselines on the non domain-specific context training corpus and perform the same evaluation as in RQ-1.  
}

\subsubsection{\rev{Experimental Results}} 
\rev{
The experimental results of our \textit{QueryRewriter} and other baselines for Python and Java dataset are presented in Table~\ref{tab:context_py} and Table~\ref{tab:context_java} respectively.
From the tables, we can deduce the following key findings: 
\begin{enumerate}
    \item The performance of all models decreases on the training set without domain-specific context. 
    This suggests that the domain-specific context has a major influence on the overall performance. 
    We further counted the unique tokens of the training corpus with and without the domain-specific context, for Python dataset, only 4,756 out of 19,208 tokens are remained; for Java dataset, only 4,874 out of 22,344 tokens are remained.
    The large proportion of domain-specific context in Stack Overflow may also explain the performance drop in all baseline methods. 
    \item It is notable that after adding the domain-specific context, our model achieves the biggest performance rise among different models. 
    This reveals that our model is effective in learning the domain-specific features and knowledge. 
    Moreover, the performance of our proposed model still outperforms the other baseline approaches even under the training corpus without domain-specific context, which justifies the robustness of our model.  
\end{enumerate}
}

\noindent
\framebox{\parbox{\dimexpr\linewidth-2\fboxsep-2\fboxrule}{
\textbf{
\rev{\revv{Answer to RQ-3}: How effective is our approach for capturing contextual information? -- we conclude that the domain-specific context can influence the model's performance, and our approach is highly effective for learning domain-specific context information.}
}}} 

\begin{table}
\caption{\rev{Context Analysis of $P@1$ (Python)}}
\label{tab:context_py}
\begin{center}
\rev{
\begin{tabular}{|c|c|c|c|}
    \hline
    {\bf Approach} & {\bf Without Context} & {\bf With Context} & {\bf $\bigtriangleup$ Improve} \\
    \hline\hline
    IR  & $24.9\pm2.2\%$ & $32.1\pm1.2\%$ & $28.9\%$ \\
    \hline
    Word2Vec & $22.2\pm2.9\%$ & $23.9\pm2.8\%$  & $7.7\%$ \\
    \hline
    FastText & $24.4\pm1.5\%$ & $28.4\pm3.6\%$ & $16.4\%$ \\
    \hline
    Sent2Vec & $24.2\pm2.1\%$ & $26.0\pm2.6\%$ & $7.4\%$ \\
    \hline
    AnswerBot & $29.4\pm3.8\%$ & $33.9\pm2.7\%$ & $15.3\%$ \\
    \hline
    CROKAGE & $26.9\pm2.3\%$  & $36.7\pm2.6\%$ & $36.4\%$ \\
    \hline
    Ours & $31.1\pm2.4\%$ & $45.8\pm3.2\%$ & $47.3\%$ \\
    \hline
\end{tabular}
}
\end{center}
\end{table}

\begin{table}
\caption{\rev{Context Analysis of $P@1$ (Java)}}
\label{tab:context_java}
\begin{center}
\rev{
\begin{tabular}{|c|c|c|c|}
    \hline
    {\bf Approach} & {\bf Without Context} & {\bf With Context} & {\bf $\bigtriangleup$ Improve} \\
    \hline\hline
    IR  & $26.9\pm2.1\%$ & $31.1\pm2.0\%$ & $15.6\%$ \\
    \hline
    Word2Vec & $24.5\pm2.4\%$ & $26.4\pm3.2\%$  & $7.6\%$ \\
    \hline
    FastText & $25.5\pm2.6\%$ & $29.8\pm1.9\%$ & $16.7\%$ \\
    \hline
    Sent2Vec & $23.4\pm2.5\%$ & $24.8\pm2.2\%$ & $6.0\%$ \\
    \hline
    AnswerBot & $26.8\pm3.3\%$ & $33.3\pm3.0\%$ & $24.3\%$ \\
    \hline
    CROKAGE & $28.2\pm2.2\%$  & $38.9\pm4.3\%$ & $37.9\%$ \\
    \hline
    Ours & $37.7\pm4.2\%$ & $58.6\pm4.3\%$ & $55.4\%$ \\
    \hline
\end{tabular}
}
\end{center}
\end{table}

\subsection{\revv{RQ-4:} Ablation Analysis}
\label{sec:rq2}
\rev{
Ablation analysis is a common method to estimate the contribution of a component to the overall system~\cite{gao2020generating}. 
It studies the performance of a system by removing certain components.
Our \textit{QueryRewriter} learns to encode the duplicate question pairs from Stack Overflow, so that two semantically-equivalent questions are close in terms of vector representation. 
When we perform question retrieval tasks, a main novelty of our approach is adding paraphrase questions generated by \textit{QueryRewriter}. 
In this research question, we perform an ablation analysis to investigate if the novel aspect that we introduce helps. 
To be more specific, we investigate the effectiveness of the added paraphrase question to our model. 
}

\begin{table}
\caption{Ablation Analysis}
\label{tab:ab_eval}
\vspace*{-10pt}
\begin{center}
\rev{
\begin{tabular}{|c||c|c||c|c|}
    \hline
    \multirow{2}{*}{Measure} & \multicolumn{2}{c||}{Python} & \multicolumn{2}{c|}{Java} \\\cline{2-5} 
    & {\bf Drop-PQ} & {\bf Ours} & {\bf Drop-PQ} & {\bf Ours} \\
    \hline\hline
    P@1 & $36.9\%$ & $\mathbf{45.8\%}$ & $49.5\%$ & $\mathbf{58.6\%}$ \\
    \hline
    P@2 & $50.0\%$ & $\mathbf{60.5\%}$ & $65.3\%$ & $\mathbf{73.9\%}$ \\
    \hline
    P@3 & $62.1\%$ & $\mathbf{72.3\%}$ & $76.9\%$ & $\mathbf{84.3\%}$ \\
    \hline
    P@4 & $74.5\%$ & $\mathbf{85.1\%}$ & $87.5\%$ & $\mathbf{92.9\%}$ \\
    \hline
    DCG@2 & $45.1\%$ & $\mathbf{55.0\%}$ & $59.4\%$ & $\mathbf{68.2\%}$ \\
    \hline
    DCG@3 & $51.2\%$ & $\mathbf{61.0\%}$ & $65.3\%$ & $\mathbf{73.4\%}$ \\
    \hline
    DCG@4 & $56.5\%$ & $\mathbf{66.5\%}$ & $69.8\%$ & $\mathbf{77.1\%}$ \\
    \hline
    DCG@5 & $66.4\%$ & $\mathbf{72.2\%}$ & $74.7\%$ & $\mathbf{79.9\%}$ \\
    \hline
\end{tabular}
}
\end{center}
\end{table}

\subsubsection{Experimental Setup}
For the ablation analysis, 
we compare our approach with one of its incomplete variants, named \textbf{Drop-PQ}.
Different from our proposed model, \textbf{Drop-PQ} removes all the generated paraphrase questions added to our model, and only keeps the embedding of the original testing question. 
By going through the same steps as our approach in Section~\ref{sec:que_ret}, we can evaluate \textbf{Drop-PQ} model for the semantic-equivalent question retrieval task. 

\subsubsection{Experimental Results}
The comparison results between \textbf{Drop-PQ} and our approach are displayed in Table~\ref{tab:ab_eval}.
We observe the following points from the table: 
\begin{enumerate}
    \item  By comparing the results of our approach and \textbf{Drop-PQ}, it is clear that \textbf{incorporating the paraphrase questions improves the overall performance}.
    When adding the paraphrase questions to our model, the $P@1$ score is improved by 24.1\% in Python and 18.3\% in Java dataset.
    We attribute this to the ability of paraphrase questions to reduce the lexical gap between the semantically-equivalent questions.
    \item By comparing the results of \textbf{Drop-PQ} and our previous baselines in RQ-1, we can see that \textbf{even by dropping the paraphrase questions, \textbf{Drop-PQ} still achieves better or comparable results than other baselines}. 
    This is because, even removing the paraphrase questions, the question embeddings are generated from the same Encoder of our text-to-text transformer. This further verifies the importance and necessity of the embeddings of our approach.
\end{enumerate}

\noindent
\framebox{\parbox{\dimexpr\linewidth-2\fboxsep-2\fboxrule}{
\textbf{\revv{Answer to \textit{RQ-4:}} How effective is the paraphrase generation component added to our \textit{QueryRewriter}? - We conclude that adding paraphrase questions significantly improves the overall performance of our model}}}.

\subsection{RQ-5: Effectiveness of CodeSelector}

When trying to solve daily coding problems,
developers often formulate their problems as a question and/or a few keywords to some search engines. 
The search engine returns a list of potential posts which may contain useful answers. 
Due to the complexity of the online CQA forums and the large volume of information generated from it, software developers may encounter the {\em information overload} problem wherein the massive amounts of information makes it hard to be aware of the most relevant resources to meet the information needs of the developers.  
To alleviate this {\em information overload} problem, 
we propose a \textit{CodeSelector} to rank the code snippets candidates via pairwise comparisons. 
To evaluate our approach, we conducted a large-scale automatic evaluation experiment to evaluate the effectiveness of our approach to identify the best code solution for a technical problem.

\subsubsection{Data Preparation}
To train the \textit{CodeSelector}, we first constructed positive and negative training samples in terms of preference pairs. 
For each Stack Overflow post, we extracted the code snippets (using $\langle code \rangle$ tags) within the post's question body and corresponding post question title. 
In order to avoid being context-specific, numbers and strings within a code snippet are replaced with special tokens ``NUMBER'' and ``STRING'' respectively. 
\rev{
We first adopted the NLTK~\cite{bird2004nltk} toolkit to tokenize the code snippets,
we removed the code snippets that are too long (more than 512 tokens) or too short (less than 5 tokens). 
This is because for a given code snippet, it is unable to capture the code semantics if it is too short. 
We set the 512 as the maximum number of code snippet tokens since the maximum input sequence length of BERT~\cite{devlin2018bert} is restricted to 512 tokens, and this setting is sufficient for most cases of code snippets in Stack Overflow~\cite{gao2020generating}. 
}
For question titles, we only preserved the ``how'' related questions in Stack Overflow. 
The resulting $\langle question, code~snippet \rangle$ pairs are added to our corpus. 
Towards this end, we collected 218K QC pairs for Python dataset and 272K QC pairs for Java dataset.

For each question in the corpus, we make the code snippet associated with the accepted answer or the highest-vote answer as the \emph{best code snippet}, the code snippet associated with the non-accepted answers as the \emph{non-best code snippet}, and randomly select the code snippet from other questions as the \emph{non-relevant code snippet}.  
According to our heuristic rules described above in Section~\ref{sec:code_sel}, 
we constructed our dataset with balanced positive and negative preference pairs. 
We randomly sampled 5,000 samples for validation and 5,000 pairs for testing respectively, and kept the rest for training. 
The details of the statistics of our collected dataset are summarized in Table~\ref{tab:QCdata}.

\begin{table} 
\caption{QC Dataset Statistics}
\begin{center}
\begin{tabular}{||c|l|c||}
    \hline
    \multirow{6}{*}{Python}   & \# $\langle q, cs \rangle$ Pairs & 218,717 \\ \cline{2-3}
                              & \# best code snippet  & 112,447 \\ \cline{2-3}
                              & \# non-best code snippet  & 106,270 \\ \cline{2-3}
                              & \# non-relevant code snippet  & 112,447 \\ \cline{2-3}
                              & \# Positive Samples & 141,074  \\ \cline{2-3}
                              & \# Negative Samples & 141,224  \\ \cline{2-3}
    \hline\hline
    \multirow{6}{*}{Java}   & \# $\langle q, cs \rangle$ Pairs & 272,120 \\ \cline{2-3}
                              & \# best code snippet  & 134,993 \\ \cline{2-3}
                              & \# non-best code snippet  & 137,127 \\ \cline{2-3}
                              & \# non-relevant code snippet  & 134,993 \\ \cline{2-3}
                              & \# Positive Samples & 177,340  \\ \cline{2-3}
                              & \# Negative Samples & 176,942 \\ \cline{2-3}
                                  
    \hline
\end{tabular}
\label{tab:QCdata}
\end{center}
\vspace{-0.0cm}
\end{table}

\subsubsection{Experimental Setup}
To evaluate our \textit{CodeSelector} performance for identifying the best code snippet for a technical question,  
for each question $q$ in the testing set, we employ the same KNN strategy in RQ-1 to search its top-K similar questions over the whole dataset.
Undoubtedly, the testing question itself can be found.
We then constructed a code snippet candidates pool $\mathcal{C}$ by gathering all the code snippets associated with the returned questions. 
In the light of this, we can ensure the ground truth code snippet (\emph{best code snippet}) is in the code snippet candidates pool $\mathcal{C}$. 
Following that, we pair the given question $q$ with each of the code snippet in $\mathcal{C}$ to make a QC pair. 
Hereafter, by doing a pairwise comparison between each two QC pairs, we can generate a ranking list of preference scores for each code snippet.
All the code snippet candidates can be ranked by their preference scores. 
For this code selection task, we also employ the same automatic evaluation metric $P@K$ and $DCG@K$ used in RQ-1. 
$P@K$ and $DCG@K$ stands for the proportion of the selected code snippets in the top-K that are the ground truth.

\subsubsection{Experimental Baselines}
To demonstrate the effectiveness of our proposed approach, we compare it with several competitive baseline approaches. 
We adapt these approaches slightly for our specific task, i.e., selecting the best code snippet from a pool of code snippet candidates.
We briefly introduce these approaches and our evaluation experimental settings below. 
For each method below, the involved parameters are carefully tuned, and the best performance of each approach is used to report the final results. 

\begin{itemize}
    \item \textbf{Traditional Classifiers} 
    Considering that our \textit{CodeSelector} ranks the code snippet candidates by doing classification between QC pairs, it is hence natural to compare our approach with traditional classifiers. 
    Recently Calefato et al.~\cite{calefato2019empirical} proposed an approach for best answer prediction problem by formulating it as a a binary-classification task. 
    The binary-classification methods output a score referring to a probability of relevance.
    They assessed 26 traditional classifiers for predicting the best answer in Stack Overflow.  
    We choose the two most effective traditional classifiers, xgbTree and RandomForest, to apply to our code snippet recommendation task. 
    In our experiments, we treated the pair of $\langle question, best~code~snippet \rangle$ as positive sample and the pair of $\langle question, non-best~code~snippet \rangle$ as negative sample, we then train the traditional classifiers with these training samples. 
    Thereafter, we rank the code snippet candidates by the relevance scores generated by the above trained classifiers.
    
    \item \textbf{Answer Ranking Methods}
    The code snippet selection need of our task is similar to the answer ranking problem in CQA forums. 
    Hence our task is transformed to find an optimal ranking order of the code snippet candidates according to the their relevance to the given query question. 
    Two answer ranking methods, i.e., AnswerBot~\cite{xu2017answerbot} and DeepAns~\cite{gao2020deepans} are chosen as baselines.
    Regarding the AnswerBot baseline, their user study showed a promising performance for selecting salient answers in the second stage of their approach. 
    Regarding the DeepAns baseline, they calculated a matching score via a deep neural network between each answer and the question title; the experimental results show that DeepAns is effective for selecting the most relevant answer compared with several state-of-the-art benchmarks. For both of these answer ranking methods, an overall score is computed to estimate the relevance between each answer and the question title. 
    For our task, we can replace the answer with the code snippet and thus adapt their answer ranking methods to our task of ranking code snippets among a set of code snippet candidates.

    \item \textbf{DL-based Code Search Methods}
    Another thread of similar research that is relevant to our work is code search. 
    A plethora of approaches have been investigated for searching code snippet in software repositories, and recent DL (deep learning) - based approaches have achieved promising results for this task. 
    The DL-based code search methods advocate the idea of mapping and matching data in a high-dimensional vector space.
    \rev{
    Three state-of-the-art DL-based code search methods, i.e., NCS~\cite{sachdev2018retrieval}, DeepCS~\cite{gu2018deep} and CROKAGE~\cite{da2020crokage} are chosen for our study.
    } 
    NCS is an unsupervised technique for neural code search proposed by Facebook~\cite{sachdev2018retrieval}. 
    They combine the word embeddings and TF-IDF weighting derived from a code corpus. 
    In our experiment, we used all the collected QC pairs as the code corpus for training the word embeddings and TF-IDF weightings. 
    DeepCS is a supervised technique which jointly embeds code and natural language description into a high-dimensional vector space proposed by Gu et al~\cite{gu2018deep}.
    The author constructed the $\langle C, D+, D- \rangle$ triples to train their model for minimizing the ranking loss, where $C$ is the code snippet, $D+$ and $D-$ are the correct and incorrect description respectively. 
    In our experiment, for each code snippet $C$, we treat the associated question title as the correct description $D+$, and treat a randomly selected question title as the incorrect description $D-$. 
    \rev{
    CROKAGE~\cite{da2020crokage} is a tool to deliver a comprehensible solution for a programming task.
    It chooses the top quality answers related to the task. 
    To mitigate the lexical gap problem, CROKAGE calculates the scores for candidate QA pairs using four similarity factors (e.g., \textit{lexical score}, \textit{semantic score}, \textit{API method score} and \textit{API class score}).  
    Specifically, they use TF-IDF and fastText embedding to calculate the \textit{lexical scores} and \textit{semantic scores} respectively.  
    They also reward the QA pairs which contain the top API methods and relevant API classes with \textit{API method scores} and \textit{API class scores}. 
    Their final outputs contain both code examples and code explanations.   
    Because our study mainly focuses on the code snippet recommendation task, we only retain the code examples part and remove the code explanation part. 
    }
\end{itemize}

\begin{table*}[htb]
\centering
\caption{Effectiveness evaluation of CodeSelector (Python)}
\label{tab:effect_eval_code_py}
\resizebox{0.99\textwidth}{!}{
    \begin{tabular}{||l|cccc|cccc||} 
      \hline
      Model & P@1  & P@2 & P@3 & P@4 & DCG@2 & DCG@3 & DCG@4 & DCG@5 \\
      \hline
      RandomForest & $26.6\pm1.6\%$ 
                  & $49.6\pm2.6\%$ 
                  & $69.7\pm2.0\%$ 
                  & $86.4\pm1.4\%$ 
                  & $41.1\pm2.1\%$ 
                  & $51.1\pm1.8 \%$ 
                  & $58.3\pm1.5\%$ 
                  & $63.5\pm1.0\%$ \\
      xgbTree & $24.3\pm1.5\%$ 
            & $47.3\pm2.4\%$ 
            & $69.1\pm2.5\%$ 
            & $85.8\pm1.8\%$ 
            & $38.8\pm1.9\%$ 
            & $49.7\pm1.8\%$ 
            & $56.9\pm1.4\%$
            & $62.4\pm0.9\%$ \\ 
      AnswerBot & $31.0\pm1.5\%$ 
                  & $51.1\pm2.3\%$ 
                  & $70.4\pm2.1\%$ 
                  & $87.4\pm1.5\%$ 
                  & $43.7\pm1.8\%$ 
                  & $53.3\pm1.4\%$ 
                  & $60.6\pm1.1\%$ 
                  & $65.5\pm0.8\%$ \\
      DeepAns & $29.6\pm2.2\%$ 
            & $52.3\pm1.9\%$ 
            & $71.2\pm1.2\%$ 
            & $88.5\pm1.2\%$ 
            & $43.9\pm1.9\%$ 
            & $53.4\pm1.5\%$ 
            & $60.8\pm1.3\%$
            & $65.3\pm1.1\%$ \\
      NCS & $29.8\pm1.6\%$ 
            & $52.7\pm2.8\%$ 
            & $71.3\pm2.0\%$ 
            & $88.3\pm1.9\%$ 
            & $44.2\pm2.2\%$ 
            & $53.5\pm1.6\%$ 
            & $60.9\pm1.6\%$
            & $65.3\pm1.0\%$ \\
      DeepCS & $29.5\pm2.2\%$ 
            & $51.3\pm2.3\%$ 
            & $70.3\pm1.5\%$ 
            & $86.7\pm1.4\%$ 
            & $43.3\pm2.2\%$ 
            & $52.8\pm1.7\%$ 
            & $59.8\pm1.4\%$
            & $64.9\pm1.2\%$ \\
      \rev{CROKAGE} & $ \rev{33.3\pm2.3\%} $ 
            & $ \rev{55.0\pm2.0\%} $ 
            & $ \rev{71.9\pm1.3\%} $ 
            & $ \rev{86.5\pm1.1\%} $ 
            & $ \rev{47.0\pm1.9\%} $ 
            & $ \rev{55.4\pm1.5\%} $ 
            & $ \rev{61.7\pm1.2\%} $
            & $ \rev{66.9\pm1.1\%} $ \\      
      \hline
      \textbf{Ours}  & $\mathbf{42.6\pm2.5\%}$ 
                     & $\mathbf{64.6\pm1.1\%}$ 
                     & $\mathbf{80.0\pm1.0\%}$ 
                     & $\mathbf{92.3\pm0.9\%}$ 
                     & $\mathbf{56.5\pm1.5\%}$ 
                     & $\mathbf{64.2\pm1.2\%}$
                     & $\mathbf{69.5\pm1.0\%}$
                     & $\mathbf{72.5\pm1.0\%}$ \\ 
      \hline
\end{tabular}
}
\end{table*}

\begin{table*}[htb]
\centering
\caption{Effectiveness evaluation of CodeSelector (Java)}
\label{tab:effect_eval_code_java}
\resizebox{0.99\textwidth}{!}{
    \begin{tabular}{||l|cccc|cccc||} 
      \hline
      Model & P@1  & P@2 & P@3 & P@4 & DCG@2 & DCG@3 & DCG@4 & DCG@5 \\
      \hline
      RandomForest & $24.8\pm2.9\%$ 
                  & $47.5\pm2.3\%$ 
                  & $67.8\pm2.1\%$ 
                  & $85.0\pm1.4\%$ 
                  & $39.1\pm2.2\%$ 
                  & $49.2\pm1.8\%$ 
                  & $57.0\pm1.7\%$ 
                  & $62.5\pm1.3\%$ \\
      xgbTree & $25.8\pm1.7\%$ 
            & $47.5\pm2.6\%$ 
            & $68.1\pm2.4\%$ 
            & $85.0\pm1.5\%$ 
            & $39.5\pm2.1\%$ 
            & $49.8\pm2.0\%$ 
            & $57.1\pm1.5\%$
            & $62.8\pm1.1\%$ \\ 
      AnswerBot & $31.4\pm2.1\%$ 
                  & $52.1\pm1.3\%$ 
                  & $70.9\pm1.1\%$ 
                  & $86.9\pm1.7\%$ 
                  & $44.5\pm1.4\%$ 
                  & $53.9\pm1.0\%$ 
                  & $60.7\pm0.8\%$ 
                  & $65.8\pm0.8\%$ \\
      DeepAns & $29.1\pm2.3\%$ 
            & $52.5\pm2.3\%$ 
            & $71.3\pm2.3\%$ 
            & $86.7\pm1.4\%$ 
            & $43.9\pm2.2\%$ 
            & $53.2\pm1.9\%$ 
            & $59.9\pm1.6\%$
            & $65.1\pm1.2\%$ \\
      NCS & $30.4\pm1.8\%$ 
            & $52.1\pm1.7\%$ 
            & $71.4\pm1.5\%$ 
            & $86.9\pm1.4\%$ 
            & $44.1\pm1.6\%$ 
            & $53.7\pm1.3\%$ 
            & $60.4\pm1.2\%$
            & $65.5\pm0.9\%$ \\
      DeepCS & $28.5\pm3.9\%$ 
            & $48.2\pm3.4\%$ 
            & $65.8\pm4.6\%$ 
            & $81.6\pm3.5\%$ 
            & $40.9\pm3.3\%$ 
            & $49.7\pm3.6\%$ 
            & $56.5\pm2.9\%$
            & $63.6\pm2.0\%$ \\
      \rev{CROKAGE} & $ \rev{33.6\pm1.8\%} $ 
            & $ \rev{54.6\pm1.8\%} $ 
            & $ \rev{72.3\pm1.9\%} $ 
            & $ \rev{86.6\pm1.1\%} $ 
            & $ \rev{46.8\pm1.6\%} $ 
            & $ \rev{55.7\pm1.4\%} $ 
            & $ \rev{61.8\pm1.1\%} $
            & $ \rev{67.0\pm0.8\%} $ \\  
      \hline
      \textbf{Ours}  & $\mathbf{42.4\pm1.9\%}$ 
                     & $\mathbf{66.1\pm1.8\%}$ 
                     & $\mathbf{81.6\pm1.5\%}$ 
                     & $\mathbf{92.6\pm1.6\%}$ 
                     & $\mathbf{57.3\pm1.6\%}$ 
                     & $\mathbf{65.1\pm1.5\%}$
                     & $\mathbf{69.8\pm1.5\%}$
                     & $\mathbf{72.7\pm0.9\%}$ \\ 
      \hline
\end{tabular}
}
\end{table*}

\subsubsection{Experimental Results}
The experimental results of our proposed model and the above baselines over our Python and Java datasets are summarized in Table~\ref{tab:effect_eval_code_py} and Table~\ref{tab:effect_eval_code_java} respectively. 
From the table, we can observe the following points:
\begin{enumerate}
    \item The performance of \textbf{traditional classifiers are comparatively suboptimal}.
    This indicates that traditional classifiers are unable to capture the semantics between the code snippets and the questions.
    
    \item The \textbf{answer ranking methods and the DL-based code search methods achieve similar results}. The underlying idea of these two kinds of methods is similar, namely the application of applying the embedding technique to map the raw data (including the questions and code snippets) into a high-dimensional space and then estimate the match score between them. 
    This may explain the reason why their performances are comparable with each other.
    \rev{
    CROKAGE has its advantage as compared to other benchmarks.
    This is caused by several reasons: First, it combines the lexical features and semantic features (using lexical score and semantic score) that is why it is superior to the traditional classifiers.
    Second, in addition to lexical features and semantic features, it also incorporates the API related features, this results in its superior to the other answer ranking methods and DL-based code search methods. 
    }
    
    \item \textbf{Our proposed model is substantially better than all of the baseline methods}. 
    We attribute this to the following reasons: 
    First, all of the baseline approach (including traditional classifiers, answer ranking methods as well as the DL-based code search methods) can be viewed as pointwise approaches. Pointwise approaches transform the task of ranking into classification or regression on single QC pairs.
    They are thus unable to consider the relative order or preference between different code snippets. 
    Nevertheless, ranking is more about predicting relative orders rather than precise relevance scores. 
    In light of this, we propose a pairwise approach to judge the preference relationship between any two given QC pairs rather than the absolute relevance value of a single QC pair. 
    Compared with the pointwise approaches, our model not only considers the relevance between a query and a code snippet, but also investigates the relevance preference of two QC pairs. 
    This is why it is superior to other pointwise baselines. Second, in addition to constructing the preference pairs, our model also incorporates the BERT model to embed the QC pairs. 
    The attention mechanism behind BERT makes it possible to express sophisticated functions beyond the simple weighted average, which results in its superior to the DL-based code search methods. 
    This also signals that the embeddings produced by BERT convey much valuable information, which can better capture the semantics between the user query question and the code snippets. 
    \item By comparing the evaluation results of the different datasets (i.e., Python and Java), we can see that our proposed model behaves consistently across different programming languages. 
    This also indicates the generalization ability of our approach. 
\end{enumerate}

\noindent
\framebox{\parbox{\dimexpr\linewidth-2\fboxsep-2\fboxrule}{
\textbf{Answer to \textit{RQ-5: How effective is our \textit{CodeSelector} for best code snippet selection?} - we conclude that our \textit{CodeSelector} is effective for selecting the correct code snippet for a given technical question. }}}

\begin{table}
\caption{Component-Wise Evaluation (Python)}
\label{tab:cw_eval_py}
\begin{center}
\begin{tabular}{|c|c|c|c|}
    \hline
    {\bf Measure} & {\bf Drop-Pairwise} & {\bf Drop-Bert} & {\bf Ours} \\
    \hline\hline
    P@1  & $36.4\pm1.8\%$ & $33.9\pm1.2\%$ & $\mathbf{42.6\pm2.5\%}$ \\
    \hline
    P@2 & $59.7\pm2.1\%$ & $57.5\pm1.1\%$  & $\mathbf{64.6\pm1.1\%}$ \\
    \hline
    P@3 & $78.1\pm1.7\%$ & $76.7\pm1.9\%$ & $\mathbf{80.0\pm1.0\%}$ \\
    \hline
    P@4 & $91.5\pm1.0\%$ & $90.5\pm1.2\%$ & $\mathbf{92.3\pm0.9\%}$ \\
    \hline
    DCG@2 & $51.1\pm1.8\%$ & $48.8\pm1.0\%$ & $\mathbf{56.5\pm1.5\%}$ \\
    \hline
    DCG@3 & $60.3\pm1.5\%$ & $58.4\pm1.2\%$ & $\mathbf{64.2\pm1.2\%}$ \\
    \hline
    DCG@4 & $66.1\pm1.2\%$ & $64.4\pm0.9\%$ & $\mathbf{69.5\pm1.0\%}$ \\
    \hline
    DCG@5 & $69.4\pm1.0\%$ & $68.0\pm0.6\%$ & $\mathbf{72.5\pm1.0\%}$  \\
    \hline
\end{tabular}
\end{center}
\end{table}

\begin{table}
\caption{Component-Wise Evaluation (Java)}
\label{tab:cw_eval_java}
\begin{center}
\begin{tabular}{|c|c|c|c|}
    \hline
    {\bf Measure} & {\bf Drop-Pairwise} & {\bf Drop-Bert} & {\bf Ours} \\
    \hline\hline
    P@1  & $36.9\pm2.2\%$ & $34.6\pm1.7\%$ & $\mathbf{42.4\pm1.9\%}$ \\
    \hline
    P@2 & $60.9\pm2.6\%$ & $59.5\pm2.6\%$  & $\mathbf{66.1\pm1.8\%}$ \\
    \hline
    P@3 & $78.4\pm1.8\%$ & $78.6\pm1.8\%$ & $\mathbf{81.6\pm1.5\%}$ \\
    \hline
    P@4 & $91.4\pm1.2\%$ & $91.1\pm1.0\%$ & $\mathbf{92.6\pm1.6\%}$ \\
    \hline
    DCG@2 & $52.0\pm2.3\%$ & $50.3\pm2.1\%$ & $\mathbf{57.3\pm1.6\%}$ \\
    \hline
    DCG@3 & $60.8\pm1.8\%$  & $59.8\pm1.6\%$ & $\mathbf{65.1\pm1.5\%}$ \\
    \hline
    DCG@4 & $66.4\pm1.6\%$ & $65.3\pm1.2\%$ & $\mathbf{69.8\pm1.5\%}$ \\
    \hline
    DCG@5 & $69.7\pm1.2\%$ & $68.7\pm0.9\%$ & $\mathbf{72.7\pm0.9\%}$  \\
    \hline
\end{tabular}
\end{center}
\end{table}

\subsection{RQ-6: Component-Wise Evaluation}
\label{sec:rq5}
Compared with other methods, the key advantages of our \textit{CodeSelector} are its two sub-components: incorporating the BERT model and employing the pairwise comparisons. 
To verify the effectiveness of both aforementioned components,
we conduct a component-wise evaluation to evaluate their performance one by one.

\subsubsection{Experimental Setup}
For our component-wise evaluation, we compare our model with two of its incomplete versions:
\begin{itemize}
    \item \textbf{Drop-Pairwise} removes the pairwise comparison component from our \textit{CodeSelector}. 
    In this experiment, for a given question, we drop the process of constructing the positive and negative preference pairs. 
    To do this, we reconstruct the best QC pairs as positive samples, and make the nonbest and nonrelevant QC pairs as negative samples. 
    The \textbf{Drop-Pairwise} model is then trained as a binary classification model same as ours. 
    \item \textbf{Drop-BERT} removes the BERT component from our \textit{CodeSelector}. 
    In this experiment, we keep the preference pairs construction process but drop the BERT embedding process. 
    To to this, we replace the BERT embedding layers (described in Section~\ref{sec:code_sel}) with the traditional Word2Vec embedding layers.
    The \textbf{Drop-BERT} can then be trained with the preference QC pairs in just the same way. 
\end{itemize}

\subsubsection{Experimental Results}
The evaluation results of \textbf{Drop-Pairwise} and \textbf{Drop-BERT} are displayed in Table~\ref{tab:cw_eval_py} and Table~\ref{tab:cw_eval_java} respectively.
In can be seen that: 
\begin{enumerate}
    \item \textbf{Dropping either component does hurt the performance of \textit{CodeSelector}}. This justifies the importance and effectiveness of both components.
    
    \item \textbf{Drop-BERT achieves the worst performance.}
    This indicates that a good embedding technique has a major influence on the overall performance.
    For example, when adding BERT component for embedding the QC pairs, the $P@1$ score is improved by 20.4\% and 18.2\% on Python and Java dataset respectively. 
    We attribute this to the advantage of BERT for capturing the intent of the query question as well as the program code.
    This is why our model outperforms other deep learning-based code searching methods. 
    
    \item By comparing the results of \textbf{Drop-Pairwise} and \textbf{Ours}, we can measure the performance improvement achieved due to the employment of pairwise comparison component. 
    It is clear that by \textbf{removing the pairwise comparison component, there is a significant drop with respect to different  metrics}. This shows that the pairwise comparison component has a significant contribution to the overall performance of our model.
    This is the reason why our model outperforms other pointwise baselines.
\end{enumerate}

\noindent
\framebox{\parbox{\dimexpr\linewidth-2\fboxsep-2\fboxrule}{
\textbf{Answer to \textit{RQ-6: How effective is the BERT component and preference pairs added to
our \textit{CodeSelector}?} - we conclude that both the BERT component and the preference pairs are effective and helpful to enhance the performance of our \textit{CodeSelector}.}}} 



\begin{figure}\vspace*{-0.0cm}
\centerline{\includegraphics[width=0.45\textwidth]{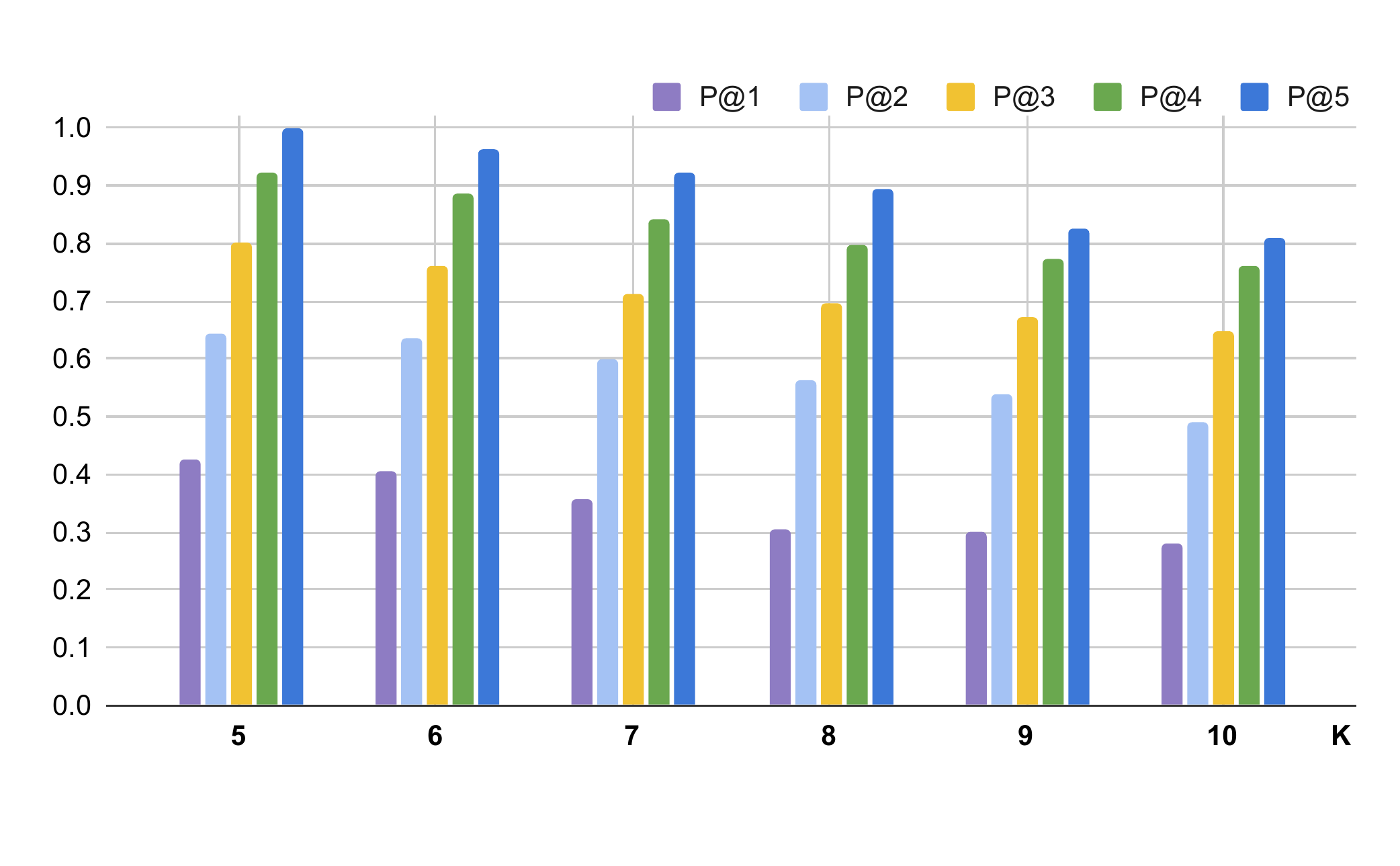}
			\includegraphics[width=0.45\textwidth]{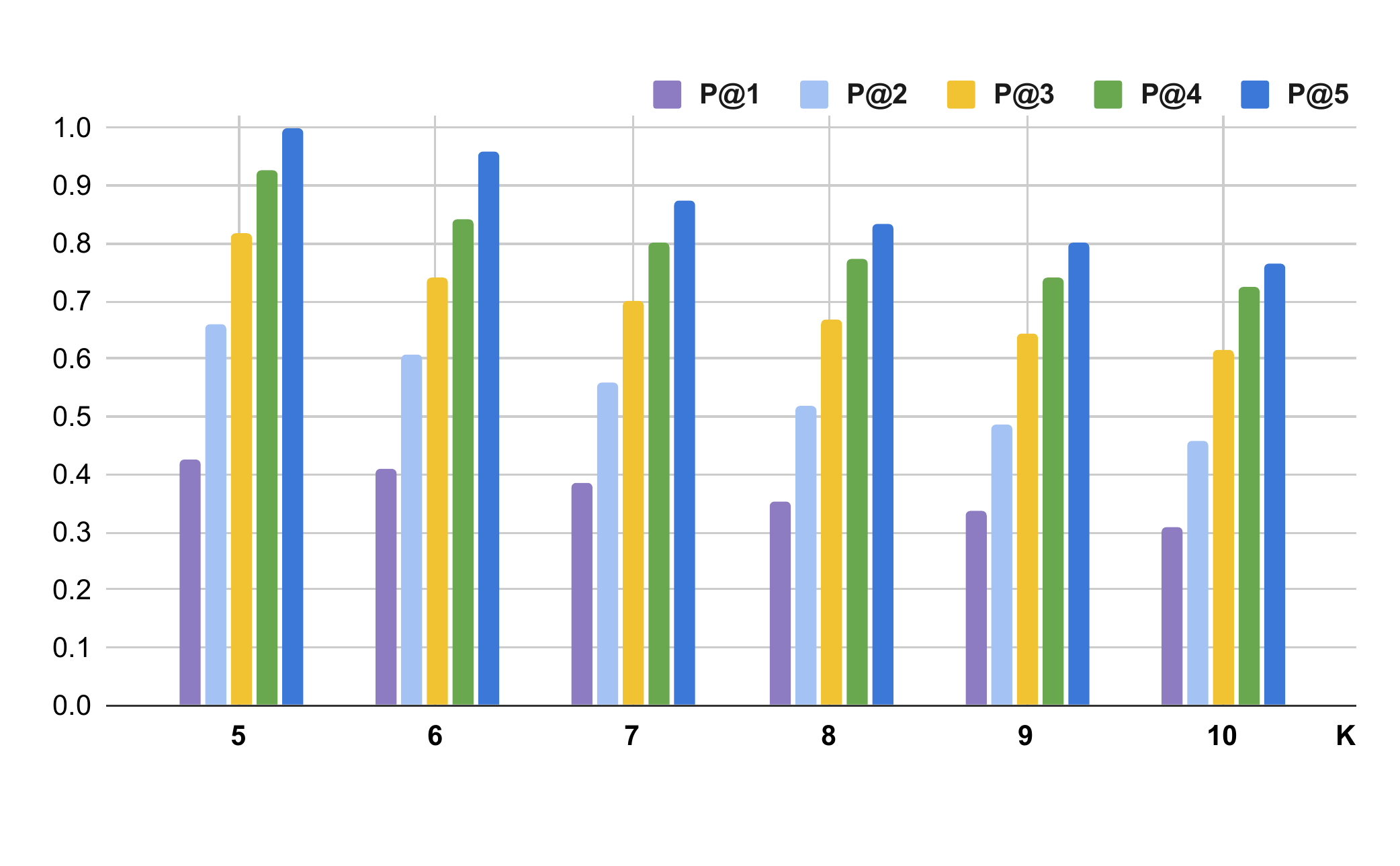}}
\caption{Robustness Analysis on Python (left) and Java (right)}
\label{fig:robust}
\end{figure}

\subsection{RQ-7: Robustness Analysis}
Considering the complexity and diversity of the CQA sites, there is little chance to find the past solved questions that exactly match the user query questions.
We thus have to enlarge $K$ - the number of retrieved questions - to improve the recall of the relevant questions as well as the potential code snippets within these questions.
However, a larger $K$ may often bring more noise into the code snippet candidates pool. 
This increases the complexity and difficulty for recommending the potential code snippet. 
We conduct a robustness analysis to investigate our model's performance with respect to different number of retrieved questions. 

\subsubsection{Experimental Setup}
To verify the robustness of our proposed model, we set different thresholds for the number of returned questions. 
In particular, we increase the number of returned questions $k$ from 5 to 10 (k=5 corresponds to our model described in RQ-4), and then evaluate the performance of our \textit{CodeSelector} with respect to different parameter settings of $K$.

\subsubsection{Experimental Results}
The average $P@1$-$5$ over Python and Java datasets are shown in Fig.~\ref{fig:robust}. 
By jointly analyzing these two figures, we can have the following observation:
\begin{enumerate}
    \item The \textbf{overall performance trend of our model goes down as $k$ increases}. This justifies our previous concerns that more noise are introduced when enlarging $k$, which incurs bigger challenges for our task. 
    \item The \textbf{performance of our model on the Java dataset decreases faster than that on Python dataset}. 
    The reason for this phenomenon may be that the Java code snippets are more complex, and contain more noise compared with Python code snippets even under the same $k$ settings. 
    \item The \textbf{performance drop of our model with increasing $k$ is not very large.} 
    For example, when we set $k$ to 10, the performance of our model on $P@1$ is still better than or comparable with the best performance on several baselines. 
    This further shows the robustness of our model.
\end{enumerate}

\noindent
\framebox{\parbox{\dimexpr\linewidth-2\fboxsep-2\fboxrule}{
\textbf{Answer to \textit{RQ-7: How robust is our \textit{CodeSelector} with different parameter settings?} - we conclude that our approach is robust to noises. }}}

\section{Human Evaluation}
\label{sec:human_eval}
The goal of our tool is to recommend the best code snippets that most closely match a developer's intent from past solved questions in Stack Overflow.
We perform a user study to measure how developers actually perceive the results produced by our approach.


\subsection{\rev{Human Evaluation Preliminary}}
\subsubsection{\rev{Participants Selection}}
\rev{
Since we are recommending code snippets for newly posted queries, we thus sampled 50 unanswered Python posts from Stack overflow for our human evaluation.  
We recruited 10 participants to join our human evaluation, which is larger than or comparable to the size of the user study participants in previous studies~\cite{liu2018neural, xu2017answerbot, gao2020deepans}.
Our user study includes 1 postdoctoral fellow and 4 Ph.D. Students majoring in Computer Science and 5 software developers from industry. 
All of these selected participants have working experience with Python development.
The years of their working experience on Python range from 3 to 10 years. 
In practice, our tool aims to help different levels of practitioners, from novice to senior developers. 
The diversity of their background (i.e., from academic and industry) and their working years can improve the generality of our user study. 
Our human evaluation includes two user studies: the user study on question relevance and the user study on code usefulness. 
All the selected evaluators are asked to participate in the above two user studies.
}

\subsubsection{\rev{Human Evaluation Baselines}}
\rev{We use the following baselines for our human evaluation:
\begin{itemize}
    \item \textbf{Google Search Engine.} 
    Considering developers usually search for technical help using the Google search engine, we employ the Google search engine as one of our baselines.
    For the Google search engine, we use the question title of the post as the search query, we then add the ``site:stackoverflow.com'' to the end of the search query so that it only searches on Stack Overflow. 
    We treat the first ranked question returned by Google search engine as the most relevant question, we treat the code fragment within its best answer (the accept answer or the highest vote answer) as the most relevant code snippet. 
    \item \textbf{Stack Overflow Search Engine.} 
    Similar to Google search engine, Stack Overflow also provide a service to search relevant posts.
    For the Stack Overflow search engine engine, we refer to the first ranked related question suggested by the Stack Overflow as the most relevant question, and the code snippet within its best answer as the recommended solution. 
    \item \textbf{CROKAGE.}
    Since \textbf{CROKAGE} outperforms the other baselines in both stage one (semantically-equivalent question retrieval) and stage two (best code snippet recommendation).
    Therefore we also employ the \textbf{CROKAGE} to find the relevant questions and code snippets for a given query in our human evaluation.
    \item \rev{$\textbf{1stRanked}$.
    Given a user query, our approach first retrieves the semantically-equivalent questions, then it reranks all the code snippets associated with these questions. 
    Different from our approach, the \textbf{1stRanked} method drops the second stage of the code reranking process and simply uses the code snippet from the first ranked question as the recommended solution. 
    }
    \item {\sc \textbf{Que2Code}}. For our approach, we perform an end-to-end evaluation using our {\sc Que2Code} tool.
    In particular, we use the \textit{QueryRewriter} to retrieve the semantic relevant questions in Stack Overflow, and then use the \textit{CodeSelector} to select the best code snippet for the unanswered question. 
    After \textit{CodeSelector} reranks all the code snippets, we treat the first ranked code snippet and its associated question as the recommended solution.
\end{itemize}
}

\begin{figure}
\vspace{0.0cm}
\centerline{\includegraphics[width=0.80\textwidth]{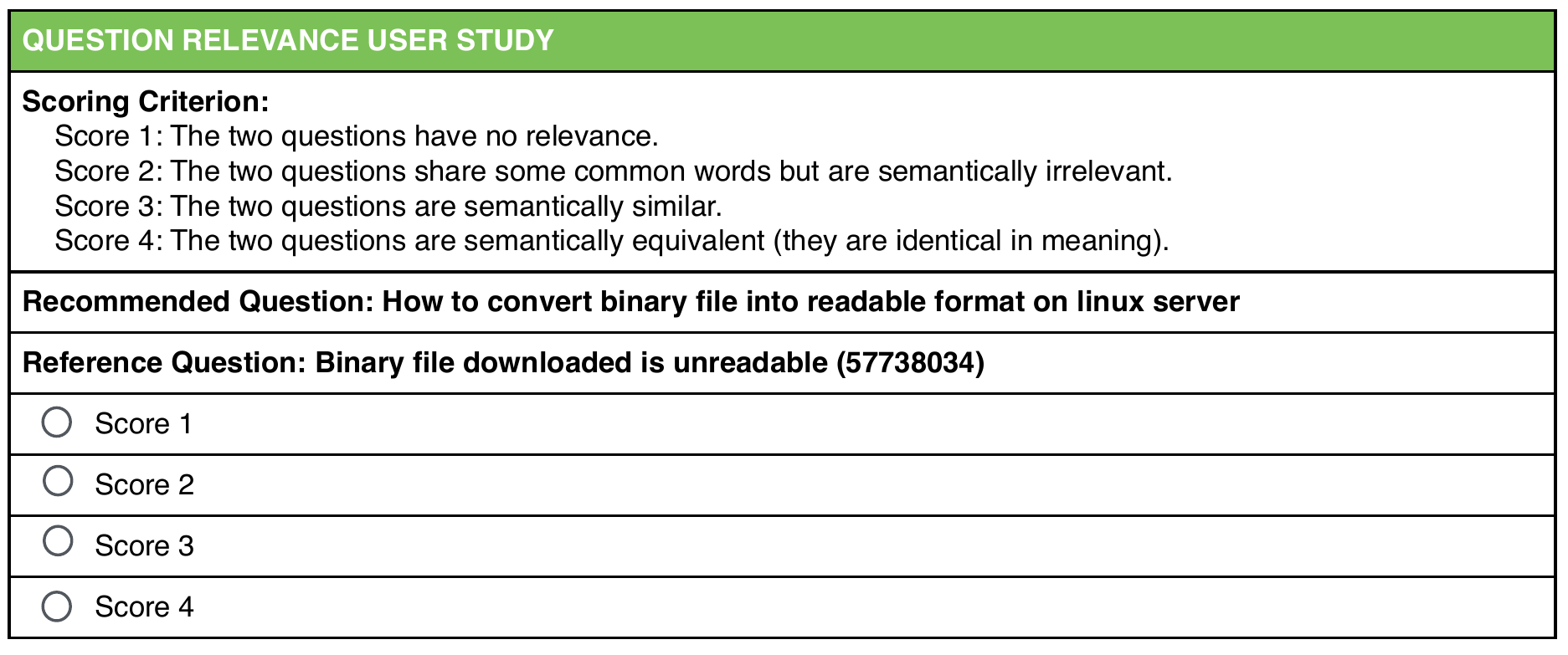}}
\vspace*{-0pt}
\caption{Example of Question Relevance User Study}
\label{fig:question_relevance_example}
\vspace{-0pt}
\end{figure}

\subsection{\rev{User Study on Question Relevance}}
\subsubsection{\rev{Experimental Setup}}
\rev{
Since we are recommending code snippets from semantically-equivalent questions, if the retrieved questions are not relevant to the user query question, it is unlikely that we can select the appropriate code snippet from the answer candidates pool. 
Therefore we first conduct a user study to measure how humans perceive the question retrieval results. 
To do this, we consider the \textit{question relevance} modality for this user study. 
The \textit{question relevance} metric measures how relevant is the retrieved question to the user query question. 
To be more specific, we asked participants to do a web questionnaire.
For each unanswered question, the evaluator is displayed with this user query question along with 5 retrieved questions from the above baselines. 
For each retrieved question, the participant is asked to give a score between 1 to 4 to measure the relevance between the user query question and the retrieved question. 
We define the scoring criterion as follows:
\begin{itemize}
    \item Score 1: The two questions have no relevance.  
    \item Score 2: The two questions share some common words but are semantically irrelevant. 
    \item Score 3: The two questions are semantically similar. 
    \item Score 4: The two questions are semantically equivalent (they are identical in meaning). 
\end{itemize}
We provide the scoring criterion in the beginning of each questionnaire to guide participants. 
Fig.~\ref{fig:question_relevance_example} shows one example in our survey.
It is worth mentioning that the order of the 5 retrieved questions is randomly decided, so the participants do not know which question is generated by our approach. 
}

\subsubsection{\rev{Experimental Results.}}
\rev{
We obtained 500 groups of scores from the above user study. 
Each group contains 5 scores for 5 retrieved questions respectively. 
\revv{
We regard a score of 1 as $low$ quality, a score of 2 as $medium-$ quality, a score of 3 as $medium+$ quality, a score of 4 as $high$ quality. 
}
The score distribution and the mean score of \textit{question relevance} across baselines are presented in Table~\ref{tab:question_relevnce_human}. 
We also evaluate whether the differences between our approach and the baselines are statistically significant by performing Wilcoxon signed-rank test~\cite{wilcoxon1992individual}. 
From the table, we can see that: 
\begin{enumerate}
    \item \textbf{The Stack Overflow search engine achieved the worst performance among all the baselines.} 
    The very large proportion of low quality questions reflects that the Stack Overflow search engine is ineffective for searching relevant questions for newly posted questions.
    We manually checked the 50 questions produced by Stack Overflow search engine, it recommended the same common question 15 times, which explains its weak performance in question retrieval tasks. 
    \item \textbf{Our approach outperforms the CROKAGE baseline regarding the \textit{question relevance} metric.} 
    The results of human evaluation are consistent with large-scale automatic evaluation results, which further justifies the effectiveness of our approach for retrieving relevant questions. 
    \item \textbf{The 1stRanked method has its advantages as compared to other baselines (i.e., Stack Overflow and CROKAGE).}
    This is reasonable because both the 1stRanked method and our approach employs \textit{QueryRewriter} for embedding relevant questions. 
    The \textit{QueryRewriter} incorporates historical duplicate question pairs from Stack Overflow, such that two semantically-equivalent questions are close in terms of vector representations. 
    \item \textbf{Google search engine performs better than our approach regarding the $high$ quality questions.}
    Considering Google's capability, i.e., the larger searching database and accumulated user searching histories, it is not surprising that the Google search engine can identify the high quality relevant questions for the user query. 
    \textbf{However, our approach still achieves comparable mean score regarding the \textit{question relevance} metric, and the difference between our approach and Google search engine is also not statistically significant.} 
    This is because the proportion of relevant questions (including the $medium+$ and $high$ quality questions) outnumber those of the Google search engine. 
    This reflects that, for a given unseen question, our approach is more likely to retrieve relevant questions in general. 
\end{enumerate}
}

\begin{table}
\caption{\rev{User Study on Question Relevance}}
\label{tab:question_relevnce_human}
\begin{center}
\rev{
\begin{tabular}{|c|c|c|c|c|c|c|}
    \hline
    {\bf Measure} & {\bf Low} & {\bf Medium-} & {\bf Medium+} & {\bf High} & {\bf Mean Score} & {\bf P-value} \\
    \hline\hline
    Google  & $3.20\%$ & $31.60\%$ & $36.80\%$ & $28.40\%$ & $2.91$ & $>0.5$  \\
    \hline
    Stack Overflow  & $58.40\%$ & $28.80\%$ & $6.80\%$ & $6.00\%$ & $1.60$ & $<0.01$ \\
    \hline
    CROKAGE  & $12.80\%$ & $40.00\%$ & $36.40\%$ & $10.80\%$ & $2.45$ & $<0.01$ \\
    \hline
    1stRanked  & $8.00\%$ & $33.20\%$ & $45.60\%$ & $13.20\%$ & $2.64$ & $<0.01$ \\
    \hline
    Ours  & $2.80\%$ & $27.20\%$ & $47.20\%$ & $22.80\%$ & $2.90$ & $-$ \\
    \hline
\end{tabular}
}
\end{center}
\end{table}


\begin{figure}
\vspace{0.0cm}
\centerline{\includegraphics[width=0.80\textwidth]{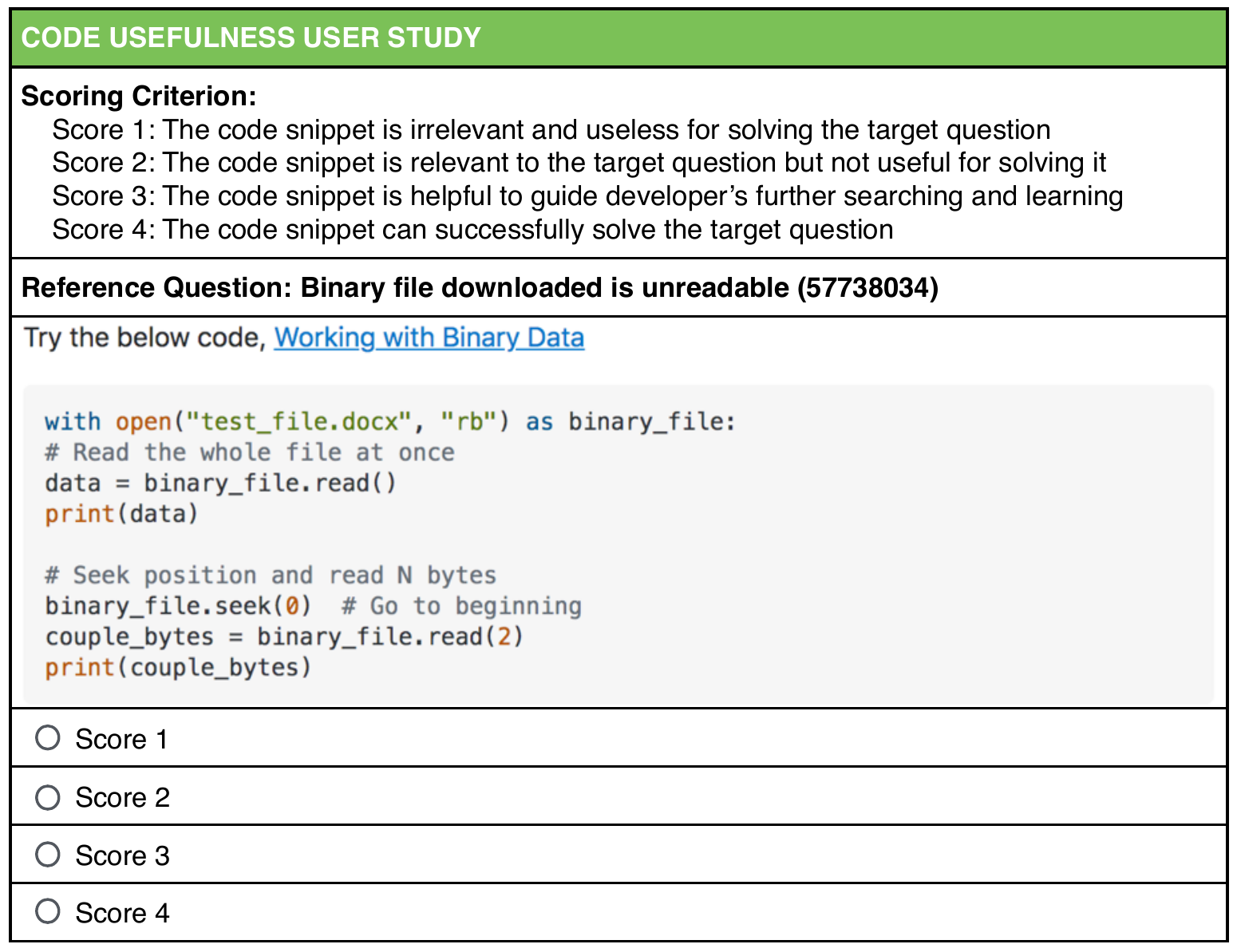}}
\vspace*{-0pt}
\caption{Example of Code Usefulness User Study}
\label{fig:code_useful_example}
\vspace{-0pt}
\end{figure}

\subsection{\rev{User Study on Code Usefulness}}
\rev{
\subsubsection{Experimental Setup.}
Our final goal is searching for useful code snippets to help developers solve unanswered questions. 
We also conduct a user study to measure the \textit{code usefulness} of our recommended code snippets. 
The \textit{code usefulness} metric refers to how useful the recommended code snippet is for solving the user query questions. 
Similar to our previous user study, for each unanswered question, we provide 5 code snippets candidates recommended by 5 baselines respectively.
After that, each evaluator was asked to rate 5 code snippets from 1 to 4 according to its \textit{code usefulness} with respect to the following scoring criterion:
\begin{itemize}
    \item Score 1: the code snippet is irrelevant and useless for solving the target question.
    \item Score 2: the code snippet is relevant to the target question, but not useful for solving it.  
    \item Score 3: the code snippet can guide developer's further searching and learning, which is useful to solve the target question.
    \item Score 4: the code snippet can successfully solve the target question.
\end{itemize}
We provide the scoring criterion in the beginning of the questionnaire to guide the evaluators. 
Fig.~\ref{fig:code_useful_example} demonstrates one example of our survey. 
The evaluators were blinded to which code snippet is generated by our approach. 
\subsubsection{Experimental Results.}
Same as for the user study on \textit{question relevance}, after obtaining the evaluator's feedback, we regard a score of 1 as $low$ quality, a score of 2 as $medium-$ quality, a score of 3 as $medium+$ quality, and a score of 4 as $high$ quality. 
The score distribution and the mean score of \textit{code usefulness} across different baselines are presented in Table~\ref{tab:code_usefulness_human}. 
The Wilcoxon signed-rank test~\cite{wilcoxon1992individual} is also performed between our approach and each baseline method, the results are displayed in the last column of Table~\ref{tab:code_usefulness_human}.
From the table, we can see that: 
\begin{enumerate}
    \item \textbf{Our model significantly outperforms all the baselines (including the Google search engine) regarding the \textit{code usefulness} metric.} 
    This suggests that the code snippets recommended by our approach are considered to be more useful to the given query question compared with baselines. 
    The reason may be due to the two stages of our approach, i.e., \textit{semantically-equivalent question retrieval} and \textit{best code snippet recommendation}. 
    The first stage focuses on retrieving as many as possible relevant questions to construct the candidate set. 
    The more relevant the retrieved question is to the query, the more likely the code snippet associated with the question is helpful to solve the problem. 
    The second stage tries to rank the useful code snippet to the top of the recommendation result. 
    \item \textbf{Compared with the 1stRanked method, our approach has its advantages.} 
    Even though the 1stranked method retrieves the same relevant question candidates as ours, it naively picks the code snippet from the most relevant question. 
    However, considering the complexity of the technical queries, it is very hard, if not possible, to find identical questions to the given query. 
    Therefore it is necessary to consider the correlation between the code snippet and the user query. 
    Different from the 1stRanked method which is solely based on \textit{question relevance}, our \textit{CodeSelector} to rerank all the code snippet candidates by performing pairwise comparisons. 
    The \textit{CodeSelector} not only considers the program semantics between the code snippet and the query question, but also investigates the relevance preference between different QC pairs.  
    Thus a useful code snippet can be ranked higher up among other candidates. 
    The superior performance of our approach regarding the mean score of \textit{code usefulness} further supports the ability of our \textit{CodeSelector} model for recommending useful code snippet.
    \item \textbf{By comparing the mean score of \textit{code usefulness} and \textit{question relevance}, there is a significant drop overall in every baseline method.} 
    This indicates that compared with relevant question retrieval tasks, identifying useful code snippets is more challenging. 
    Technical questions in Stack Overflow are rather complicated and specific, even though two technical questions are semantically relevant, the code snippet is not applicable or reusable for the programming task.
    \textbf{We also observe that the performance drop of our approach is much smaller than other baseline models.}
    We attribute this to the effectiveness of \textit{CodeSelector} for putting relevant snippet on higher positions than the irrelevant ones, this also verifies the importance and necessity of our \textit{CodeSelector} in the second stage.
\end{enumerate}
}


\begin{table}
\caption{\rev{User Study on Code Usefulness}}
\label{tab:code_usefulness_human}
\begin{center}
\rev{
\begin{tabular}{|c|c|c|c|c|c|c|}
    \hline
    {\bf Measure} & {\bf Low} & {\bf Medium-} & {\bf Medium+} & {\bf High} & {\bf Mean Score} & {\bf P-value} \\
    \hline\hline
    Google  & $17.60\%$ & $23.60\%$ & $32.40\%$ & $26.40\%$ & $2.68$ & $<0.01$  \\
    \hline
    Stack Overflow  & $73.60\%$ & $18.00\%$ & $6.40\%$ & $2.00\%$ & $1.37$ & $<0.01$ \\
    \hline
    CROKAGE  & $30.80\%$ & $26.00\%$ & $28.40\%$ & $14.80\%$ & $2.27$ & $<0.01$ \\
    \hline
    1stRanked  & $28.40\%$ & $31.60\%$ & $23.20\%$ & $16.80\%$ & $2.28$ & $<0.01$ \\
    \hline
    Ours  & $12.00\%$ & $22.00\%$ & $34.00\%$ & $32.00\%$ & $2.86$ & $-$ \\
    \hline
\end{tabular}
}
\end{center}
\end{table}

\subsection{Qualitative Analysis}
\subsubsection{Experimental Setup}
\begin{figure}
\vspace{0.0cm}
\centerline{\includegraphics[width=0.99\textwidth]{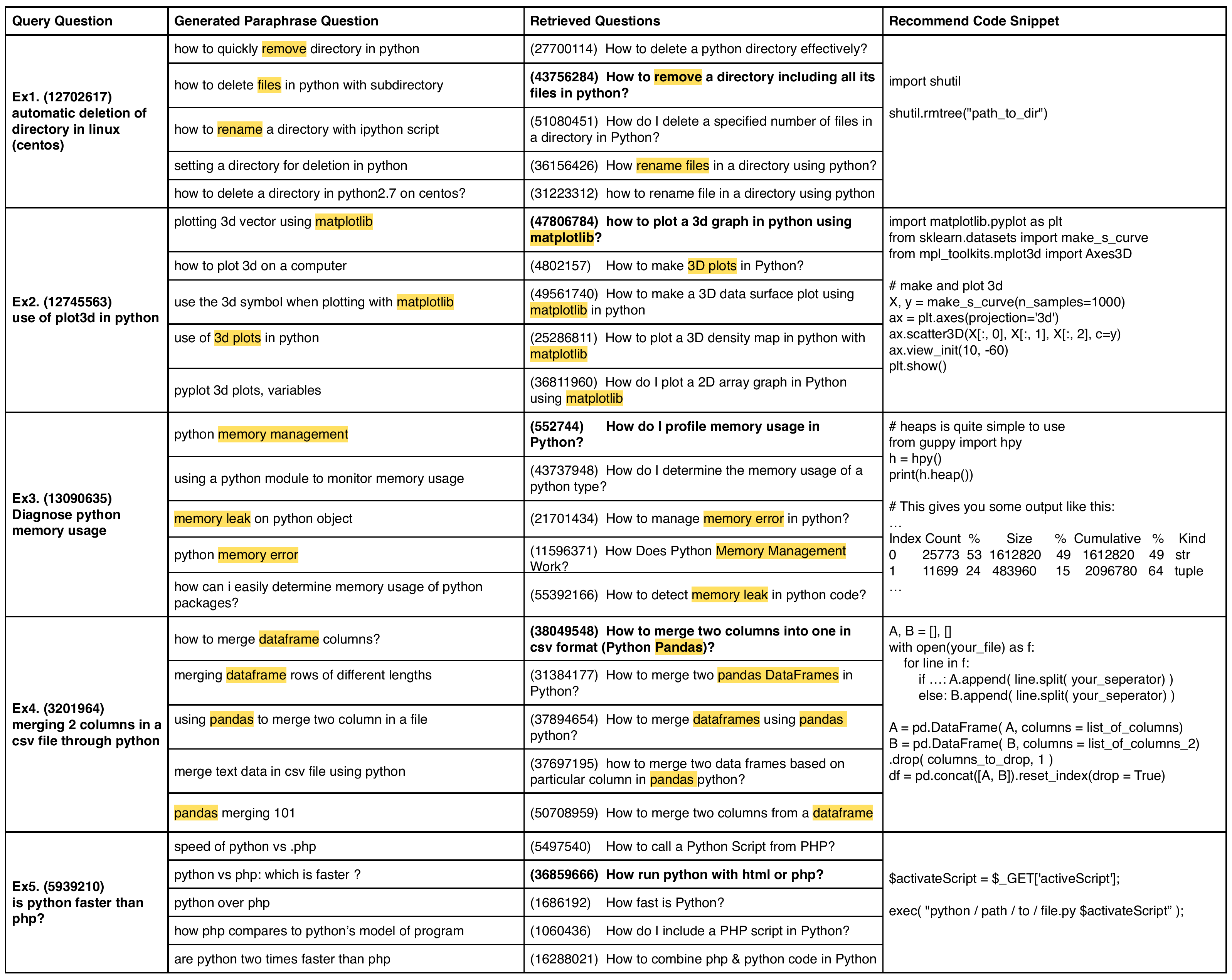}}
\vspace*{-0pt}
\caption{Qualitative Analysis}
\label{fig:case_study}
\vspace{-0pt}
\end{figure}

In this work, we aim to alleviate the {\em query mismatch} and {\em information overload} problems by using the \textit{QueryRewriter} and \textit{CodeSelector} respectively. 
\rev{
To vividly demonstrate the workflow details of our model (i.e., from generating paraphrase questions, semantically-equivalent question retrieval and best code snippet selection), we further conduct a case study to manually investigate some questions used in the previous user study. 
Fig.~\ref{fig:case_study} shows 5 examples from the previous user studies to demonstrate the detailed results.
}
For each unanswered question, we present the intermediary results of the generated paraphrase questions, the top-5 ranked retrieved question, as well as the recommended code snippets generated by our approach.
The words that do not appear in the original query question, but include both in the generated paraphrase questions and the target retrieved questions are highlighted in yellow color. 
We also highlight the question in boldface which provides the recommended code snippet. 

\subsubsection{Experimental Results}
From the cases demonstrated in Fig~\ref{fig:case_study}, we can see that: 
\begin{enumerate}
    \item \textbf{The paraphrase questions generated by our \textit{QueryRewriter} are meaningful for the given user query question.} 
    Note that the \textit{QueryRewriter} automatically transforms the user query question into different forms of semantically-equivalent user expressions.
    For example, in the third case, the original user query question is about ``\textit{Diagnose python memory usage}'', and our approach generates multiple paraphrase questions such as ``\textit{python memory management}'', ``\textit{using a python module to monitor memory usage}'', ``\textit{memory leak on python object}''.
    These generated paraphrase questions can be viewed as meaningful outputs to capture the developer's intent and used as a way of question boosting for the original user query question. 
    
    \item \textbf{It is clear that adding the paraphrase questions can reduce the lexical gap between different user expressions}, which increases the likelihood of retrieving the semantic relevant questions in Stack Overflow. 
    \rev{
    As shown in the second case, the developer formulate his/her problem as a user query ``\textit{use of plot3d in python}'', the generated paraphrase question ``\textit{plotting 3d vector using matplotlib}'' can add missing information for the user query question and better link to the target semantically-equivalent question in Stack Overflow (i.e., ``\textit{how to plot a 3d graph in python using matplotlib}''), which verify the ability of our {QueryRewriter} to alleviate the {\em query mismatch} problem.
    }
    
    \item 
    \rev{
    \textbf{A large number of code snippets recommended by our system are relevant and useful with respect to the user query question.}
    Some code snippets can well satisfy the developer's programming tasks directly. 
    For example, as shown in the first case of Fig.~\ref{fig:case_study}, the developer's query question ``\textit{automatic deletion of directory in linux (centos)}'' can be successfully solved by the code snippets within a relevant question ``\textit{How to remove a directory including all its files in python?}''.
    This verifies the effectiveness and possibility of our approach for recommending code solutions from the existing historical answers.  
    }
    
    \item 
    \rev{
    \textbf{We also notice that the recommended code snippets are not always selected from the first ranked question candidate.} 
    Instead of naively choosing the first ranked code snippet like the Google search engine and Stack Overflow search engine, our \textit{CodeSelector} selects best code snippet among a set of code snippet candidates via pairwise comparisons, which justify the ability of our \textit{CodeSelector} to alleviate the {\em information overload} problem.
    }
    \item However, \textbf{the recommended code snippets from our system are not always useful}.
    For example, in the second case, the developer would like to inquire about ``\textit{use of plot3d in python}'', our recommended code snippet is about ``plot 3d graph using matplotlib'', which can be viewed as helpful by looking at the intent of the developer. 
    In the fourth sample, even though the recommended code snippet can not be applied directly, it can be easily adapted to the user query question with minor modification. 
    
    \item Also, \textbf{the recommended code snippets from our system are not always relevant}. 
    For example, in the last sample, even though the generated paraphrase questions are meaningful and relevant to the user query question, the final recommended code snippet is still irrelevant to the problem described in the query question. 
    This is because some user queries posted by developers are often complex and sophisticated; there may not exist semantically-equivalent questions that to the given one. 
    It is thus very hard to search the query-specific code snippet to solve the corresponding problem. 
    
\end{enumerate}


\noindent
\framebox{\parbox{\dimexpr\linewidth-2\fboxsep-2\fboxrule}{
\textbf{\rev{
Overall, our approach is more effective for retrieving relevant questions and searching useful code snippets compared with other baselines under human evaluation.}
}}}

\section{Practical Usage}
\label{sec:practical}
\rev{
The experiment was conducted on an Nvidia GeForce GTX 2080 GPU with 12GB memory.
The time cost of our approach is mostly for the training process which takes approximately 20 to 24 hours for training Python and Java datasets respectively. 
However, after finishing the training process, each question title and code snippet in our data base can be converted into vector representations, which is highly efficient for later computing and searching processes. 
For example, the searching process on 5,000 examples takes five to eight minutes, while searching a single code snippet only costs 60 to 80ms. 
}

\begin{figure}
\vspace{0.0cm}
\centerline{\includegraphics[width=0.90\textwidth]{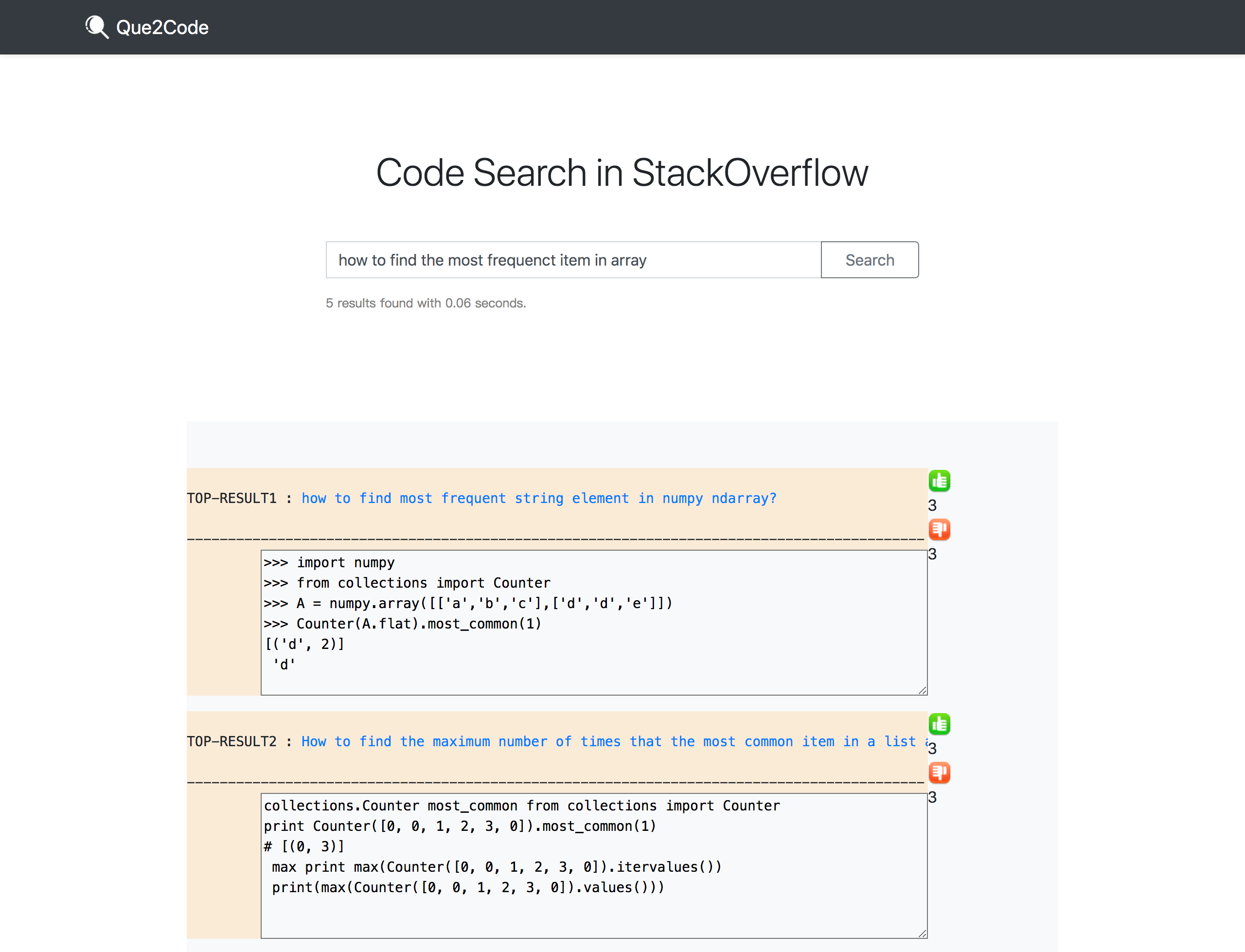}}
\vspace*{-0pt}
\caption{Prototype of Que2Code}
\label{fig:prototype}
\vspace{-0pt}
\end{figure}

\rev{
Considering that searching code snippets in Stack Overflow with our approach is efficient, we have implemented {\sc Que2Code} as a prototype web-based tool, which can facilitate developers in using our approach and inspire follow up research. 
Fig.~\ref{fig:prototype} shows the web interface of {\sc Que2Code}. 
Developers can type or paste their query question into our web application, after that {\sc Que2Code} goes through the question retrieval and code snippet reranking process, and recommends the top ranked questions and code snippets to the developers. 
We below describe the details of the input and output of our tool. 
\begin{itemize}
    \item \textbf{Input:} the input to the {\sc Que2Code} is a user query question, which is a sequence of tokens. 
    The input box in Fig.~\ref{fig:prototype} shows an example of the user query question, i.e., ``\textit{how to find the most frequent item in array}''.
    After inputting the user query question, the developer can click the ``Search'' button to submit their query.
    \item \textbf{Output:} the output of the {\sc Que2Code} is two folds: relevant questions and code snippets. 
    After the developer submits his/her query to the server, the {\sc Que2Code} searches through the codebase and returns top 5 relevant code snippets with their associated question titles.  
    The link to these question posts on Stack Overflow are also provided for user references.
    For example, the relevant question post ``\textit{how to find the most frequent string element in numpy array}'' and its code snippet are retrieved from our database to guide developers for solving their problems. 
    Developers can use our tool to quickly locate the potential solutions to their query programming tasks and have a better understanding of their problems. 
\end{itemize}
The goal of our tool is helping developers effectively search code snippets from Stack Overflow to their programming tasks and saving their time to do so. 
This is by no means these code snippets can perfectly solve the developers' problem. 
After browsing these code snippets, developers still need to manually modify these code snippets for further refactoring and testing. 
}

\noindent
\framebox{\parbox{\dimexpr\linewidth-2\fboxsep-2\fboxrule}{
\textbf{\rev{Overall, our approach is efficient enough for practical use and we have implemented a web service tool, {\sc Que2Code}, to apply our approach for practical use.}}}}

\section{Discussion}
\label{sec:discussion}

We discuss the strengths and weaknesses of our approach as well as the threats to validity for our experiments. 

\subsection{\revv{Strengths and Limitations of Our Approach}}
\subsubsection{Strengths of Our approach.}
To address the {\em query mismatch} and {\em information overload} problem in the Stack Overflow community, we proposed a novel query-driven code snippet recommendation model. Its key strengths are summarised below.
\begin{enumerate}
    \item Paraphrase Question Generation. A key advantage of our model is training a text-to-text transformer, named \textit{QueryRewriter}, for generating paraphrase questions as a way of question boosting. This greatly improves the likelihood of our approach of retrieving semantically-equivalent questions for a user query question. By training on the duplicate question pairs in Stack Overflow, for a user query question, \textit{QueryRewriter} is able to generate different user descriptions for the same problem, which is helpful for our semantically-equivalent question retrieval tasks. The experimental results in Section~\ref{sec:rq2} verify the effectiveness of adding paraphrased questions to our model. 
    \item Pairwise Learning to Rank. To recommend the most relevant code snippets in Stack Overflow, we propose a novel pairwise learning to rank model in our work. Guided by our three heuristic rules, we can automatically construct the preference QC pairs and transform the code snippet recommendation task to a binary classification task. Rather than calculating a precise relevance score for a single QC pair, we estimate the preference relationship between two QC pairs. Ranking code snippets by pairwise comparison is more suitable for the code recommendation task in our study.
    \item BERT Model for Embeddings. BERT is designed to pre-train deep bidirectional representations from unlabeled text. It is conceptually simple and empirically powerful, which obtains new state-of-the-art results on eleven natural language processing tasks, such as question answering, language inference, etc. In our research, we investigated the BERT model for embedding query and code snippet pairs, and the ablation analysis in Section~\ref{sec:rq5} demonstrates that it greatly enhanced the performance of our proposed model. 
\end{enumerate}
\subsubsection{\revv{Limitations of Our approach.}}
\revv{
We have identified the following limitations of our approach: 
\begin{enumerate}
    \item 
    Model Limitations. We have collected our dataset from official data dump from Stack Overflow. 
    Our work is up to the release date of the official data dump that we used, we can't automatically update our model with the newly added data after the release date. 
    However, this limitation can be supported by adding extra engineering work (e.g., automatically updating the model every month by checking the newest version of Stack Overflow data dump), we will try to expand our approach to handle the newly added data in our future work. 
    \minor{
    In terms of practical usage, the experimental results show that our model outperforms Google search for low-quality questions, while our work does not provide mechanisms to estimate the question quality currently. 
    However, this limitation can be alleviated by incorporating a question quality prediction model, which has been widely studied in prior works~\cite{ponzanelli2014improving, baltadzhieva2015predicting, toth2019towards}.
    For example, a developer can first leverage the quality prediction model to estimate the question quality, then our tool can be used to search suitable code snippets from Stack Overflow for low-quality questions. 
    }
    \item 
    Query Limitations. Our model aims to identify the best code solutions for a user query from Stack Overflow posts. 
    However, there are different types of user queries that are relevant to different search tasks (e.g., explanations for unknown terminologies, explanations for exceptions/error messages), which may not always be searching for code snippets. 
    The Que2Code model currently recommends code snippet for a user query if it is asking about code solutions, we will try to focus on other types of questions in our future work. 
\end{enumerate}
}

\subsection{Threats to Validity}
We have identified the following threats to validity:
\subsubsection{\rev{Internal Validity}} 
Internal validity relate to potential errors in our model implementation and experimental settings.
To reduce errors in automatic evaluation, we have carefully tuned the parameters of the baseline approaches and used them in their highest performing settings for comparison, but there may still exist errors that we did not notice. 
Considering such cases, we have released the code and data of our research to facilitate other researchers to repeat our work and verify their ideas. 
\minor{
Regarding the distance metric, we choose Euclidean distance to estimate the semantic distance between different questions, we did not consider other distance metrics (e.g., cosine distance) in this preliminary study. 
We believe our model will generalize to other distance metrics due to the valuable information of our embeddings.
We will explore the effectiveness of different distance metrics in future studies. 
}


\subsubsection{External Validity} 
Threats to external validity are concerned with the generalizability of our dataset.
Our dataset is collected from the official Stack Overflow data dump. 
We focus on two popular programming languages, i.e., Python and Java for our experiment. 
However, there are still many other programming languages which are not considered in our study.
We believe that our model will generalize to other programming languages due to the effectiveness and robustness of our approach. 
We will try to extend our approach to other programming languages to benefit more developers in future studies.

\subsubsection{Model Validity}
The model validity relates to model structure that could affect the learning performance of our approach. 
\rev{
In the first stage, we choose an encoder-decoder architecture for our \textit{QueryRewriter}. 
Such an encoder-decoder architecture targets the sequence-to-sequence learning problem, which requires a large amount of manually labeled duplicate question pairs. 
However, our model may not generalize to other technical Q\&A sites if the training set is limited.
}
In the second stage, we choose the basic BERT model as our embedding layer due to its promising results on a wide range of NLP tasks~\cite{devlin2018bert, gao2021automating}.
Recent research has proposed new models, such as GPT~\cite{radford2018improving}, RoBERTa~\cite{liu2019roberta}, DistilBERT~\cite{sanh2019distilbert}, ALBERT~\cite{lan2019albert} that can achieve better performance than BERT and/or similar performance with much less parameters. 
However, our results do not shed light on the effectiveness of employing other deep learning models with respect to different structures and new advanced features.
We will try to use other deep learning models for our tasks in future work and compare them to the results that we report in this paper.

\section{Related Work}
\label{sec:related}

\subsection{Code Search in Software Engineering} 
The goal of code search is to find code fragments from a large code repository that most closely match a developer’s intent. 
Many code search methods have been proposed in the literature~\cite{gu2018deep, cambronero2019deep, sachdev2018retrieval, ye2016word, lu2015query, lv2015codehow, moreno2015can, ponzanelli2016codetube}.
Existing code search methods can be classified into two mainstreams: Information Retrieval-based methods and Deep Learning-based methods. 

Ye et al.~\cite{ye2016word} proposed a model to fill the gap between natural-language queries and code snippets by projecting them into the same high dimensional vector space. 
Sachdev et al.~\cite{sachdev2018retrieval} proposed a neural code search method which combined the of token-level embeddings and conventional information retrieval techniques TF-IDF. They found that the basic word embedding techniques can achieve good performance on code search task. 
Gu et al.~\cite{gu2018deep} proposed a supervised technique, named DeepCS, for code searching using deep neural networks. They used multiple sequence-to-sequence-based networks to capture the features of the natural language queries and the code snippets. 

The above code search methods are designed to measure the relevance degree between an individual QC (natural language query-code snippet) pair. 
However, in our study, rather than modeling a single QC pair to predict the precise relevance score, we model the preference relationship between two QC pairs. In other words, our model not only considers the relevance between a query and a code fragment, but also investigates the preference relationship between different QC pairs.

\subsection{Duplicated Questions in Stack Overflow}
The quality of the user-generated content is a key factor to attract users to visit the CQA sites, such as Stack Overflow. 
Prior work suggests that a quality decay problem occurs in these CQA community due to the growth in the number of duplicate questions~\cite{srba2016stack}.   This makes finding answers to a question harder and may dilute quality of answers.
To maintain the quality of posts in Stack overflow, many studies have investigated the duplicate questions in Stack Overflow~\cite{zhang2015multi, silva2018duplicate, ahasanuzzaman2016mining, wang2020duplicate, mizobuchi2017two, zhang2017detecting}. 

Zhang et al.~\cite{zhang2015multi} proposed an approach, named DupPredictor, to predict whether a question is duplicate question in Stack Overflow. 
They considered multiple factors, such as similarity scores of topics, titles, descriptions and tags for each question pair and calculated an overall similarity score by combining these features. 
Followed by their research, Ahasanuzzaman et al.~\cite{ahasanuzzaman2016mining} first manually investigated why duplicate questions are asked by users in Stack Overflow, then proposed Dupe, which extracted features from question corpus and then built a binary classifier to judge if a question pair is duplicated or not. 
More recently, Wang et al.~\cite{wang2020duplicate} presented a deep-learning based approach to detect duplicate questions in Stack Overflow, which can capture the document-level and word-level semantic information respectively. 

In our work, instead of considering the negative aspect of the duplicated questions in Stack Overflow, we consider duplicated question as semantically-equivalent questions pairs. We then train a query rewriting model for retrieving relevant questions in Stack Overflow. 

\subsection{\rev{Query Reformulation in Software Engineering}}
\rev{
The effectiveness of code search heavily relies on the quality of the search query. 
If a query performs poorly, searching useful code snippets become increasingly difficult. 
Therefore it is necessary to reformulate and/or improve the user query when the query is poorly expressed.  
This need has motivated researchers to investigate the query reformulation (or expansion) approaches for software engineering tasks~\cite{sirres2018augmenting, rahman2018effective, rahman2019automatic, cao2021automated, shepherd2007using, haiduc2013automatic, hill2014nl, lu2015query, nie2016query, jiang2016rosf}. 
}

\rev{
Shepherd et al.~\cite{shepherd2007using} presented an approach, V-DO, that automatically extracts verbs and objects from source code comments for misspelled query terms. 
Haiduc et al.~\cite{haiduc2013automatic} developed a query reformulation strategy by performing machine learning on a set of historical queries and relevant results. 
Following that, Hill et al.~\cite{hill2014nl} proposed a query expansion tool, named \textit{Conquer}, which combines the V-DO and contextual searching technique to suggest alternative query words. 
Lu et al.~\cite{lu2015query} implemented an approach to expand a query by using synonyms with the help of Wordnet. 
Nie et al.~\cite{nie2016query} proposed a model, named \textit{QECK}, to identify the software-specific expansion words from the high quality pseudo feedback on Stack Overflow and generate expansion queries. 
After that, Rahman et al.~\cite{rahman2019automatic} proposed a query reformulation approach that suggests a list of relevant API classes for code search. 
Most recently, Cao et al.~\cite{cao2021automated} proposed an automated deep-learning based query reformulation approach by using the query logs in Stack Overflow. 
}

\rev{
Different from the existing query expansion approaches, in this study, we first investigate the possibility of using duplicate question pairs from Stack Overflow for query rewriting. 
Our \textit{QueryRewriter} can capture features between semantically equivalent questions and address the query mismatch problem. 
}

\subsection{Question Answering in CQA Sites}
Finding similar questions and/or appropriate answers from historical archives has been applied in CQA sites. Great effort has been dedicated to various tasks such as question retrieval~\cite{wang2009syntactic, cao2010generalized, ganguly2015partially, ye2014interrogative, zou2015learning, xu2018domain}, answer selection~\cite{xu2017answerbot, gao2020deepans, singh2016using, nie2017data}, tagging~\cite{zhou2019deep, wang2015tagcombine, gonzalez2015multi},  and expert identification~\cite{pal2012exploring, tian2013predicting, kumar2016mining}.

Conventional techniques for retrieving answers primarily focus on complementary features of the CQA sites.
Calefato et al.~\cite{calefato2019empirical} transform the answer selection task to a binary classification problem, they empirically evaluated 26 answer prediction model in Stack Overflow. 
Xu et al.~\cite{xu2017answerbot} proposed a novel framework for generating relevant, useful and diverse answer summary for technical questions in Stack Overflow. 
Rather than directly ranking community answers, Tian et al.~\cite{tian2013predicting} predicted the best answerer for a technical question in Stack Overflow by assuming that good respondents will give better answers. 
More recently, Gao et al.~\cite{gao2020generating} proposed a model for generating good question titles for developers by mining the code snippets in Stack Overflow. 

Different from the aforementioned studies, we aim to search the best code fragment from the historical data in CQA database. 
We frame this task as a query-driven code recommendation task, and we propose a two stage framework to address the semantic-equivalent question retrieval and best code recommendation task respectively.

\section{Conclusion and Future work}
\label{sec:con}

We have presented a fully data-driven approach, named {\sc Que2Code}, for recommending the best code snippet in Stack Overflow for a user query question. 
We formulate this task as a query-driven code recommendation problem.
Our proposed {\sc Que2Code} model contains two components: \textit{QueryRewriter} and \textit{CodeSelector}. 
In particular, we proposed a \textit{QueryRewriter} for retrieving semantically-equivalent questions (as the first stage) and a \textit{CodeSelector} for selecting the best code fragment in Stack Overflow (as the second stage). 
We have conducted extensive experiments to evaluate our approach on Stack Overflow dataset.
Compared with several existing baselines, experimental results have comparatively demonstrated the effectiveness and superiority of our proposed model in both evaluation and human evaluation.

\section{Acknowledgements}
\label{sec:ack}
This research was partially supported by ARC Laureate Fellowship FL190100035, and National Research Foundation, Singapore, under its Industry Alignment Fund – Pre-positioning (IAF-PP) Funding Initiative. Any opinions, findings and conclusions or recommendations expressed in this material are those of the author(s) and do not reflect the views of National Research Foundation, Singapore.
The authors would like to thank the reviewers for the insightful and constructive feedback.

\balance
\bibliographystyle{ACM-Reference-Format}
\bibliography{samples}


\begin{thebibliography}{78}


\ifx \showCODEN    \undefined \def \showCODEN     #1{\unskip}     \fi
\ifx \showDOI      \undefined \def \showDOI       #1{#1}\fi
\ifx \showISBNx    \undefined \def \showISBNx     #1{\unskip}     \fi
\ifx \showISBNxiii \undefined \def \showISBNxiii  #1{\unskip}     \fi
\ifx \showISSN     \undefined \def \showISSN      #1{\unskip}     \fi
\ifx \showLCCN     \undefined \def \showLCCN      #1{\unskip}     \fi
\ifx \shownote     \undefined \def \shownote      #1{#1}          \fi
\ifx \showarticletitle \undefined \def \showarticletitle #1{#1}   \fi
\ifx \showURL      \undefined \def \showURL       {\relax}        \fi
\providecommand\bibfield[2]{#2}
\providecommand\bibinfo[2]{#2}
\providecommand\natexlab[1]{#1}
\providecommand\showeprint[2][]{arXiv:#2}

\bibitem[\protect\citeauthoryear{Ahasanuzzaman, Asaduzzaman, Roy, and
  Schneider}{Ahasanuzzaman et~al\mbox{.}}{2016}]%
        {ahasanuzzaman2016mining}
\bibfield{author}{\bibinfo{person}{Muhammad Ahasanuzzaman},
  \bibinfo{person}{Muhammad Asaduzzaman}, \bibinfo{person}{Chanchal~K Roy},
  {and} \bibinfo{person}{Kevin~A Schneider}.} \bibinfo{year}{2016}\natexlab{}.
\newblock \showarticletitle{Mining duplicate questions of stack overflow}. In
  \bibinfo{booktitle}{\emph{2016 IEEE/ACM 13th Working Conference on Mining
  Software Repositories (MSR)}}. IEEE, \bibinfo{pages}{402--412}.
\newblock


\bibitem[\protect\citeauthoryear{Ahmed and Bagherzadeh}{Ahmed and
  Bagherzadeh}{2018}]%
        {ahmed2018concurrency}
\bibfield{author}{\bibinfo{person}{Syed Ahmed} {and} \bibinfo{person}{Mehdi
  Bagherzadeh}.} \bibinfo{year}{2018}\natexlab{}.
\newblock \showarticletitle{What do concurrency developers ask about? a
  large-scale study using stack overflow}. In
  \bibinfo{booktitle}{\emph{Proceedings of the 12th ACM/IEEE international
  symposium on empirical software engineering and measurement}}.
  \bibinfo{pages}{1--10}.
\newblock


\bibitem[\protect\citeauthoryear{Ba, Kiros, and Hinton}{Ba
  et~al\mbox{.}}{2016}]%
        {ba2016layer}
\bibfield{author}{\bibinfo{person}{Jimmy~Lei Ba}, \bibinfo{person}{Jamie~Ryan
  Kiros}, {and} \bibinfo{person}{Geoffrey~E Hinton}.}
  \bibinfo{year}{2016}\natexlab{}.
\newblock \showarticletitle{Layer normalization}.
\newblock \bibinfo{journal}{\emph{arXiv preprint arXiv:1607.06450}}
  (\bibinfo{year}{2016}).
\newblock


\bibitem[\protect\citeauthoryear{Bajaj, Pattabiraman, and Mesbah}{Bajaj
  et~al\mbox{.}}{2014}]%
        {bajaj2014mining}
\bibfield{author}{\bibinfo{person}{Kartik Bajaj}, \bibinfo{person}{Karthik
  Pattabiraman}, {and} \bibinfo{person}{Ali Mesbah}.}
  \bibinfo{year}{2014}\natexlab{}.
\newblock \showarticletitle{Mining questions asked by web developers}. In
  \bibinfo{booktitle}{\emph{Proceedings of the 11th Working conference on
  mining software repositories}}. \bibinfo{pages}{112--121}.
\newblock


\bibitem[\protect\citeauthoryear{Baltadzhieva and Chrupa{\l}a}{Baltadzhieva and
  Chrupa{\l}a}{2015}]%
        {baltadzhieva2015predicting}
\bibfield{author}{\bibinfo{person}{Antoaneta Baltadzhieva} {and}
  \bibinfo{person}{Grzegorz Chrupa{\l}a}.} \bibinfo{year}{2015}\natexlab{}.
\newblock \showarticletitle{Predicting the quality of questions on
  stackoverflow}. In \bibinfo{booktitle}{\emph{Proceedings of the international
  conference recent advances in natural language processing}}.
  \bibinfo{pages}{32--40}.
\newblock


\bibitem[\protect\citeauthoryear{Bird and Loper}{Bird and Loper}{2004}]%
        {bird2004nltk}
\bibfield{author}{\bibinfo{person}{Steven Bird} {and} \bibinfo{person}{Edward
  Loper}.} \bibinfo{year}{2004}\natexlab{}.
\newblock \showarticletitle{NLTK: the natural language toolkit}. In
  \bibinfo{booktitle}{\emph{Proceedings of the ACL 2004 on Interactive poster
  and demonstration sessions}}. Association for Computational Linguistics,
  \bibinfo{pages}{31}.
\newblock


\bibitem[\protect\citeauthoryear{Bojanowski, Grave, Joulin, and
  Mikolov}{Bojanowski et~al\mbox{.}}{2017}]%
        {bojanowski2017enriching}
\bibfield{author}{\bibinfo{person}{Piotr Bojanowski}, \bibinfo{person}{Edouard
  Grave}, \bibinfo{person}{Armand Joulin}, {and} \bibinfo{person}{Tomas
  Mikolov}.} \bibinfo{year}{2017}\natexlab{}.
\newblock \showarticletitle{Enriching word vectors with subword information}.
\newblock \bibinfo{journal}{\emph{Transactions of the Association for
  Computational Linguistics}}  \bibinfo{volume}{5} (\bibinfo{year}{2017}),
  \bibinfo{pages}{135--146}.
\newblock


\bibitem[\protect\citeauthoryear{Brandt, Guo, Lewenstein, Dontcheva, and
  Klemmer}{Brandt et~al\mbox{.}}{2009}]%
        {brandt2009two}
\bibfield{author}{\bibinfo{person}{Joel Brandt}, \bibinfo{person}{Philip~J
  Guo}, \bibinfo{person}{Joel Lewenstein}, \bibinfo{person}{Mira Dontcheva},
  {and} \bibinfo{person}{Scott~R Klemmer}.} \bibinfo{year}{2009}\natexlab{}.
\newblock \showarticletitle{Two studies of opportunistic programming:
  interleaving web foraging, learning, and writing code}. In
  \bibinfo{booktitle}{\emph{Proceedings of the SIGCHI Conference on Human
  Factors in Computing Systems}}. \bibinfo{pages}{1589--1598}.
\newblock


\bibitem[\protect\citeauthoryear{Calefato, Lanubile, and Novielli}{Calefato
  et~al\mbox{.}}{2019}]%
        {calefato2019empirical}
\bibfield{author}{\bibinfo{person}{Fabio Calefato}, \bibinfo{person}{Filippo
  Lanubile}, {and} \bibinfo{person}{Nicole Novielli}.}
  \bibinfo{year}{2019}\natexlab{}.
\newblock \showarticletitle{An empirical assessment of best-answer prediction
  models in technical Q\&A sites}.
\newblock \bibinfo{journal}{\emph{Empirical Software Engineering}}
  \bibinfo{volume}{24}, \bibinfo{number}{2} (\bibinfo{year}{2019}),
  \bibinfo{pages}{854--901}.
\newblock


\bibitem[\protect\citeauthoryear{Cambronero, Li, Kim, Sen, and
  Chandra}{Cambronero et~al\mbox{.}}{2019}]%
        {cambronero2019deep}
\bibfield{author}{\bibinfo{person}{Jose Cambronero}, \bibinfo{person}{Hongyu
  Li}, \bibinfo{person}{Seohyun Kim}, \bibinfo{person}{Koushik Sen}, {and}
  \bibinfo{person}{Satish Chandra}.} \bibinfo{year}{2019}\natexlab{}.
\newblock \showarticletitle{When deep learning met code search}. In
  \bibinfo{booktitle}{\emph{Proceedings of the 2019 27th ACM Joint Meeting on
  European Software Engineering Conference and Symposium on the Foundations of
  Software Engineering}}. \bibinfo{pages}{964--974}.
\newblock


\bibitem[\protect\citeauthoryear{Cao, Chen, Baltes, Treude, and Chen}{Cao
  et~al\mbox{.}}{2021}]%
        {cao2021automated}
\bibfield{author}{\bibinfo{person}{Kaibo Cao}, \bibinfo{person}{Chunyang Chen},
  \bibinfo{person}{Sebastian Baltes}, \bibinfo{person}{Christoph Treude}, {and}
  \bibinfo{person}{Xiang Chen}.} \bibinfo{year}{2021}\natexlab{}.
\newblock \showarticletitle{Automated Query Reformulation for Efficient Search
  based on Query Logs From Stack Overflow}. In \bibinfo{booktitle}{\emph{2021
  IEEE/ACM 43rd International Conference on Software Engineering (ICSE)}}.
  IEEE, \bibinfo{pages}{1273--1285}.
\newblock


\bibitem[\protect\citeauthoryear{Cao, Cong, Cui, and Jensen}{Cao
  et~al\mbox{.}}{2010}]%
        {cao2010generalized}
\bibfield{author}{\bibinfo{person}{Xin Cao}, \bibinfo{person}{Gao Cong},
  \bibinfo{person}{Bin Cui}, {and} \bibinfo{person}{Christian~S Jensen}.}
  \bibinfo{year}{2010}\natexlab{}.
\newblock \showarticletitle{A generalized framework of exploring category
  information for question retrieval in community question answer archives}. In
  \bibinfo{booktitle}{\emph{Proceedings of the 19th international conference on
  World wide web}}. \bibinfo{pages}{201--210}.
\newblock


\bibitem[\protect\citeauthoryear{da~Silva, Roy, Rahman, Schneider, Paix{\~a}o,
  de~Carvalho~Dantas, and de~Almeida~Maia}{da~Silva et~al\mbox{.}}{2020}]%
        {da2020crokage}
\bibfield{author}{\bibinfo{person}{Rodrigo Fernandes~Gomes da Silva},
  \bibinfo{person}{Chanchal~K Roy}, \bibinfo{person}{Mohammad~Masudur Rahman},
  \bibinfo{person}{Kevin~A Schneider}, \bibinfo{person}{Kl{\'e}risson
  Paix{\~a}o}, \bibinfo{person}{Carlos~Eduardo de Carvalho~Dantas}, {and}
  \bibinfo{person}{Marcelo de Almeida~Maia}.} \bibinfo{year}{2020}\natexlab{}.
\newblock \showarticletitle{CROKAGE: effective solution recommendation for
  programming tasks by leveraging crowd knowledge}.
\newblock \bibinfo{journal}{\emph{Empirical Software Engineering}}
  \bibinfo{volume}{25}, \bibinfo{number}{6} (\bibinfo{year}{2020}),
  \bibinfo{pages}{4707--4758}.
\newblock


\bibitem[\protect\citeauthoryear{Devlin, Chang, Lee, and Toutanova}{Devlin
  et~al\mbox{.}}{2018}]%
        {devlin2018bert}
\bibfield{author}{\bibinfo{person}{Jacob Devlin}, \bibinfo{person}{Ming-Wei
  Chang}, \bibinfo{person}{Kenton Lee}, {and} \bibinfo{person}{Kristina
  Toutanova}.} \bibinfo{year}{2018}\natexlab{}.
\newblock \showarticletitle{Bert: Pre-training of deep bidirectional
  transformers for language understanding}.
\newblock \bibinfo{journal}{\emph{arXiv preprint arXiv:1810.04805}}
  (\bibinfo{year}{2018}).
\newblock


\bibitem[\protect\citeauthoryear{Ganguly and Jones}{Ganguly and Jones}{2015}]%
        {ganguly2015partially}
\bibfield{author}{\bibinfo{person}{Debasis Ganguly} {and}
  \bibinfo{person}{Gareth~JF Jones}.} \bibinfo{year}{2015}\natexlab{}.
\newblock \showarticletitle{Partially labeled supervised topic models for
  RetrievingSimilar questions in CQA forums}. In
  \bibinfo{booktitle}{\emph{Proceedings of the 2015 International Conference on
  The Theory of Information Retrieval}}. \bibinfo{pages}{161--170}.
\newblock


\bibitem[\protect\citeauthoryear{Gao, Jiang, Xia, Lo, and Grundy}{Gao
  et~al\mbox{.}}{2020a}]%
        {gao2020checking}
\bibfield{author}{\bibinfo{person}{Zhipeng Gao}, \bibinfo{person}{Lingxiao
  Jiang}, \bibinfo{person}{Xin Xia}, \bibinfo{person}{David Lo}, {and}
  \bibinfo{person}{John Grundy}.} \bibinfo{year}{2020}\natexlab{a}.
\newblock \showarticletitle{Checking smart contracts with structural code
  embedding}.
\newblock \bibinfo{journal}{\emph{IEEE Transactions on Software Engineering}}
  (\bibinfo{year}{2020}).
\newblock


\bibitem[\protect\citeauthoryear{Gao, Xia, Grundy, Lo, and Li}{Gao
  et~al\mbox{.}}{2020b}]%
        {gao2020generating}
\bibfield{author}{\bibinfo{person}{Zhipeng Gao}, \bibinfo{person}{Xin Xia},
  \bibinfo{person}{John Grundy}, \bibinfo{person}{David Lo}, {and}
  \bibinfo{person}{Yuan-Fang Li}.} \bibinfo{year}{2020}\natexlab{b}.
\newblock \showarticletitle{Generating question titles for stack overflow from
  mined code snippets}.
\newblock \bibinfo{journal}{\emph{ACM Transactions on Software Engineering and
  Methodology (TOSEM)}} \bibinfo{volume}{29}, \bibinfo{number}{4}
  (\bibinfo{year}{2020}), \bibinfo{pages}{1--37}.
\newblock


\bibitem[\protect\citeauthoryear{Gao, Xia, Lo, and Grundy}{Gao
  et~al\mbox{.}}{2020c}]%
        {gao2020deepans}
\bibfield{author}{\bibinfo{person}{Zhipeng Gao}, \bibinfo{person}{Xin Xia},
  \bibinfo{person}{David Lo}, {and} \bibinfo{person}{John Grundy}.}
  \bibinfo{year}{2020}\natexlab{c}.
\newblock \showarticletitle{Technical Q8A Site Answer Recommendation via
  Question Boosting}.
\newblock \bibinfo{journal}{\emph{ACM Transactions on Software Engineering and
  Methodology (TOSEM)}} \bibinfo{volume}{30}, \bibinfo{number}{1}
  (\bibinfo{year}{2020}), \bibinfo{pages}{1--34}.
\newblock


\bibitem[\protect\citeauthoryear{Gao, Xia, Lo, Grundy, and Zimmermann}{Gao
  et~al\mbox{.}}{2021}]%
        {gao2021automating}
\bibfield{author}{\bibinfo{person}{Zhipeng Gao}, \bibinfo{person}{Xin Xia},
  \bibinfo{person}{David Lo}, \bibinfo{person}{John Grundy}, {and}
  \bibinfo{person}{Thomas Zimmermann}.} \bibinfo{year}{2021}\natexlab{}.
\newblock \showarticletitle{Automating the removal of obsolete TODO comments}.
  In \bibinfo{booktitle}{\emph{Proceedings of the 29th ACM Joint Meeting on
  European Software Engineering Conference and Symposium on the Foundations of
  Software Engineering}}. \bibinfo{pages}{218--229}.
\newblock


\bibitem[\protect\citeauthoryear{Gonz{\'a}lez, Romero, Guerrero, and
  Calder{\'o}n}{Gonz{\'a}lez et~al\mbox{.}}{2015}]%
        {gonzalez2015multi}
\bibfield{author}{\bibinfo{person}{Jos{\'e} R~Cedeno Gonz{\'a}lez},
  \bibinfo{person}{Juan J~Flores Romero}, \bibinfo{person}{Mario~Graff
  Guerrero}, {and} \bibinfo{person}{Felix Calder{\'o}n}.}
  \bibinfo{year}{2015}\natexlab{}.
\newblock \showarticletitle{Multi-class multi-tag classifier system for
  stackoverflow questions}. In \bibinfo{booktitle}{\emph{2015 IEEE
  International Autumn Meeting on Power, Electronics and Computing (ROPEC)}}.
  IEEE, \bibinfo{pages}{1--6}.
\newblock


\bibitem[\protect\citeauthoryear{Gu, Zhang, and Kim}{Gu et~al\mbox{.}}{2018}]%
        {gu2018deep}
\bibfield{author}{\bibinfo{person}{Xiaodong Gu}, \bibinfo{person}{Hongyu
  Zhang}, {and} \bibinfo{person}{Sunghun Kim}.}
  \bibinfo{year}{2018}\natexlab{}.
\newblock \showarticletitle{Deep code search}. In
  \bibinfo{booktitle}{\emph{2018 IEEE/ACM 40th International Conference on
  Software Engineering (ICSE)}}. IEEE, \bibinfo{pages}{933--944}.
\newblock


\bibitem[\protect\citeauthoryear{Haiduc, Bavota, Marcus, Oliveto, De~Lucia, and
  Menzies}{Haiduc et~al\mbox{.}}{2013}]%
        {haiduc2013automatic}
\bibfield{author}{\bibinfo{person}{Sonia Haiduc}, \bibinfo{person}{Gabriele
  Bavota}, \bibinfo{person}{Andrian Marcus}, \bibinfo{person}{Rocco Oliveto},
  \bibinfo{person}{Andrea De~Lucia}, {and} \bibinfo{person}{Tim Menzies}.}
  \bibinfo{year}{2013}\natexlab{}.
\newblock \showarticletitle{Automatic query reformulations for text retrieval
  in software engineering}. In \bibinfo{booktitle}{\emph{2013 35th
  International Conference on Software Engineering (ICSE)}}. IEEE,
  \bibinfo{pages}{842--851}.
\newblock


\bibitem[\protect\citeauthoryear{Hashemi, Aliannejadi, Zamani, and
  Croft}{Hashemi et~al\mbox{.}}{2020}]%
        {hashemi2020antique}
\bibfield{author}{\bibinfo{person}{Helia Hashemi}, \bibinfo{person}{Mohammad
  Aliannejadi}, \bibinfo{person}{Hamed Zamani}, {and} \bibinfo{person}{W~Bruce
  Croft}.} \bibinfo{year}{2020}\natexlab{}.
\newblock \showarticletitle{ANTIQUE: A non-factoid question answering
  benchmark}. In \bibinfo{booktitle}{\emph{European Conference on Information
  Retrieval}}. Springer, \bibinfo{pages}{166--173}.
\newblock


\bibitem[\protect\citeauthoryear{He, Zhang, Ren, and Sun}{He
  et~al\mbox{.}}{2016}]%
        {he2016deep}
\bibfield{author}{\bibinfo{person}{Kaiming He}, \bibinfo{person}{Xiangyu
  Zhang}, \bibinfo{person}{Shaoqing Ren}, {and} \bibinfo{person}{Jian Sun}.}
  \bibinfo{year}{2016}\natexlab{}.
\newblock \showarticletitle{Deep residual learning for image recognition}. In
  \bibinfo{booktitle}{\emph{Proceedings of the IEEE conference on computer
  vision and pattern recognition}}. \bibinfo{pages}{770--778}.
\newblock


\bibitem[\protect\citeauthoryear{He, Liao, Zhang, Nie, Hu, and Chua}{He
  et~al\mbox{.}}{2017}]%
        {he2017neural}
\bibfield{author}{\bibinfo{person}{Xiangnan He}, \bibinfo{person}{Lizi Liao},
  \bibinfo{person}{Hanwang Zhang}, \bibinfo{person}{Liqiang Nie},
  \bibinfo{person}{Xia Hu}, {and} \bibinfo{person}{Tat-Seng Chua}.}
  \bibinfo{year}{2017}\natexlab{}.
\newblock \showarticletitle{Neural collaborative filtering}. In
  \bibinfo{booktitle}{\emph{Proceedings of the 26th international conference on
  world wide web}}. \bibinfo{pages}{173--182}.
\newblock


\bibitem[\protect\citeauthoryear{Hill, Roldan-Vega, Fails, and Mallet}{Hill
  et~al\mbox{.}}{2014}]%
        {hill2014nl}
\bibfield{author}{\bibinfo{person}{Emily Hill}, \bibinfo{person}{Manuel
  Roldan-Vega}, \bibinfo{person}{Jerry~Alan Fails}, {and} \bibinfo{person}{Greg
  Mallet}.} \bibinfo{year}{2014}\natexlab{}.
\newblock \showarticletitle{NL-based query refinement and contextualized code
  search results: A user study}. In \bibinfo{booktitle}{\emph{2014 Software
  Evolution Week-IEEE Conference on Software Maintenance, Reengineering, and
  Reverse Engineering (CSMR-WCRE)}}. IEEE, \bibinfo{pages}{34--43}.
\newblock


\bibitem[\protect\citeauthoryear{Huang, Cui, Sun, and Towey}{Huang
  et~al\mbox{.}}{2020}]%
        {huang2020poster}
\bibfield{author}{\bibinfo{person}{Rubing Huang}, \bibinfo{person}{Chenhui
  Cui}, \bibinfo{person}{Weifeng Sun}, {and} \bibinfo{person}{Dave Towey}.}
  \bibinfo{year}{2020}\natexlab{}.
\newblock \showarticletitle{Poster: Is Euclidean Distance the best Distance
  Measurement for Adaptive Random Testing?}. In \bibinfo{booktitle}{\emph{2020
  IEEE 13th International Conference on Software Testing, Validation and
  Verification (ICST)}}. IEEE, \bibinfo{pages}{406--409}.
\newblock


\bibitem[\protect\citeauthoryear{Jiang, Nie, Sun, Ren, Kong, Zhang, and
  Luo}{Jiang et~al\mbox{.}}{2016}]%
        {jiang2016rosf}
\bibfield{author}{\bibinfo{person}{He Jiang}, \bibinfo{person}{Liming Nie},
  \bibinfo{person}{Zeyi Sun}, \bibinfo{person}{Zhilei Ren},
  \bibinfo{person}{Weiqiang Kong}, \bibinfo{person}{Tao Zhang}, {and}
  \bibinfo{person}{Xiapu Luo}.} \bibinfo{year}{2016}\natexlab{}.
\newblock \showarticletitle{Rosf: Leveraging information retrieval and
  supervised learning for recommending code snippets}.
\newblock \bibinfo{journal}{\emph{IEEE Transactions on Services Computing}}
  \bibinfo{volume}{12}, \bibinfo{number}{1} (\bibinfo{year}{2016}),
  \bibinfo{pages}{34--46}.
\newblock


\bibitem[\protect\citeauthoryear{Koehn}{Koehn}{2004}]%
        {koehn2004pharaoh}
\bibfield{author}{\bibinfo{person}{Philipp Koehn}.}
  \bibinfo{year}{2004}\natexlab{}.
\newblock \showarticletitle{Pharaoh: a beam search decoder for phrase-based
  statistical machine translation models}. In
  \bibinfo{booktitle}{\emph{Conference of the Association for Machine
  Translation in the Americas}}. Springer, \bibinfo{pages}{115--124}.
\newblock


\bibitem[\protect\citeauthoryear{Kumar and Pedanekar}{Kumar and
  Pedanekar}{2016}]%
        {kumar2016mining}
\bibfield{author}{\bibinfo{person}{Varun Kumar} {and} \bibinfo{person}{Niranjan
  Pedanekar}.} \bibinfo{year}{2016}\natexlab{}.
\newblock \showarticletitle{Mining shapes of expertise in online social Q\&A
  communities}. In \bibinfo{booktitle}{\emph{Proceedings of the 19th ACM
  conference on computer supported cooperative work and social computing
  companion}}. \bibinfo{pages}{317--320}.
\newblock


\bibitem[\protect\citeauthoryear{Lan, Chen, Goodman, Gimpel, Sharma, and
  Soricut}{Lan et~al\mbox{.}}{2019}]%
        {lan2019albert}
\bibfield{author}{\bibinfo{person}{Zhenzhong Lan}, \bibinfo{person}{Mingda
  Chen}, \bibinfo{person}{Sebastian Goodman}, \bibinfo{person}{Kevin Gimpel},
  \bibinfo{person}{Piyush Sharma}, {and} \bibinfo{person}{Radu Soricut}.}
  \bibinfo{year}{2019}\natexlab{}.
\newblock \showarticletitle{Albert: A lite bert for self-supervised learning of
  language representations}.
\newblock \bibinfo{journal}{\emph{arXiv preprint arXiv:1909.11942}}
  (\bibinfo{year}{2019}).
\newblock


\bibitem[\protect\citeauthoryear{Le and Mikolov}{Le and Mikolov}{2014}]%
        {le2014distributed}
\bibfield{author}{\bibinfo{person}{Quoc Le} {and} \bibinfo{person}{Tomas
  Mikolov}.} \bibinfo{year}{2014}\natexlab{}.
\newblock \showarticletitle{Distributed representations of sentences and
  documents}. In \bibinfo{booktitle}{\emph{International conference on machine
  learning}}. \bibinfo{pages}{1188--1196}.
\newblock


\bibitem[\protect\citeauthoryear{Liu, Ott, Goyal, Du, Joshi, Chen, Levy, Lewis,
  Zettlemoyer, and Stoyanov}{Liu et~al\mbox{.}}{2019}]%
        {liu2019roberta}
\bibfield{author}{\bibinfo{person}{Yinhan Liu}, \bibinfo{person}{Myle Ott},
  \bibinfo{person}{Naman Goyal}, \bibinfo{person}{Jingfei Du},
  \bibinfo{person}{Mandar Joshi}, \bibinfo{person}{Danqi Chen},
  \bibinfo{person}{Omer Levy}, \bibinfo{person}{Mike Lewis},
  \bibinfo{person}{Luke Zettlemoyer}, {and} \bibinfo{person}{Veselin
  Stoyanov}.} \bibinfo{year}{2019}\natexlab{}.
\newblock \showarticletitle{Roberta: A robustly optimized bert pretraining
  approach}.
\newblock \bibinfo{journal}{\emph{arXiv preprint arXiv:1907.11692}}
  (\bibinfo{year}{2019}).
\newblock


\bibitem[\protect\citeauthoryear{Liu, Xia, Hassan, Lo, Xing, and Wang}{Liu
  et~al\mbox{.}}{2018}]%
        {liu2018neural}
\bibfield{author}{\bibinfo{person}{Zhongxin Liu}, \bibinfo{person}{Xin Xia},
  \bibinfo{person}{Ahmed~E Hassan}, \bibinfo{person}{David Lo},
  \bibinfo{person}{Zhenchang Xing}, {and} \bibinfo{person}{Xinyu Wang}.}
  \bibinfo{year}{2018}\natexlab{}.
\newblock \showarticletitle{Neural-machine-translation-based commit message
  generation: how far are we?}. In \bibinfo{booktitle}{\emph{Proceedings of the
  33rd ACM/IEEE International Conference on Automated Software Engineering}}.
  \bibinfo{pages}{373--384}.
\newblock


\bibitem[\protect\citeauthoryear{Lu, Sun, Wang, Lo, and Duan}{Lu
  et~al\mbox{.}}{2015}]%
        {lu2015query}
\bibfield{author}{\bibinfo{person}{Meili Lu}, \bibinfo{person}{Xiaobing Sun},
  \bibinfo{person}{Shaowei Wang}, \bibinfo{person}{David Lo}, {and}
  \bibinfo{person}{Yucong Duan}.} \bibinfo{year}{2015}\natexlab{}.
\newblock \showarticletitle{Query expansion via wordnet for effective code
  search}. In \bibinfo{booktitle}{\emph{2015 IEEE 22nd International Conference
  on Software Analysis, Evolution, and Reengineering (SANER)}}. IEEE,
  \bibinfo{pages}{545--549}.
\newblock


\bibitem[\protect\citeauthoryear{Lv, Zhang, Lou, Wang, Zhang, and Zhao}{Lv
  et~al\mbox{.}}{2015}]%
        {lv2015codehow}
\bibfield{author}{\bibinfo{person}{Fei Lv}, \bibinfo{person}{Hongyu Zhang},
  \bibinfo{person}{Jian-guang Lou}, \bibinfo{person}{Shaowei Wang},
  \bibinfo{person}{Dongmei Zhang}, {and} \bibinfo{person}{Jianjun Zhao}.}
  \bibinfo{year}{2015}\natexlab{}.
\newblock \showarticletitle{Codehow: Effective code search based on api
  understanding and extended boolean model (e)}. In
  \bibinfo{booktitle}{\emph{2015 30th IEEE/ACM International Conference on
  Automated Software Engineering (ASE)}}. IEEE, \bibinfo{pages}{260--270}.
\newblock


\bibitem[\protect\citeauthoryear{Manning, Raghavan, and Sch{\"u}tze}{Manning
  et~al\mbox{.}}{2005}]%
        {Manning2005IntroductionTI}
\bibfield{author}{\bibinfo{person}{Christopher~D. Manning}, \bibinfo{person}{P.
  Raghavan}, {and} \bibinfo{person}{Hinrich Sch{\"u}tze}.}
  \bibinfo{year}{2005}\natexlab{}.
\newblock \showarticletitle{Introduction to information retrieval}.
\newblock


\bibitem[\protect\citeauthoryear{Mikolov, Sutskever, Chen, Corrado, and
  Dean}{Mikolov et~al\mbox{.}}{2013}]%
        {mikolov2013distributed}
\bibfield{author}{\bibinfo{person}{Tomas Mikolov}, \bibinfo{person}{Ilya
  Sutskever}, \bibinfo{person}{Kai Chen}, \bibinfo{person}{Greg~S Corrado},
  {and} \bibinfo{person}{Jeff Dean}.} \bibinfo{year}{2013}\natexlab{}.
\newblock \showarticletitle{Distributed representations of words and phrases
  and their compositionality}. In \bibinfo{booktitle}{\emph{Advances in neural
  information processing systems}}. \bibinfo{pages}{3111--3119}.
\newblock


\bibitem[\protect\citeauthoryear{Mizobuchi and Takayama}{Mizobuchi and
  Takayama}{2017}]%
        {mizobuchi2017two}
\bibfield{author}{\bibinfo{person}{Yuji Mizobuchi} {and}
  \bibinfo{person}{Kuniharu Takayama}.} \bibinfo{year}{2017}\natexlab{}.
\newblock \showarticletitle{Two improvements to detect duplicates in Stack
  Overflow}. In \bibinfo{booktitle}{\emph{2017 IEEE 24th International
  Conference on Software Analysis, Evolution and Reengineering (SANER)}}. IEEE,
  \bibinfo{pages}{563--564}.
\newblock


\bibitem[\protect\citeauthoryear{Moreno, Bavota, Di~Penta, Oliveto, and
  Marcus}{Moreno et~al\mbox{.}}{2015}]%
        {moreno2015can}
\bibfield{author}{\bibinfo{person}{Laura Moreno}, \bibinfo{person}{Gabriele
  Bavota}, \bibinfo{person}{Massimiliano Di~Penta}, \bibinfo{person}{Rocco
  Oliveto}, {and} \bibinfo{person}{Andrian Marcus}.}
  \bibinfo{year}{2015}\natexlab{}.
\newblock \showarticletitle{How can I use this method?}. In
  \bibinfo{booktitle}{\emph{2015 IEEE/ACM 37th IEEE International Conference on
  Software Engineering}}, Vol.~\bibinfo{volume}{1}. IEEE,
  \bibinfo{pages}{880--890}.
\newblock


\bibitem[\protect\citeauthoryear{Nie, Jiang, Ren, Sun, and Li}{Nie
  et~al\mbox{.}}{2016}]%
        {nie2016query}
\bibfield{author}{\bibinfo{person}{Liming Nie}, \bibinfo{person}{He Jiang},
  \bibinfo{person}{Zhilei Ren}, \bibinfo{person}{Zeyi Sun}, {and}
  \bibinfo{person}{Xiaochen Li}.} \bibinfo{year}{2016}\natexlab{}.
\newblock \showarticletitle{Query expansion based on crowd knowledge for code
  search}.
\newblock \bibinfo{journal}{\emph{IEEE Transactions on Services Computing}}
  \bibinfo{volume}{9}, \bibinfo{number}{5} (\bibinfo{year}{2016}),
  \bibinfo{pages}{771--783}.
\newblock


\bibitem[\protect\citeauthoryear{Nie, Wei, Zhang, Wang, Gao, and Yang}{Nie
  et~al\mbox{.}}{2017}]%
        {nie2017data}
\bibfield{author}{\bibinfo{person}{Liqiang Nie}, \bibinfo{person}{Xiaochi Wei},
  \bibinfo{person}{Dongxiang Zhang}, \bibinfo{person}{Xiang Wang},
  \bibinfo{person}{Zhipeng Gao}, {and} \bibinfo{person}{Yi Yang}.}
  \bibinfo{year}{2017}\natexlab{}.
\newblock \showarticletitle{Data-driven answer selection in community QA
  systems}.
\newblock \bibinfo{journal}{\emph{IEEE transactions on knowledge and data
  engineering}} \bibinfo{volume}{29}, \bibinfo{number}{6}
  (\bibinfo{year}{2017}), \bibinfo{pages}{1186--1198}.
\newblock


\bibitem[\protect\citeauthoryear{Pal, Harper, and Konstan}{Pal
  et~al\mbox{.}}{2012}]%
        {pal2012exploring}
\bibfield{author}{\bibinfo{person}{Aditya Pal}, \bibinfo{person}{F~Maxwell
  Harper}, {and} \bibinfo{person}{Joseph~A Konstan}.}
  \bibinfo{year}{2012}\natexlab{}.
\newblock \showarticletitle{Exploring question selection bias to identify
  experts and potential experts in community question answering}.
\newblock \bibinfo{journal}{\emph{ACM Transactions on Information Systems
  (TOIS)}} \bibinfo{volume}{30}, \bibinfo{number}{2} (\bibinfo{year}{2012}),
  \bibinfo{pages}{1--28}.
\newblock


\bibitem[\protect\citeauthoryear{Ponzanelli, Bavota, Mocci, Di~Penta, Oliveto,
  Russo, Haiduc, and Lanza}{Ponzanelli et~al\mbox{.}}{2016}]%
        {ponzanelli2016codetube}
\bibfield{author}{\bibinfo{person}{Luca Ponzanelli}, \bibinfo{person}{Gabriele
  Bavota}, \bibinfo{person}{Andrea Mocci}, \bibinfo{person}{Massimiliano
  Di~Penta}, \bibinfo{person}{Rocco Oliveto}, \bibinfo{person}{Barbara Russo},
  \bibinfo{person}{Sonia Haiduc}, {and} \bibinfo{person}{Michele Lanza}.}
  \bibinfo{year}{2016}\natexlab{}.
\newblock \showarticletitle{CodeTube: extracting relevant fragments from
  software development video tutorials}. In \bibinfo{booktitle}{\emph{2016
  IEEE/ACM 38th International Conference on Software Engineering Companion
  (ICSE-C)}}. IEEE, \bibinfo{pages}{645--648}.
\newblock


\bibitem[\protect\citeauthoryear{Ponzanelli, Mocci, Bacchelli, Lanza, and
  Fullerton}{Ponzanelli et~al\mbox{.}}{2014}]%
        {ponzanelli2014improving}
\bibfield{author}{\bibinfo{person}{Luca Ponzanelli}, \bibinfo{person}{Andrea
  Mocci}, \bibinfo{person}{Alberto Bacchelli}, \bibinfo{person}{Michele Lanza},
  {and} \bibinfo{person}{David Fullerton}.} \bibinfo{year}{2014}\natexlab{}.
\newblock \showarticletitle{Improving low quality stack overflow post
  detection}. In \bibinfo{booktitle}{\emph{2014 IEEE international conference
  on software maintenance and evolution}}. IEEE, \bibinfo{pages}{541--544}.
\newblock


\bibitem[\protect\citeauthoryear{Radford, Narasimhan, Salimans, and
  Sutskever}{Radford et~al\mbox{.}}{2018}]%
        {radford2018improving}
\bibfield{author}{\bibinfo{person}{Alec Radford}, \bibinfo{person}{Karthik
  Narasimhan}, \bibinfo{person}{Tim Salimans}, {and} \bibinfo{person}{Ilya
  Sutskever}.} \bibinfo{year}{2018}\natexlab{}.
\newblock \bibinfo{title}{Improving language understanding by generative
  pre-training}.
\newblock
\newblock


\bibitem[\protect\citeauthoryear{Raffel, Shazeer, Roberts, Lee, Narang, Matena,
  Zhou, Li, and Liu}{Raffel et~al\mbox{.}}{2019}]%
        {raffel2019exploring}
\bibfield{author}{\bibinfo{person}{Colin Raffel}, \bibinfo{person}{Noam
  Shazeer}, \bibinfo{person}{Adam Roberts}, \bibinfo{person}{Katherine Lee},
  \bibinfo{person}{Sharan Narang}, \bibinfo{person}{Michael Matena},
  \bibinfo{person}{Yanqi Zhou}, \bibinfo{person}{Wei Li}, {and}
  \bibinfo{person}{Peter~J Liu}.} \bibinfo{year}{2019}\natexlab{}.
\newblock \showarticletitle{Exploring the limits of transfer learning with a
  unified text-to-text transformer}.
\newblock \bibinfo{journal}{\emph{arXiv preprint arXiv:1910.10683}}
  (\bibinfo{year}{2019}).
\newblock


\bibitem[\protect\citeauthoryear{Rahman, Barson, Paul, Kayani, Lois, Quezada,
  Parnin, Stolee, and Ray}{Rahman et~al\mbox{.}}{2018}]%
        {rahman2018evaluating}
\bibfield{author}{\bibinfo{person}{Md~Masudur Rahman}, \bibinfo{person}{Jed
  Barson}, \bibinfo{person}{Sydney Paul}, \bibinfo{person}{Joshua Kayani},
  \bibinfo{person}{Federico~Andr{\'e}s Lois},
  \bibinfo{person}{Sebasti{\'a}n~Fernandez Quezada},
  \bibinfo{person}{Christopher Parnin}, \bibinfo{person}{Kathryn~T Stolee},
  {and} \bibinfo{person}{Baishakhi Ray}.} \bibinfo{year}{2018}\natexlab{}.
\newblock \showarticletitle{Evaluating how developers use general-purpose
  web-search for code retrieval}. In \bibinfo{booktitle}{\emph{Proceedings of
  the 15th International Conference on Mining Software Repositories}}.
  \bibinfo{pages}{465--475}.
\newblock


\bibitem[\protect\citeauthoryear{Rahman and Roy}{Rahman and Roy}{2018}]%
        {rahman2018effective}
\bibfield{author}{\bibinfo{person}{Mohammad~Masudur Rahman} {and}
  \bibinfo{person}{Chanchal Roy}.} \bibinfo{year}{2018}\natexlab{}.
\newblock \showarticletitle{Effective reformulation of query for code search
  using crowdsourced knowledge and extra-large data analytics}. In
  \bibinfo{booktitle}{\emph{2018 IEEE International Conference on Software
  Maintenance and Evolution (ICSME)}}. IEEE, \bibinfo{pages}{473--484}.
\newblock


\bibitem[\protect\citeauthoryear{Rahman, Roy, and Lo}{Rahman
  et~al\mbox{.}}{2019}]%
        {rahman2019automatic}
\bibfield{author}{\bibinfo{person}{Mohammad~M Rahman},
  \bibinfo{person}{Chanchal~K Roy}, {and} \bibinfo{person}{David Lo}.}
  \bibinfo{year}{2019}\natexlab{}.
\newblock \showarticletitle{Automatic query reformulation for code search using
  crowdsourced knowledge}.
\newblock \bibinfo{journal}{\emph{Empirical Software Engineering}}
  \bibinfo{volume}{24}, \bibinfo{number}{4} (\bibinfo{year}{2019}),
  \bibinfo{pages}{1869--1924}.
\newblock


\bibitem[\protect\citeauthoryear{{\v R}eh{\r u}{\v r}ek and Sojka}{{\v R}eh{\r
  u}{\v r}ek and Sojka}{2010}]%
        {rehurek_lrec}
\bibfield{author}{\bibinfo{person}{Radim {\v R}eh{\r u}{\v r}ek} {and}
  \bibinfo{person}{Petr Sojka}.} \bibinfo{year}{2010}\natexlab{}.
\newblock \showarticletitle{{Software Framework for Topic Modelling with Large
  Corpora}}. In \bibinfo{booktitle}{\emph{{Proceedings of the LREC 2010
  Workshop on New Challenges for NLP Frameworks}}}. \bibinfo{publisher}{ELRA},
  \bibinfo{address}{Valletta, Malta}, \bibinfo{pages}{45--50}.
\newblock
\newblock
\shownote{\url{http://is.muni.cz/publication/884893/en}.}


\bibitem[\protect\citeauthoryear{Rosen and Shihab}{Rosen and Shihab}{2016}]%
        {rosen2016mobile}
\bibfield{author}{\bibinfo{person}{Christoffer Rosen} {and}
  \bibinfo{person}{Emad Shihab}.} \bibinfo{year}{2016}\natexlab{}.
\newblock \showarticletitle{What are mobile developers asking about? a large
  scale study using stack overflow}.
\newblock \bibinfo{journal}{\emph{Empirical Software Engineering}}
  \bibinfo{volume}{21}, \bibinfo{number}{3} (\bibinfo{year}{2016}),
  \bibinfo{pages}{1192--1223}.
\newblock


\bibitem[\protect\citeauthoryear{Sachdev, Li, Luan, Kim, Sen, and
  Chandra}{Sachdev et~al\mbox{.}}{2018}]%
        {sachdev2018retrieval}
\bibfield{author}{\bibinfo{person}{Saksham Sachdev}, \bibinfo{person}{Hongyu
  Li}, \bibinfo{person}{Sifei Luan}, \bibinfo{person}{Seohyun Kim},
  \bibinfo{person}{Koushik Sen}, {and} \bibinfo{person}{Satish Chandra}.}
  \bibinfo{year}{2018}\natexlab{}.
\newblock \showarticletitle{Retrieval on source code: a neural code search}. In
  \bibinfo{booktitle}{\emph{Proceedings of the 2nd ACM SIGPLAN International
  Workshop on Machine Learning and Programming Languages}}.
  \bibinfo{pages}{31--41}.
\newblock


\bibitem[\protect\citeauthoryear{Sanh, Debut, Chaumond, and Wolf}{Sanh
  et~al\mbox{.}}{2019}]%
        {sanh2019distilbert}
\bibfield{author}{\bibinfo{person}{Victor Sanh}, \bibinfo{person}{Lysandre
  Debut}, \bibinfo{person}{Julien Chaumond}, {and} \bibinfo{person}{Thomas
  Wolf}.} \bibinfo{year}{2019}\natexlab{}.
\newblock \showarticletitle{DistilBERT, a distilled version of BERT: smaller,
  faster, cheaper and lighter}.
\newblock \bibinfo{journal}{\emph{arXiv preprint arXiv:1910.01108}}
  (\bibinfo{year}{2019}).
\newblock


\bibitem[\protect\citeauthoryear{Shepherd, Fry, Hill, Pollock, and
  Vijay-Shanker}{Shepherd et~al\mbox{.}}{2007}]%
        {shepherd2007using}
\bibfield{author}{\bibinfo{person}{David Shepherd}, \bibinfo{person}{Zachary~P
  Fry}, \bibinfo{person}{Emily Hill}, \bibinfo{person}{Lori Pollock}, {and}
  \bibinfo{person}{K Vijay-Shanker}.} \bibinfo{year}{2007}\natexlab{}.
\newblock \showarticletitle{Using natural language program analysis to locate
  and understand action-oriented concerns}. In
  \bibinfo{booktitle}{\emph{Proceedings of the 6th international conference on
  Aspect-oriented software development}}. \bibinfo{pages}{212--224}.
\newblock


\bibitem[\protect\citeauthoryear{Silva, Paix{\~a}o, and de~Almeida~Maia}{Silva
  et~al\mbox{.}}{2018}]%
        {silva2018duplicate}
\bibfield{author}{\bibinfo{person}{Rodrigo~FG Silva},
  \bibinfo{person}{Kl{\'e}risson Paix{\~a}o}, {and} \bibinfo{person}{Marcelo de
  Almeida~Maia}.} \bibinfo{year}{2018}\natexlab{}.
\newblock \showarticletitle{Duplicate question detection in stack overflow: A
  reproducibility study}. In \bibinfo{booktitle}{\emph{2018 IEEE 25th
  International Conference on Software Analysis, Evolution and Reengineering
  (SANER)}}. IEEE, \bibinfo{pages}{572--581}.
\newblock


\bibitem[\protect\citeauthoryear{Singh and Simperl}{Singh and Simperl}{2016}]%
        {singh2016using}
\bibfield{author}{\bibinfo{person}{Priyanka Singh} {and} \bibinfo{person}{Elena
  Simperl}.} \bibinfo{year}{2016}\natexlab{}.
\newblock \showarticletitle{Using semantics to search answers for unanswered
  questions in q\&a forums}. In \bibinfo{booktitle}{\emph{Proceedings of the
  25th International Conference Companion on World Wide Web}}.
  \bibinfo{pages}{699--706}.
\newblock


\bibitem[\protect\citeauthoryear{Sirres, Bissyand{\'e}, Kim, Lo, Klein, Kim,
  and Le~Traon}{Sirres et~al\mbox{.}}{2018}]%
        {sirres2018augmenting}
\bibfield{author}{\bibinfo{person}{Raphael Sirres},
  \bibinfo{person}{Tegawend{\'e}~F Bissyand{\'e}}, \bibinfo{person}{Dongsun
  Kim}, \bibinfo{person}{David Lo}, \bibinfo{person}{Jacques Klein},
  \bibinfo{person}{Kisub Kim}, {and} \bibinfo{person}{Yves Le~Traon}.}
  \bibinfo{year}{2018}\natexlab{}.
\newblock \showarticletitle{Augmenting and structuring user queries to support
  efficient free-form code search}.
\newblock \bibinfo{journal}{\emph{Empirical Software Engineering}}
  \bibinfo{volume}{23}, \bibinfo{number}{5} (\bibinfo{year}{2018}),
  \bibinfo{pages}{2622--2654}.
\newblock


\bibitem[\protect\citeauthoryear{Song, Ren, Liang, Li, Ma, and de~Rijke}{Song
  et~al\mbox{.}}{2017}]%
        {song2017summarizing}
\bibfield{author}{\bibinfo{person}{Hongya Song}, \bibinfo{person}{Zhaochun
  Ren}, \bibinfo{person}{Shangsong Liang}, \bibinfo{person}{Piji Li},
  \bibinfo{person}{Jun Ma}, {and} \bibinfo{person}{Maarten de Rijke}.}
  \bibinfo{year}{2017}\natexlab{}.
\newblock \showarticletitle{Summarizing answers in non-factoid community
  question-answering}. In \bibinfo{booktitle}{\emph{Proceedings of the Tenth
  ACM International Conference on Web Search and Data Mining}}.
  \bibinfo{pages}{405--414}.
\newblock


\bibitem[\protect\citeauthoryear{Srba and Bielikova}{Srba and
  Bielikova}{2016}]%
        {srba2016stack}
\bibfield{author}{\bibinfo{person}{Ivan Srba} {and} \bibinfo{person}{Maria
  Bielikova}.} \bibinfo{year}{2016}\natexlab{}.
\newblock \showarticletitle{Why is stack overflow failing? preserving
  sustainability in community question answering}.
\newblock \bibinfo{journal}{\emph{IEEE Software}} \bibinfo{volume}{33},
  \bibinfo{number}{4} (\bibinfo{year}{2016}), \bibinfo{pages}{80--89}.
\newblock


\bibitem[\protect\citeauthoryear{Tian, Kochhar, Lim, Zhu, and Lo}{Tian
  et~al\mbox{.}}{2013}]%
        {tian2013predicting}
\bibfield{author}{\bibinfo{person}{Yuan Tian}, \bibinfo{person}{Pavneet~Singh
  Kochhar}, \bibinfo{person}{Ee-Peng Lim}, \bibinfo{person}{Feida Zhu}, {and}
  \bibinfo{person}{David Lo}.} \bibinfo{year}{2013}\natexlab{}.
\newblock \showarticletitle{Predicting best answerers for new questions: An
  approach leveraging topic modeling and collaborative voting}. In
  \bibinfo{booktitle}{\emph{International Conference on Social Informatics}}.
  Springer, \bibinfo{pages}{55--68}.
\newblock


\bibitem[\protect\citeauthoryear{T{\'o}th, Nagy, Janth{\'o}, Vid{\'a}cs, and
  Gyim{\'o}thy}{T{\'o}th et~al\mbox{.}}{2019}]%
        {toth2019towards}
\bibfield{author}{\bibinfo{person}{L{\'a}szl{\'o} T{\'o}th},
  \bibinfo{person}{Bal{\'a}zs Nagy}, \bibinfo{person}{D{\'a}vid Janth{\'o}},
  \bibinfo{person}{L{\'a}szl{\'o} Vid{\'a}cs}, {and} \bibinfo{person}{Tibor
  Gyim{\'o}thy}.} \bibinfo{year}{2019}\natexlab{}.
\newblock \showarticletitle{Towards an Accurate Prediction of the Question
  Quality on Stack Overflow using a Deep-Learning-Based NLP Approach.}. In
  \bibinfo{booktitle}{\emph{ICSOFT}}. \bibinfo{pages}{631--639}.
\newblock


\bibitem[\protect\citeauthoryear{Vaswani, Shazeer, Parmar, Uszkoreit, Jones,
  Gomez, Kaiser, and Polosukhin}{Vaswani et~al\mbox{.}}{2017}]%
        {vaswani2017attention}
\bibfield{author}{\bibinfo{person}{Ashish Vaswani}, \bibinfo{person}{Noam
  Shazeer}, \bibinfo{person}{Niki Parmar}, \bibinfo{person}{Jakob Uszkoreit},
  \bibinfo{person}{Llion Jones}, \bibinfo{person}{Aidan~N Gomez},
  \bibinfo{person}{{\L}ukasz Kaiser}, {and} \bibinfo{person}{Illia
  Polosukhin}.} \bibinfo{year}{2017}\natexlab{}.
\newblock \showarticletitle{Attention is all you need}. In
  \bibinfo{booktitle}{\emph{Advances in neural information processing
  systems}}. \bibinfo{pages}{5998--6008}.
\newblock


\bibitem[\protect\citeauthoryear{Wang, Ming, and Chua}{Wang
  et~al\mbox{.}}{2009}]%
        {wang2009syntactic}
\bibfield{author}{\bibinfo{person}{Kai Wang}, \bibinfo{person}{Zhaoyan Ming},
  {and} \bibinfo{person}{Tat-Seng Chua}.} \bibinfo{year}{2009}\natexlab{}.
\newblock \showarticletitle{A syntactic tree matching approach to finding
  similar questions in community-based qa services}. In
  \bibinfo{booktitle}{\emph{Proceedings of the 32nd international ACM SIGIR
  conference on Research and development in information retrieval}}.
  \bibinfo{pages}{187--194}.
\newblock


\bibitem[\protect\citeauthoryear{Wang, Zhang, and Jiang}{Wang
  et~al\mbox{.}}{2020}]%
        {wang2020duplicate}
\bibfield{author}{\bibinfo{person}{Liting Wang}, \bibinfo{person}{Li Zhang},
  {and} \bibinfo{person}{Jing Jiang}.} \bibinfo{year}{2020}\natexlab{}.
\newblock \showarticletitle{Duplicate Question Detection With Deep Learning in
  Stack Overflow}.
\newblock \bibinfo{journal}{\emph{IEEE Access}}  \bibinfo{volume}{8}
  (\bibinfo{year}{2020}), \bibinfo{pages}{25964--25975}.
\newblock


\bibitem[\protect\citeauthoryear{Wang, Xia, and Lo}{Wang et~al\mbox{.}}{2015}]%
        {wang2015tagcombine}
\bibfield{author}{\bibinfo{person}{Xin-Yu Wang}, \bibinfo{person}{Xin Xia},
  {and} \bibinfo{person}{David Lo}.} \bibinfo{year}{2015}\natexlab{}.
\newblock \showarticletitle{Tagcombine: Recommending tags to contents in
  software information sites}.
\newblock \bibinfo{journal}{\emph{Journal of Computer Science and Technology}}
  \bibinfo{volume}{30}, \bibinfo{number}{5} (\bibinfo{year}{2015}),
  \bibinfo{pages}{1017--1035}.
\newblock


\bibitem[\protect\citeauthoryear{Wilcoxon}{Wilcoxon}{1992}]%
        {wilcoxon1992individual}
\bibfield{author}{\bibinfo{person}{Frank Wilcoxon}.}
  \bibinfo{year}{1992}\natexlab{}.
\newblock \showarticletitle{Individual comparisons by ranking methods}.
\newblock In \bibinfo{booktitle}{\emph{Breakthroughs in statistics}}.
  \bibinfo{publisher}{Springer}, \bibinfo{pages}{196--202}.
\newblock


\bibitem[\protect\citeauthoryear{Xia, Bao, Lo, Kochhar, Hassan, and Xing}{Xia
  et~al\mbox{.}}{2017}]%
        {xia2017developers}
\bibfield{author}{\bibinfo{person}{Xin Xia}, \bibinfo{person}{Lingfeng Bao},
  \bibinfo{person}{David Lo}, \bibinfo{person}{Pavneet~Singh Kochhar},
  \bibinfo{person}{Ahmed~E Hassan}, {and} \bibinfo{person}{Zhenchang Xing}.}
  \bibinfo{year}{2017}\natexlab{}.
\newblock \showarticletitle{What do developers search for on the web?}
\newblock \bibinfo{journal}{\emph{Empirical Software Engineering}}
  \bibinfo{volume}{22}, \bibinfo{number}{6} (\bibinfo{year}{2017}),
  \bibinfo{pages}{3149--3185}.
\newblock


\bibitem[\protect\citeauthoryear{Xu, Xing, Xia, and Lo}{Xu
  et~al\mbox{.}}{2017}]%
        {xu2017answerbot}
\bibfield{author}{\bibinfo{person}{Bowen Xu}, \bibinfo{person}{Zhenchang Xing},
  \bibinfo{person}{Xin Xia}, {and} \bibinfo{person}{David Lo}.}
  \bibinfo{year}{2017}\natexlab{}.
\newblock \showarticletitle{AnswerBot: Automated generation of answer summary
  to developers' technical questions}. In \bibinfo{booktitle}{\emph{2017 32nd
  IEEE/ACM International Conference on Automated Software Engineering (ASE)}}.
  IEEE, \bibinfo{pages}{706--716}.
\newblock


\bibitem[\protect\citeauthoryear{Xu, Xing, Xia, Lo, and Li}{Xu
  et~al\mbox{.}}{2018}]%
        {xu2018domain}
\bibfield{author}{\bibinfo{person}{Bowen Xu}, \bibinfo{person}{Zhenchang Xing},
  \bibinfo{person}{Xin Xia}, \bibinfo{person}{David Lo}, {and}
  \bibinfo{person}{Shanping Li}.} \bibinfo{year}{2018}\natexlab{}.
\newblock \showarticletitle{Domain-specific cross-language relevant question
  retrieval}.
\newblock \bibinfo{journal}{\emph{Empirical Software Engineering}}
  \bibinfo{volume}{23}, \bibinfo{number}{2} (\bibinfo{year}{2018}),
  \bibinfo{pages}{1084--1122}.
\newblock


\bibitem[\protect\citeauthoryear{Yang, Lo, Xia, Bao, and Sun}{Yang
  et~al\mbox{.}}{2016a}]%
        {yang2016combining}
\bibfield{author}{\bibinfo{person}{Xinli Yang}, \bibinfo{person}{David Lo},
  \bibinfo{person}{Xin Xia}, \bibinfo{person}{Lingfeng Bao}, {and}
  \bibinfo{person}{Jianling Sun}.} \bibinfo{year}{2016}\natexlab{a}.
\newblock \showarticletitle{Combining word embedding with information retrieval
  to recommend similar bug reports}. In \bibinfo{booktitle}{\emph{2016 IEEE
  27Th international symposium on software reliability engineering (ISSRE)}}.
  IEEE, \bibinfo{pages}{127--137}.
\newblock


\bibitem[\protect\citeauthoryear{Yang, Lo, Xia, Wan, and Sun}{Yang
  et~al\mbox{.}}{2016b}]%
        {yang2016security}
\bibfield{author}{\bibinfo{person}{Xin-Li Yang}, \bibinfo{person}{David Lo},
  \bibinfo{person}{Xin Xia}, \bibinfo{person}{Zhi-Yuan Wan}, {and}
  \bibinfo{person}{Jian-Ling Sun}.} \bibinfo{year}{2016}\natexlab{b}.
\newblock \showarticletitle{What security questions do developers ask? a
  large-scale study of stack overflow posts}.
\newblock \bibinfo{journal}{\emph{Journal of Computer Science and Technology}}
  \bibinfo{volume}{31}, \bibinfo{number}{5} (\bibinfo{year}{2016}),
  \bibinfo{pages}{910--924}.
\newblock


\bibitem[\protect\citeauthoryear{Ye, Xie, Zou, and Chen}{Ye
  et~al\mbox{.}}{2014}]%
        {ye2014interrogative}
\bibfield{author}{\bibinfo{person}{Ting Ye}, \bibinfo{person}{Bing Xie},
  \bibinfo{person}{Yanzhen Zou}, {and} \bibinfo{person}{Xiuzhao Chen}.}
  \bibinfo{year}{2014}\natexlab{}.
\newblock \showarticletitle{Interrogative-guided re-ranking for
  question-oriented software text retrieval}. In
  \bibinfo{booktitle}{\emph{Proceedings of the 29th ACM/IEEE international
  conference on Automated software engineering}}. \bibinfo{pages}{115--120}.
\newblock


\bibitem[\protect\citeauthoryear{Ye, Shen, Ma, Bunescu, and Liu}{Ye
  et~al\mbox{.}}{2016}]%
        {ye2016word}
\bibfield{author}{\bibinfo{person}{Xin Ye}, \bibinfo{person}{Hui Shen},
  \bibinfo{person}{Xiao Ma}, \bibinfo{person}{Razvan Bunescu}, {and}
  \bibinfo{person}{Chang Liu}.} \bibinfo{year}{2016}\natexlab{}.
\newblock \showarticletitle{From word embeddings to document similarities for
  improved information retrieval in software engineering}. In
  \bibinfo{booktitle}{\emph{Proceedings of the 38th international conference on
  software engineering}}. \bibinfo{pages}{404--415}.
\newblock


\bibitem[\protect\citeauthoryear{Zhang, Sheng, Lau, and Abebe}{Zhang
  et~al\mbox{.}}{2017}]%
        {zhang2017detecting}
\bibfield{author}{\bibinfo{person}{Wei~Emma Zhang}, \bibinfo{person}{Quan~Z
  Sheng}, \bibinfo{person}{Jey~Han Lau}, {and} \bibinfo{person}{Ermyas Abebe}.}
  \bibinfo{year}{2017}\natexlab{}.
\newblock \showarticletitle{Detecting duplicate posts in programming QA
  communities via latent semantics and association rules}. In
  \bibinfo{booktitle}{\emph{Proceedings of the 26th International Conference on
  World Wide Web}}. \bibinfo{pages}{1221--1229}.
\newblock


\bibitem[\protect\citeauthoryear{Zhang, Lo, Xia, and Sun}{Zhang
  et~al\mbox{.}}{2015}]%
        {zhang2015multi}
\bibfield{author}{\bibinfo{person}{Yun Zhang}, \bibinfo{person}{David Lo},
  \bibinfo{person}{Xin Xia}, {and} \bibinfo{person}{Jian-Ling Sun}.}
  \bibinfo{year}{2015}\natexlab{}.
\newblock \showarticletitle{Multi-factor duplicate question detection in stack
  overflow}.
\newblock \bibinfo{journal}{\emph{Journal of Computer Science and Technology}}
  \bibinfo{volume}{30}, \bibinfo{number}{5} (\bibinfo{year}{2015}),
  \bibinfo{pages}{981--997}.
\newblock


\bibitem[\protect\citeauthoryear{Zhou, Liu, Liu, Yang, and Grundy}{Zhou
  et~al\mbox{.}}{2019}]%
        {zhou2019deep}
\bibfield{author}{\bibinfo{person}{P. Zhou}, \bibinfo{person}{J. Liu},
  \bibinfo{person}{X. Liu}, \bibinfo{person}{Z. Yang}, {and}
  \bibinfo{person}{John~C. Grundy}.} \bibinfo{year}{2019}\natexlab{}.
\newblock \showarticletitle{Is Deep Learning Better than Traditional Approaches
  in Tag Recommendation for Software Information Sites?}
\newblock \bibinfo{journal}{\emph{Information and Software Technology}}
  \bibinfo{volume}{109} (\bibinfo{year}{2019}), \bibinfo{pages}{1--13}.
\newblock


\bibitem[\protect\citeauthoryear{Zou, Ye, Lu, Mylopoulos, and Zhang}{Zou
  et~al\mbox{.}}{2015}]%
        {zou2015learning}
\bibfield{author}{\bibinfo{person}{Yanzhen Zou}, \bibinfo{person}{Ting Ye},
  \bibinfo{person}{Yangyang Lu}, \bibinfo{person}{John Mylopoulos}, {and}
  \bibinfo{person}{Lu Zhang}.} \bibinfo{year}{2015}\natexlab{}.
\newblock \showarticletitle{Learning to rank for question-oriented software
  text retrieval (t)}. In \bibinfo{booktitle}{\emph{2015 30th IEEE/ACM
  International Conference on Automated Software Engineering (ASE)}}. IEEE,
  \bibinfo{pages}{1--11}.
\newblock


\end{thebibliography}

\end{document}